\documentclass[%
 reprint,
 amsmath,amssymb,
 aps,
]{revtex4-2}

\usepackage{float}
\usepackage{graphicx}
\graphicspath{{figures/}}
\usepackage{dcolumn}
\usepackage{amsmath}
\usepackage{bm}
\usepackage{appendix}
\usepackage[english]{babel}
\makeatletter
\adddialect\l@en\l@english
\makeatother
\usepackage{hyperref}
\usepackage{color}

\begin{document}

\title{Revealing topological protection through unprotected operations in a Majorana parity qubit}

\author{Sebastian Miles}
\email{sebastian.miles@rwth-aachen.de}
\author{Juan Daniel Torres Luna}
\author{Francesco Zatelli}
\author{Nick van Loo}
\author{Antonio Lombardi}
\author{Vincent P. M. Sietses}
\author{Florian J. Bennebroek Evertsz'}
\author{Bart Roovers}
\author{Leo P. Kouwenhoven}
\author{Michael Wimmer}
\email{m.t.wimmer@tudelft.nl}
\affiliation{QuTech and Kavli Institute of Nanoscience, Delft University of Technology, P.O. Box 4056, 2600 GA Delft, The Netherlands}

\date{October 12,  2026}

\begin{abstract}

A recent experiment reported coherent parity oscillations between
minimal Kitaev chains, demonstrating a
Majorana parity qubit~\cite{zatelli_majorana_2026}. Operations on this qubit
are controlled by coupling two Kitaev chains and are intrinsically unprotected
against noise in the coupling strength. Here we show how the exponential protection in the
underlying chains can nevertheless be revealed through unprotected qubit
operations. To this end, we develop an analytical theory of quasistatic
dephasing for two coupled $N$-site Kitaev chains at the sweet spot. In Rabi
experiments, protection can be resolved by adapting the operating point with
chain length: depending on the coupling architecture, it appears as an
exponentially increasing optimized number of coherent oscillations or as a
longer coherence time. Ramsey experiments show an exponentially increasing
coherence time, provided the chains can be sufficiently decoupled. Our
analytical theory takes into account the non-Gaussian splitting statistics of
the Kitaev chains, allowing us to quantitatively predict the scaling with
chain length. Moreover, we determine how residual noise and imperfections
limit the range of chain lengths over which this scaling can be observed. Our
results show how qubit dynamics can be used to probe Majorana protection in
experimentally realistic devices.
\end{abstract} 
\maketitle

\section{Introduction}

Encoding quantum information in spatially separated Majorana zero modes offers a
route to suppressing errors from local
perturbations~\cite{kitaev_unpaired_2001,nayak_non-abelian_2008,alicea_new_2012,beenakker_search_2013}.
In a finite wire supporting a topological phase, this protection manifests itself
in an overlap-induced energy splitting that is exponentially suppressed with the
separation of the states localized at the ends~\cite{kitaev_unpaired_2001,das_sarma_splitting_2012,rainis_towards_2013}.

Hybrid semiconductor-superconductor quantum dot arrays provide a tunable platform for investigating
this protection. These arrays implement the Kitaev
chain~\cite{sau_realizing_2012,leijnse_parity_2012,liu_tunable_2022} and allow
the chain length and local parameters to be controlled individually. Even a
two-site chain can host spatially separated Majorana modes at the so-called
sweet spot, although these ``poor man's Majoranas'' lack the same exponential 
protection against detuning as long topological wires~\cite{leijnse_parity_2012,tsintzis_majorana_2024}.
Experiments have realized such minimal chains in nanowires and two-dimensional
electron
gases~\cite{dvir_realization_2023,ten_haaf_two-site_2024,zatelli_robust_2024}.
Extending the chains to three sites has revealed enhanced stability against
local perturbations~\cite{bordin_enhanced_2025} and localized zero-energy end
states separated by a gapped bulk~\cite{ten_haaf_observation_2025}. 
Small energy splittings can remain unresolved in transport measurements because of, for example, thermal broadening~\cite{ten_haaf_observation_2025,bordin_probing_2026}.
This limits their ability to quantify the exactness of the degeneracy. 
Coherent qubit dynamics offers a complementary probe: small energy differences accumulate phase over time and can therefore become visible through dephasing.

The recent observation of coherent parity oscillations between two coupled
minimal Kitaev chains demonstrates that this approach is experimentally
accessible~\cite{zatelli_majorana_2026}.
In this parity-qubit layout, the joint
parity states of the chains encode the qubit, and a controllable interchain
coupling generates rotations between them~\cite{leijnse_parity_2012,tsintzis_majorana_2024,pan_rabi_2025}. 
The induced oscillations are sensitive to fluctuations of the coupling parameter which can obscure the increasing protection when making the chains longer. 
This motivates the central question of this work: \emph{Can qubit dynamics reveal signatures of topological protection in the underlying Kitaev chains, despite using unprotected operations?}

The suppression of dephasing with increasing Majorana separation has been
studied theoretically for low-frequency charge noise~\cite{knapp_dephasing_2018} and
quasistatic disorder~\cite{boross_dephasing_2022}. Proposals for measuring
Majorana-qubit coherence~\cite{aasen_milestones_2016}, including Ramsey-type
tests that distinguish topological from Andreev-state
qubits~\cite{mishmash_dephasing_2020}, establish how coherent evolution can
probe the protection of stored information. Rabi coherence has also been studied theoretically 
for rotations generated by tunneling between
nanowires~\cite{sau_capacitance-based_2025}. 
In Kitaev chain based parity qubits, finite-length effects and noise in the control junction must be treated
together. Ref.~\cite{pan_rabi_2025} addressed these effects in Rabi and Ramsey
dynamics for tunnel coupled two-site chains, including charge noise, leakage, and fluctuations
of the interchain coupling. 
Very recently, Pan and Das Sarma showed that extending two-site
chains to three sites does not automatically improve coherence~\cite{pan_scaling_2026},
emphasizing the importance of our central question.

In this work, we show how unprotected qubit operations can reveal the 
exponential protection of the underlying Kitaev chains, and identify the 
operating conditions needed to observe it.
To this end, we develop an analytical theory of quasistatic dephasing that connects the
experimentally realized two-site parity qubit to chains of arbitrary length,
tuned to the sweet spot. After introducing the model and noise framework in
Secs.~\ref{sec:model} and~\ref{sec:noise}, we analytically describe the
competing dephasing mechanisms in the minimal qubit across the relevant coupling
regimes (Sec.~\ref{sec:minimal}). This analytical theory reproduces key behavior
of the experiment~\cite{zatelli_majorana_2026}. For longer chains, we obtain
a semi-analytical description of the splitting distribution
(Sec.~\ref{sec:splitting_statistics}), accounting for the non-Gaussian
statistics needed to predict the length dependence of coherence.

We show that observables which quantify the protection of a chain depend on the
specific coupling mechanism and operating conditions. In Rabi experiments, protection
can appear as an increasing number of coherent oscillations at the optimal
operating points or as a
longer coherence time (Sec.~\ref{sec:rabi}). Ramsey experiments probe the
protection of stored information, provided residual coupling between the chains
is sufficiently suppressed (Sec.~\ref{sec:ramsey}). We predict how these
signatures evolve with chain length and the experimental conditions needed to
resolve them.

\section{Model and projected Hamiltonian} \label{sec:model}

We consider a parity qubit formed by two coupled hybrid quantum dot-superconductor arrays.
We describe each array as an effective spinless
Kitaev chain with nearest-neighbor hopping and pairing.
We assume a fixed, uniform superconducting phase throughout the device,
so that the phase difference between the chains is $\phi=0$, and choose
a gauge in which the pairing amplitudes are real.
The resulting Hamiltonian is
\begin{align} \label{model:MZM_hamiltonian}
H &= H_L + H_R + H_C^{(i)} \\
H_\alpha
  &=\sum_{i=1}^{N}\mu_{\alpha,i}c_{\alpha,i}^\dagger c_{\alpha,i}\nonumber \\
  &+\sum_{i=1}^{N-1}
    \begin{aligned}[t]
      \Big(&t_{\alpha,i}c_{\alpha,i}^\dagger c_{\alpha,i+1} \\
      {}+{}&
      \Delta_{\alpha,i}c_{\alpha,i}^\dagger c_{\alpha,i+1}^\dagger+\text{h.c.}\Big),
    \end{aligned} \nonumber
\end{align}
where $c_{\alpha,i}$ annihilates a spinless electron on site $i$ of chain
$\alpha=L,R$, and $c_D$ annihilates an electron on the central dot.
The parameters $\mu_{\alpha,i}$ and $\mu_D$ are the corresponding chemical
potentials in the chains and central dot.
The intrachain couplings $t_{\alpha,i}$ and $\Delta_{\alpha,i}$ describe
elastic cotunneling (ECT) and crossed Andreev reflection (CAR) on the bond
between sites $i$ and $i+1$, respectively\cite{liu_tunable_2022}.
Finally, $H_C^{(i)}$ describes the coupling between the chains for which we consider two configurations.

First, we consider the device geometry of Refs.~\cite{zatelli_majorana_2026,leijnse_parity_2012}, which couples the two Kitaev chains through a normal quantum dot.
The corresponding coupling Hamiltonian is
\begin{align} \label{model:h_dot}
    H_C^{\mathrm{dot}} =  \mu_D c_D^\dagger c_D + t_{LD} c_{L,N}^\dagger c_D + t_{RD} c_D^\dagger c_{R,1} + h.c.,
\end{align}
where $\mu_D$ is the chemical potential of the central dot, and the tunneling amplitudes $t_{LD}$ and $t_{RD}$ describe normal tunneling between
the central dot and the inner dots, i.e. sites $(L,N)$ and $(R,1)$, respectively.

In Sec.~\ref{sec:rabi_barrier}, we also consider the case of normal tunnel coupling as in Ref.\cite{pan_rabi_2025} where the coupling is
\begin{align} \label{model:h_barrier}
    H_C^{\mathrm{barrier}} = \tau_0 c_{L,N}^\dagger c_{R,1} + h.c.,
\end{align}
with $\tau_0$ being the tunnel amplitude.

\begin{figure}
    \centering
    \includegraphics[width=\linewidth]{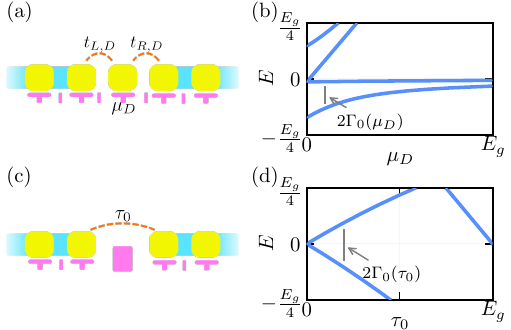}
    \caption{Schematic of the Majorana parity qubit coupled via a tunable quantum dot weak link, (a), and a tunnel barrier, (c).
    Each chain consists of $N$ quantum dots coupled via cross Andreev reflection and elastic cotunneling, and is tuned to the uniform sweet spot.
    (b) shows the spectrum of the Majorana parity qubit when coupled through a quantum dot, (a), at $t_{L,D}=t_{R,D}$; (d) shows a spectrum for the tunnel coupled system.
    Oscillations between the two lifted groundstates are induced by the $\mu_D>0$ dependent transverse coupling $\Gamma_0(\mu_D)$.}
    \label{fig:figure_1}
\end{figure}

We show schematics for both configurations in Fig.~\ref{fig:figure_1} (a) and (c).
We furthermore show spectra of the low energy states depending on the control parameters $\mu_D$ and $\tau_0$ in Fig.~\ref{fig:figure_1} (b) and (d), respectively.

\paragraph{Qubit subspace}

At the sweet spot, $\mu_{\alpha,i}=0$ and
$t_{\alpha,i}=\Delta_{\alpha,i}=t_\alpha=\Delta_\alpha>0$, each isolated
chain hosts two Majorana edge states, $\eta_\alpha$ and $\zeta_\alpha$, through two
degenerate many-body ground states with opposite fermion
parity\cite{leijnse_parity_2012,sau_realizing_2012,kitaev_unpaired_2001}.
For both chains, $\eta_\alpha$ describes the zero mode amplitude on the inner dots, while $\zeta_\alpha$ describes the amplitude of the outer Majoranas on the inner dots.
At the sweet spot, they are localized with $\eta_{L,N}=\eta_{R,1}=1$ and $\zeta_{L,N}=\zeta_{R,1}=0$.
The gap to bulk excitations is $E_{\alpha,g}=2t_\alpha$.
We denote the even- and odd-parity ground states by $|e\rangle_\alpha$
and $|o\rangle_\alpha$. For identical clean chains, we write
$E_{L,g}=E_{R,g}=E_g$, so that the nominal hopping and pairing
amplitudes are $E_g/2$.
Conservation of total fermion parity restricts the qubit to one
of two sectors, labeled by $p=+1$ (even) or $p=-1$ (odd).
In the even sector, we define the qubit states as 
\begin{align}
 |0\rangle=|e\rangle_L|e\rangle_R=|ee\rangle, \nonumber \\
 |1\rangle=|o\rangle_L|o\rangle_R=|oo\rangle,
 \label{model:states_even}
\end{align}
which, for the case of coupling through $H_C^{\mathrm{dot}}$, have to be understood as $(|ee\rangle,|oo\rangle)\otimes |0\rangle_D$.
In the odd sector, we use the analogous $(|eo\rangle,|oe\rangle)$, with the empty central dot implicit.

For weak coupling through either $H_C^{\mathrm{barrier}}$ or $H_C^{\mathrm{dot}}$ with $\mu_D$ off-resonant, we can eliminate the coupling and project the full Hamiltonian onto the qubit subspace perturbatively\cite{lowdin_note_1951,bravyi_schriefferwolff_2011,araya_day_pymablock_2025}.
Up to scalar energy shifts, the projected Hamiltonian is
\begin{equation}
 H_{\mathrm q}^{(p)}
 =\varepsilon\sigma_z+\Gamma\sigma_x.
 \label{model:h_pq_general}
\end{equation}
Here $\varepsilon$ and $\Gamma$ are the longitudinal and transverse
qubit coefficients, respectively.
We neglect Coulomb corrections in the regime discussed in
App.~\ref{app:coulomb}.

\paragraph{Longitudinal coupling}

We define the energy splitting of a single chain as
\begin{equation}
 \delta\epsilon_{\alpha,N}=E_{e,\alpha,N}-E_{o,\alpha,N}.
 \label{model:chain_splitting}
\end{equation}
It vanishes at the sweet spot, but may be nonzero under noise.
The two chains together contribute to the longitudinal qubit coefficient
\begin{equation}
\delta\varepsilon_N
 =\frac{\delta\epsilon_{L,N}+p\,\delta\epsilon_{R,N}}{2}.
 \label{model:sum_epsilon}
\end{equation}

Eliminating the central dot also induces corrections to the onsite energies,
which are suppressed at large dot detuning $\mu_D$. Their static contributions
can in principle be compensated by gate voltages when interaction effects are sufficiently negligible.
We neglect these corrections and take $\varepsilon\simeq\delta\varepsilon_N$.

\paragraph{Transverse coupling}

The transverse coupling $\Gamma$ depends on the specifics of the coupling.
In each coupling geometry, we choose the relative phase of the logical basis
states such that the reference transverse coupling is positive.

For $H_C^{\mathrm{dot}}$, the middle dot permits tunneling between the chains and acts as a
tunable weak link (Fig.~\ref{fig:figure_1} (a)). 
In the far-detuned regime, projection onto the chain ground states gives the transverse coupling as~\cite{zatelli_majorana_2026}
\begin{equation}
 \Gamma=\frac{t_{LD}t_{RD}}{2\mu_D} (\eta_L\eta_R-p\zeta_L\zeta_R),
 \label{model:transverse_couplings}
\end{equation}
where $t_{LD}$ and $t_{RD}$ are the tunneling amplitudes between the dot and the inner dots of the chains, and $\mu_D$ is the dot detuning.

Similarly, $H_C^{\mathrm{barrier}}$ describes direct tunneling between the chains.
In the weak tunneling limit, an analogous projection on the qubit states gives
\begin{align}
    \Gamma_{\mathrm{barrier}} =\frac{\tau_0}{2} (\eta_L \eta_R - p \zeta_L \zeta_R).
\end{align}
In both expressions, both $\eta_\alpha$ and $\zeta_\alpha$ enter the transverse coupling, describing the inner-inner and outer-outer Majorana couplings.
Both amplitudes depend on local chain parameters\cite{leumer_exact_2020}.
Only $\zeta_\alpha$ is suppressed with chain length when a chain is detuned from the sweet spot\cite{liu_scaling_2025,leumer_exact_2020}; see also App.~\ref{app:splitting}.
We neglect the remote-overlap term proportional to $\zeta_L\zeta_R$.
We collect the transverse coupling as
\begin{equation}
 \Gamma= \Gamma_0 \eta_L \eta_R,
 \label{model:gamma}
\end{equation}
with the length independent $\Gamma_0=t_{LD}t_{RD}/(2\mu_D)>0$ for $\Gamma_{\mathrm{dot}}$ and $\Gamma_0 = \tau_0/2>0$ for $\Gamma_{\mathrm{barrier}}$.

In the presence of noise, we write $\Gamma=\Gamma_0+\delta\Gamma$, where $\delta\Gamma$ includes both fluctuations and any noise-induced mean shift.
With these approximations, the effective Hamiltonian becomes
\begin{equation}
 H_{\mathrm q}
 =(\Gamma_0+\delta\Gamma)\sigma_x+\delta\varepsilon_N\sigma_z.
 \label{model:h_pq}
\end{equation}


\section{Dephasing under quasistatic noise} \label{sec:noise}

To analyze the impact of noise on the qubit we apply the quasistatic approximation where we neglect dynamical fluctuations during the time of the experiment~\cite{ithier_decoherence_2005,mishmash_dephasing_2020}.

\subsection{Quasistatic noise model}
\label{subsec:quasistatic_noise_model}

We model quasistatic noise by shifting the microscopic parameters,
\begin{align*}
 \mu_{\alpha,i}&\rightarrow\mu_{\alpha,i}+\delta\mu_{\alpha,i},\\
 (t_{\alpha,i}-\Delta_{\alpha,i})
 &\rightarrow (t_{\alpha,i}-\Delta_{\alpha,i})
 +\delta(t_{\alpha,i}-\Delta_{\alpha,i}),
\end{align*}
with independent Gaussian fluctuations
\begin{align*}
 \delta\mu_{\alpha,i}&\sim\mathcal N(0,\sigma_\mu^2),\\
 \delta(t_{\alpha,i}-\Delta_{\alpha,i})
 &\sim\mathcal N(0,\sigma_{t-\Delta}^2).
\end{align*}
We take the noise widths to be equal for both chains.
We furthermore restrict ourselves to weak noise much smaller than the gap $E_g$ for leakage outside the qubit subspace to be suppressed.
Aggregating the microscopic noise to the parameters of Eq.~\eqref{model:h_pq}, $\delta\varepsilon$ and $\delta\Gamma$ become non-Gaussian variables that are in principle correlated, 
e.g. through onsite fluctuations entering in $\delta \varepsilon_N$ and in $\Gamma$ through $\eta_\alpha$.
For the analytics, we neglect these correlations and treat $\delta\varepsilon$ and $\delta\Gamma$ as independent variables,
but keep them in the numerics.

\subsection{Rabi return probability}
\label{subsec:rabi_return_probability}

Starting from $|0\rangle\equiv|ee\rangle$, the return probability averaged over quasistatic trajectories is
\begin{align} \label{quasi:expintegral}
    P(t)&=\iint |\langle 0|U(t;\varepsilon,\Gamma)|0\rangle|^2
    p_{\varepsilon,\Gamma}(\delta\varepsilon,\delta\Gamma)
    \,d\delta\varepsilon\,d\delta\Gamma,
\end{align}
where $U(t;\varepsilon,\Gamma)=\exp[-iH_{\mathrm q}t/\hbar]$ and $p_{\varepsilon,\Gamma}(\delta\varepsilon,\delta\Gamma)$ is the probability distribution of $\delta\varepsilon$ and $\delta\Gamma$.

Using Eq.~\eqref{model:h_pq}, the average becomes
\begin{align} \label{quasi:abstract_return}
    P(t)=1-\frac{\langle W\rangle}{2}+\frac{1}{2}\operatorname{Re}\!\left\langle We^{-i2\Omega t/\hbar}\right\rangle,
\end{align}
where 
\begin{align}
    \Omega &= \sqrt{(\Gamma_0+\delta\Gamma)^2+\delta\varepsilon^2}, \label{quasi:omega}
\end{align}
is half the qubit-level splitting and
\begin{align}
    W &= \frac{(\Gamma_0+\delta\Gamma)^2}{(\Gamma_0+\delta\Gamma)^2+\delta\varepsilon^2} \label{quasi:weight}
\end{align}
is the return visibility of a single trajectory.

Equation~\eqref{quasi:abstract_return} motivates two quantities that can be extracted from parity-oscillation measurements:
\begin{align}
    \mathcal{C}(t) &= \frac{\left|\left\langle We^{-i2\Omega t/\hbar}\right\rangle\right|}{\langle W\rangle}, \label{quasi:contour}\\
    f_q &= \frac{\Gamma_0}{\pi\hbar}. \label{quasi:frequency}
\end{align}
The \emph{envelope} $\mathcal{C}(t)$ measures the loss of phase coherence between trajectories caused by fluctuations of both the chains and the central-dot coupling.
The \emph{frequency} $f_q$ is the oscillation frequency due to the nominal transverse coupling $\Gamma_0$.

We define the \emph{dephasing time} $T_\varphi$ as the first time at which $\mathcal{C}(t)$ crosses $e^{-1}$, and define the dimensionless \emph{number of visible return cycles} as an observable quality metric
\begin{align} \label{quasi:quality}
    N_\varphi=f_qT_\varphi.
\end{align}
It permits a direct comparison of different chain lengths and operating points while penalizing systematic detuning of the Kitaev chains away from the sweet spot.
We finally define the \emph{visibility}
\begin{align} \label{quasi:visibility}
    \mathcal{V}=\langle W\rangle 
\end{align}
as the average return weight, which quantifies the reduction of the return amplitude, for convenience.
With these definitions, the averaged return probability becomes
\begin{equation}
 P(t)=1-\frac{\mathcal V}{2}
 +\frac{\mathcal V}{2}\mathcal C(t)\cos\Phi(t),
 \label{quasi:return_envelope}
\end{equation}
where $\Phi(t)$ is the phase of the averaged oscillation, defined by $
 \left\langle We^{-2i\Omega t/\hbar}\right\rangle
 =\mathcal V\,\mathcal C(t)e^{-i\Phi(t)}.$

The resulting qubit-noise distribution determines the ensemble averages in
Eqs.~\eqref{quasi:abstract_return}, \eqref{quasi:contour}, and
\eqref{quasi:visibility}. These form the basis of our subsequent calculations.

\subsection{Ramsey return probability}
\label{subsec:ramsey_return_probability}

In a Ramsey experiment, two  $\pi/2$ pulses about $x$
enclose a waiting interval of duration $t$. 
We take the pulses to rotate the state in the same direction and set the interchain coupling to zero during the waiting period,
so that $H_{\mathrm q}=\delta\varepsilon_N\sigma_z$.
With ideal pulses and decoupled chains during the waiting interval,
every realization has unit contrast, so the Ramsey average contains 
no visibility weight $W$.
For our splitting-noise model, the phase average is real and positive
(App.~\ref{app:ramsey_signal}). The Ramsey envelope is therefore
\begin{align}
 \mathcal C_{\mathrm{Ramsey}}(t)
 &=\left|\left\langle e^{-2i\delta\varepsilon_Nt/\hbar}\right\rangle\right|
 \nonumber\\
 &=\left\langle\cos(2\delta\varepsilon_Nt/\hbar)\right\rangle,
 \label{quasi:ramsey_envelope}
\end{align}
and the averaged return probability for an initial state $|0\rangle$ is
\begin{equation}
 \overline P_{\mathrm{Ramsey}}(t)
 =\frac{1-\mathcal C_{\mathrm{Ramsey}}(t)}{2}.
 \label{quasi:ramsey_return}
\end{equation}
For our pulse sequence, the return probability is zero at $t=0$.

As for Rabi oscillations, we define the dephasing time
$T_\varphi^{\mathrm{Ramsey}}$ by the first crossing
$\mathcal C_{\mathrm{Ramsey}}(T_\varphi^{\mathrm{Ramsey}})=e^{-1}$.
Importantly, for two decoupled Kitaev chains at the sweet spot, $N_{\varphi}\equiv 0$ for Ramsey experiments, since the oscillation frequency is zero during the waiting interval.

\section{The minimal parity qubit} \label{sec:minimal}

We first consider the minimal parity qubit formed by two $N=2$ chains,
as realized in Ref.~\cite{zatelli_majorana_2026}, and study its behavior
under quasistatic noise in different coupling regimes. The derivations
are collected in App.~\ref{app:n2}.

\subsection{Splitting and noise dependence}
\label{subsec:minimal_splitting}

For a single chain at the Kitaev sweet spot with quasistatic noise, the energy splitting is given by~\cite{leijnse_parity_2012}
\begin{equation}
 \delta\epsilon_{\alpha,2}\simeq
 \delta(t_\alpha-\Delta_\alpha)
 -\frac{\delta\mu_{\alpha,1}\delta\mu_{\alpha,2}}{E_g},
 \label{minimal:h2}
\end{equation}
where $\delta\mu_{\alpha,i}$ and $\delta(t_{\alpha,i}-\Delta_{\alpha,i})$
are the noise-induced shifts discussed in Sec.~\ref{sec:noise}.
Bond mismatch enters $\delta\varepsilon_{\alpha,2}$ linearly, whereas onsite noise contributes
through the product of the two site potentials. The qubit splitting is then
$\delta \varepsilon_2 = (\delta\epsilon_{L,2}+p\,\delta\epsilon_{R,2})/2$ following Eq.~\eqref{model:sum_epsilon}.

In this expression, each bond contributes a Gaussian variable, while
each contribution from onsite fluctuations is a product of two independent Gaussian variables.
The resulting probability distribution of $\delta\varepsilon_2$ then becomes
a convolution of Gaussian and Laplace distributions, making it
explicitly non-Gaussian. We derive its characteristic function
and probability density in App.~\ref{app:n2}.
The splitting distribution is identical in both global-parity sectors.

The relative strength of the onsite and bond contributions is expressed as
\begin{equation}
 u=\frac{\sigma_\mu^2}{E_g\sigma_{t-\Delta}}.
 \label{minimal:noise_ratio}
\end{equation}
The variance of the probability distribution is
\begin{equation}
 \sigma_{\varepsilon,2}^2
 =\frac{\sigma_{t-\Delta}^2}{2}+\frac{\sigma_\mu^4}{2E_g^2}
 =\frac{\sigma_{t-\Delta}^2}{2}(1+u^2).
 \label{minimal:sigma_epsilon}
\end{equation}
We write $\delta\varepsilon=\delta\varepsilon_2$ and
$\sigma_\varepsilon=\sigma_{\varepsilon,2}$ in the following $N=2$ analysis.

We first focus on noise through $\mu_D$ alone.
Noise contributions from fluctuating interface amplitudes
$\eta_\alpha$ and tunnel couplings $t_{\alpha,D}$ will be considered systematically in Sec.~\ref{sec:rabi_optimization_limits}.
Coupling noise then arises only from the independent middle-dot fluctuation $\delta\mu_D$.
For $\sigma_{\mu_D}\ll\mu_D$, linearizing the coupling in
Eq.~\eqref{model:gamma} gives
\begin{equation}
 \delta\Gamma\simeq-\Gamma_0\frac{\delta\mu_D}{\mu_D},
 \qquad
 \sigma_\Gamma=\Gamma_0\frac{\sigma_{\mu_D}}{\mu_D},
 \label{minimal:coupling_noise}
\end{equation}
making $\delta\Gamma$ Gaussian and independent of $\delta\varepsilon$.

\subsection{Strong-coupling regime}
\label{subsec:strong}

For $\sigma_\varepsilon,\sigma_\Gamma\ll\Gamma_0$, we are in the so-called
strong-coupling regime. The coupling then dominates both the oscillation
frequency and the rotation axis and we can expand Eq.~\eqref{quasi:omega} and Eq.~\eqref{quasi:weight}

\begin{align}
    \Omega & \simeq \Gamma_0+\delta\Gamma+\frac{\delta\varepsilon^2}{2\Gamma_0}, \label{strong:omega}\\
    W &\simeq 1-\frac{\delta\varepsilon^2}{\Gamma_0^2}. \label{strong:visibility}
\end{align}

Considering Eq.~\eqref{strong:omega}, coupling noise enters linearly, whereas splitting noise enters quadratically.
This quadratic sensitivity is a consequence of the qubit sweet-spot which suppresses dephasing from low-frequency noise~\cite{makhlin_dephasing_2004,ithier_decoherence_2005}.
Increasing $\Gamma_0$ further suppresses the quadratic contribution.
Similarly, considering Eq.~\eqref{strong:visibility}, the visibility remains close to unity up to a negligible correction.
Under these conditions, the independent noise averages in Eq.~\eqref{quasi:abstract_return}
factorize as
\begin{equation}
 \mathcal C(t)\simeq
 e^{-2\sigma_\Gamma^2t^2/\hbar^2}\left|\psi_2(s;u)\right|,
 \qquad s=\frac{\sigma_\varepsilon^2t}{\Gamma_0\hbar}.
 \label{minimal:strong_envelope}
\end{equation}
The Gaussian suppression stems from coupling noise. The second factor depends
on the full splitting distribution and can also be evaluated analytically
(App.~\ref{app:n2_strong}):
\begin{align}
 \psi_2(s;u)
 &=\left\langle
 e^{-is\delta\varepsilon^2/\sigma_\varepsilon^2}
 \right\rangle \nonumber\\
 &=\sqrt{\frac{\pi(1+u^2)}{2is\,u^2}}\,
 \operatorname{erfcx}\!\left[
 \sqrt{\frac{1+u^2+2is}{2is\,u^2}}\right],
 \label{minimal:strong_psi}
\end{align}
where $\operatorname{erfcx}(z)=e^{z^2}\operatorname{erfc}(z)$ is the scaled complementary error function.
For comparison, Gaussian splitting noise of the same variance gives
$\psi_{\mathrm G}(s)=(1+2is)^{-1/2}$. The non-Gaussian contribution
from onsite noise therefore distinguishes the minimal parity qubit from
a double-dot charge qubit with Gaussian detuning
noise~\cite{petersson_quantum_2010}. For both, the present parity qubit and charge qubits, the splitting-noise envelopes nevertheless decay as $t^{-1/2}$ at long times, slower than a typical Gaussian envelope.

\subsubsection{Coupling-noise-dominated dephasing}

At small detuning, $\sigma_\Gamma\gg\sigma_\varepsilon^2/(2\Gamma_0)$ and coupling
fluctuations through Eq.~\eqref{minimal:coupling_noise} dominate the quadratic, $\delta\varepsilon^2$, contribution in
Eq.~\eqref{strong:omega}. 
We can then set $|\psi_2(s;u)|\simeq1$ in Eq.~\eqref{minimal:strong_envelope}.
Using Eq.~\eqref{minimal:coupling_noise}, the dephasing time then follows from the Gaussian integral as
\begin{equation}
 T_\varphi\simeq\frac{\hbar}{\sqrt{2}\sigma_\Gamma}
 =\frac{\hbar\mu_D}{\sqrt{2}\Gamma_0\sigma_{\mu_D}}.
 \label{minimal:strong_linear_dephasing}
\end{equation}
Since $\Gamma_0\propto\mu_D^{-1}$, increasing the middle-dot detuning reduces the effect of coupling noise and gives $T_\varphi\propto\mu_D^2$. 
Together with $f_q\propto\mu_D^{-1}$, this gives the scaling of the number of return cycles, $N_\varphi\propto\mu_D$.

\subsubsection{Splitting-noise-dominated dephasing}

Increasing $\mu_D$, the noise width in Eq.~\eqref{minimal:coupling_noise} can become smaller than the noise width of the quadratic splitting.
This corresponds to $\sigma_\Gamma\ll\sigma_\varepsilon^2/(2\Gamma_0)$ and we can approximate Eq.~\eqref{minimal:strong_envelope} as $\mathcal C(t)\simeq|\psi_2(s;u)|$. 
The dephasing time is then given by
\begin{equation}
 T_\varphi\simeq\frac{\hbar\Gamma_0}{\sigma_\varepsilon^2}
 s_{1/e}(u),
 \label{minimal:strong_quadratic_dephasing}
\end{equation}
where $s_{1/e}(u)$ is the solution of
$|\psi_2(s;u)|=e^{-1}$, which we determine numerically.
However, this dimensionless factor is independent of $\mu_D$.
Increasing the detuning reduces the coupling $\Gamma_0$ and hence
increases the effect of splitting noise.
The dephasing time then scales as $T_\varphi\propto\mu_D^{-1}$.
Consequently, the number of return cycles scales as
$N_\varphi\propto\mu_D^{-2}$.

\subsection{Crossover to weak coupling}
\label{subsec:minimal_crossover}

As the middle-dot detuning increases, $\Gamma_0$ becomes comparable to $\sigma_\varepsilon$, and the strong-coupling expansion no longer applies.
The rotation axis now varies appreciably between noise realizations, signaling the onset of the weak coupling regime.
This reduces the visibility and generally contributes an additional
time-dependent phase.
For $\sigma_\Gamma\ll\Gamma_0$, we retain the full splitting dependence
and average over the linear coupling-noise response without expanding
the accumulated phase. The resulting integral and its validity conditions
are derived in App.~\ref{app:n2_crossover}.

Dephasing due to coupling noise is negligible for times $t\ll\hbar/\sigma_\Gamma$.
The statistics are then determined by the splitting distribution.
Following Ref.~\cite{koppens_universal_2007}, the long-time behavior is then determined by realizations near $\delta\varepsilon=0$, where the oscillation frequency is stationary. 
Since $p_\varepsilon(0)$ is finite, a stationary-phase expansion gives
\begin{align}
 P(t)&\simeq1-\frac{\mathcal V}{2}
 +\frac{p_\varepsilon(0)}{2}\sqrt{\frac{\pi\Gamma_0\hbar}{t}} \cos\!\left(\frac{2\Gamma_0t}{\hbar}+\frac{\pi}{4}\right).
 \label{minimal:crossover_tail}
\end{align}
Thus, the envelope decays as $t^{-1/2}$, while the oscillations approach
the bare carrier frequency $f_q=\Gamma_0/(\pi\hbar)$, with a constant phase shift of $\pi/4$ \cite{koppens_universal_2007}.
This asymptotic frequency motivates our choice of $f_q$ as the reference frequency in
the cycle count $N_\varphi = f_q T_\varphi$ [Eq.~\eqref{quasi:quality}], even
beyond the strong-coupling regime.

The mechanism of Ref.~\cite{koppens_universal_2007} applies here despite the non-Gaussian splitting 
distribution, with the sole additional contribution of $p_\varepsilon(0)$. For finite coupling noise 
in the crossover regime, $\sigma_\varepsilon\sim\Gamma_0$, this behavior applies only in the time window
$\hbar/\Gamma_0\ll t\ll\hbar/\sigma_\Gamma$, before fluctuations due to the coupling
yield appreciable dephasing contributions.

\subsection{Weak-coupling regime}
\label{subsec:weak_coupling}

Detuning $\mu_D$ further, we eventually reach the regime where $\Gamma_0\ll\sigma_\varepsilon$.
Then, Eq.~\eqref{quasi:weight} is sharply peaked and selects nearly resonant realizations with $|\delta\varepsilon|\lesssim\Gamma_0$~\cite{koppens_universal_2007}.
The splitting density varies little across this narrow range, so we can
replace $p_\varepsilon(\delta\varepsilon)$ by $p_\varepsilon(0)$ and obtain
$\mathcal V\simeq\pi\Gamma_0p_\varepsilon(0)$, with $p_\varepsilon(0)$
given in Eq.~\eqref{n2:density_zero}.
Meanwhile, $\sigma_\Gamma/\Gamma_0=\sigma_{\mu_D}/\mu_D$ decreases with detuning,
and coupling noise becomes negligible on the timescale
$t\sim\hbar/\Gamma_0$.
The normalized envelope gives the dephasing time and cycle count
\begin{align}
 T_\varphi&\simeq\frac{\hbar\tau_{1/e}}{2\Gamma_0}
 \propto\mu_D,
 \label{minimal:weak_dephasing}\\
 N_\varphi&\simeq\frac{\tau_{1/e}}{2\pi}\approx0.703,
 \label{minimal:weak_cycles}
\end{align}
where $\tau_{1/e}\approx4.416$ is a universal constant derived in
App.~\ref{app:n2_weak}. Increasing the middle-dot detuning therefore
lengthens the dephasing time without increasing the cycle count.
At fixed chain noise, the visibility decreases as
$\mathcal V\propto\mu_D^{-1}$, so the longer lifetime comes with a
smaller oscillation amplitude.

\subsection{Overview of the coupling regimes}
\label{subsec:minimal_overview}
\label{subsec:optimal_q_2}

\begin{figure*}[t]
 \centering
 \includegraphics[width=\textwidth]{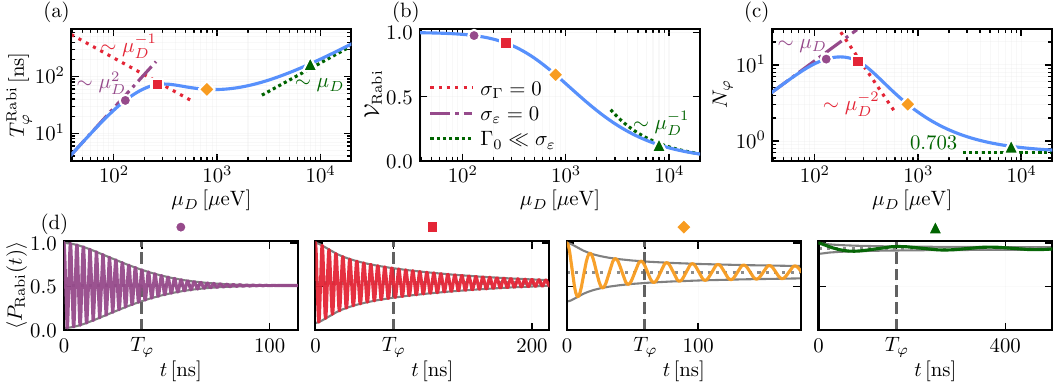}
 \caption{Noise regimes of the minimal parity qubit for $u=1.125$.
  (a) Dephasing time $T_\varphi$, (b) visibility $\mathcal V$, and
  (c) number of visible return cycles $N_\varphi$ depending on middle-dot
 detuning $\mu_D$. Solid blue curves are obtained by evaluating Eq.~\eqref{n2:crossover_average}.
(Dash-)dotted lines show the asymptotics for the $\sigma_\Gamma$ dominated, purple, and $\sigma_\varepsilon$ dominated, red, strong-coupling limits.
 Green dotted lines show the weak-coupling asymptotes for $\Gamma_0\ll\sigma_\varepsilon$. 
 (d) Noise-averaged return probabilities at the colored symbols indicated
 in (a)--(c), showing oscillations in the different coupling regimes.
 Black curves bound the oscillations as
 $1-\mathcal V/2\pm\mathcal V\mathcal C(t)/2$; horizontal dotted lines
 indicate $1-\mathcal V/2$, and vertical dashed lines mark $T_\varphi$.
 Both chains have $N=2$, $t=\Delta=40\,\mu\mathrm{eV}$
 ($E_g=80\,\mu\mathrm{eV}$), and $t_{LD}=t_{RD}=13\,\mu\mathrm{eV}$.
 The noise widths are $\sigma_\mu=3\,\mu\mathrm{eV}$,
 $\sigma_{t-\Delta}=0.1\,\mu\mathrm{eV}$, and
 $\sigma_{\mu_D}=2\,\mu\mathrm{eV}$; fluctuations of the tunnel links
 and interface amplitudes are neglected.
 }
 \label{fig:n2_overview}
\end{figure*}

Figure~\ref{fig:n2_overview} brings together the different coupling regimes. 
The solid curves are computed from the crossover average in
Eq.~\eqref{n2:crossover_average}, normalized as in Eq.~\eqref{quasi:contour},
while the asymptotes show the limiting behavior derived above.
For small $\mu_D$, middle-dot noise limits coherence, giving
$T_\varphi\propto\mu_D^2$ and $N_\varphi\propto\mu_D$.
Increasing $\mu_D$ suppresses coupling noise with respect to the
splitting noise. When splitting noise dominates, we found
$T_\varphi\propto\mu_D^{-1}$ and $N_\varphi\propto\mu_D^{-2}$.
For the parameters shown, the competition between these two limiting
behaviors produces local maxima in $T_\varphi$ and $N_\varphi$ at
intermediate detuning.
Further increasing $\mu_D$ eventually leads to the weak-coupling regime,
where $T_\varphi\propto\mu_D$ and $N_\varphi\simeq0.703$, but a strongly
reduced visibility.

For $N=2$ and experimentally relevant parameters,
it is difficult to satisfy the conditions for both strong
coupling and the splitting-noise-dominated regime,
$\sigma_\varepsilon\ll\Gamma_0$ and
$\sigma_\Gamma\ll\sigma_\varepsilon^2/(2\Gamma_0)$. Hence,
the splitting-noise-dominated strong-coupling asymptote agrees less well
with the full calculation than the other asymptotes. In particular,
$T_\varphi$ does not need to develop a clear local maximum as it eventually
increases again in the weak-coupling regime. A further example without
a clear maximum is shown in Fig.~\ref{fig:rabi_optimization}(a).
In contrast, $N_\varphi$ tends to show a local maximum, as it converges
to a constant smaller than one in the weak-coupling regime.

These findings reproduce the qualitative behavior observed in
Ref.~\cite{zatelli_majorana_2026}: the number of coherent oscillations initially
increases approximately linearly with middle-dot detuning, reaches a maximum,
and shows signs of a decrease at larger detuning.

Figure~\ref{fig:n2_overview}(d) illustrates the corresponding average
return probabilities $P(t)$ in the different regimes.
The strong-coupling traces have high contrast,
transitioning from a Gaussian decay to a slower power-law decay as splitting noise
becomes dominant (left two panels). In the crossover, both the contrast and
cycle count are reduced (third panel). The weak-coupling trace decays more slowly
in physical time but has only a small oscillating component (right panel).
Both of these traces show the characteristic power-law decay.

\section{Energy-splitting statistics}
\label{sec:splitting_statistics}

We now extend the $N=2$ analysis to longer chains. To this end,
we need the statistics of the single-chain splitting $\delta\epsilon_N$
under quasistatic noise. We first motivate why the non-Gaussian nature
of the splitting distribution becomes even more important for longer chains,
and then derive its statistics analytically.

\subsection{Behavior of energy splittings with chain length}
\label{sec:splitting_rms_typical}

\begin{figure}[t]
 \centering
 \includegraphics[width=\columnwidth]{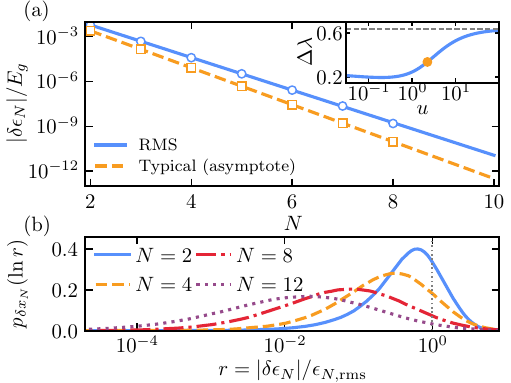}
 \caption{Splitting suppression with length and probability distributions for
 $E_g=40\,\mu\mathrm{eV}$, $\sigma_\mu=3\,\mu\mathrm{eV}$, and
 $\sigma_{t-\Delta}=0.1\,\mu\mathrm{eV}$, giving $u=2.25$. 
 (a) Root-mean-square (RMS), Eq.~\eqref{main:splitting_rms}, and geometric-mean splitting, Eq.~\eqref{main:splitting_typical}, magnitudes in units of $E_g$.
 The solid curve gives the finite-length RMS from the recurrence, and the
 dashed curve gives the typical-splitting asymptote including its boundary
 prefactor.
 Circles and squares indicate RMS and geometric means from sampling isolated-chain spectra using exact diagonalization through $N=8$.
 The inset shows the exponent difference $\Delta\lambda(u)$, Eq.~\eqref{eq:delta_lambda}, with the selected $u$ marked and the pure-onsite limit dashed.
 (b) The density of $x_N=\log r_N$, with
 $r_N=|\delta\epsilon_N|/\delta\epsilon_{N,\mathrm{rms}}$,
 approaches a Gaussian as $N$ increases. Its center shifts below the RMS
 scale, marked by the dashed line at $r_N=1$.
}
 \label{fig:main_splitting}
\end{figure}

For our analysis, we need to understand how the splitting magnitude
$|\delta\epsilon_N|$ behaves as the chain length $N$ increases.
To this end, we define the root mean square (RMS),
\begin{align}
 \delta\epsilon_{N,\mathrm{rms}}
 &=\sqrt{\langle(\delta\epsilon_N)^2\rangle}.
 \label{main:splitting_rms_definition}
 \end{align}
For a zero-mean Gaussian distribution, the RMS fully determines the distribution.
However, we have seen in Sec.~\ref{subsec:minimal_splitting} that onsite noise
produces a non-Gaussian splitting distribution already for $N=2$.
We will see that this non-Gaussianity becomes more pronounced for longer chains.
To capture the typical behavior of the splitting, we also consider
the geometric mean, which we call the typical splitting:
 \begin{align}
 |\delta\epsilon_N|_{\mathrm{typ}}
 &=E_g\exp\left\langle\log\frac{|\delta\epsilon_N|}{E_g}\right\rangle.
 \label{main:splitting_typical_definition}
\end{align}
Figure~\ref{fig:main_splitting}(a) shows both quantities computed
numerically for different chain lengths. Both decrease approximately
exponentially with $N$, but with different slopes: the typical splitting
decreases faster than the RMS. Their separation therefore grows as sites
are added.

The origin of this stark difference is visible in
Fig.~\ref{fig:main_splitting}(b). There, we show numerically sampled
distributions of
$\log(|\delta\epsilon_N|/\delta\epsilon_{N,\mathrm{rms}})$.
Normalizing each splitting by the RMS removes the overall decrease
of that scale with length. Nevertheless, the central
part of the distribution moves to the left
as $N$ increases. Most realizations
thus have splittings increasingly far below the RMS.

The RMS gives disproportionate weight to rare realizations with
large splittings, as these enter quadratically in its definition.
This explains why the RMS splitting decays slower than the
typical splitting. In contrast, the coherence average does \emph{not} give
greater weight to large splittings: phase factors have unit magnitude,
and the Rabi return weight $W$ further suppresses their contribution
(see Sec.~\ref{sec:noise}).
It is therefore important to know the distribution around the \emph{typical} splitting to determine the dephasing time.
Replacing the full distribution by a Gaussian with the same RMS can
substantially overestimate the dephasing rate.

Fig.~\ref{fig:main_splitting}(b) shows that, in addition to the shift of the maximum,
the distribution of the logarithm of the splitting broadens.
Additionally, it becomes increasingly Gaussian with increasing length, although
short chains still show an appreciable skewness. For large $N$, the
distribution of the splitting magnitude thus approaches a log-normal form.

Broad, approximately log-normal distributions of Majorana splitting
are established in the literature on disordered
wires and chains~\cite{brouwer_probability_2011,sau_realizing_2012,
hegde_majorana_2016, fortin_finite_2026}. However, to our knowledge,
the consequences of this broad, non-Gaussian distribution for qubit
dephasing have not yet been established. For this,
we need a quantitative description of the splitting distribution for
both onsite and bond disorder, including corrections
to the log-normal form for short chains. We will derive this
in Sec.~\ref{sec:splitting_distribution}.

\subsection{Analytical splitting statistics}
\label{sec:splitting_distribution}

We obtain a semi-analytical description of the splitting statistics for
arbitrary chain length $N$ within the weak-noise model of
Sec.~\ref{sec:noise}. The derivation is based on a
random recurrence relation for the splitting as sites are added to the chain
(App.~\ref{app:splitting_recurrence}).
This construction is closely related
to transfer-matrix descriptions of disordered Kitaev
chains~\cite{sau_realizing_2012,degottardi_majorana_2013,hegde_majorana_2016}.
We solve it in the general case of both onsite and bond disorder by mapping
the recurrence to a Markov chain. The technical details are given in
App.~\ref{app:splitting}. Here we summarize the main results.

The finite-length RMS splitting is
\begin{align}
 \delta\epsilon_{N,\mathrm{rms}}
 &\simeq E_g\left(\frac{\sigma_{t-\Delta}}{E_g}\right)^{N/2}
 \left[\frac{r_+^{N+1}-r_-^{N+1}}{r_+-r_-}\right]^{1/2},
 \label{main:splitting_rms}\\
 r_\pm(u)&=\frac{u\pm\sqrt{u^2+4}}2.\nonumber
\end{align}
This expression holds for arbitrary $N$ and noise ratio $u$
[Eq.~\eqref{minimal:noise_ratio}], and
includes even--odd corrections at finite length
(App.~\ref{app:splitting_moments}). It gives the solid curve in
Fig.~\ref{fig:main_splitting}(a), in good agreement with the numerical
results. For $u>0$, the larger root $r_+$ determines the asymptotic decay,
\begin{align}
 \delta\epsilon_{N,\mathrm{rms}}
 &\propto e^{\lambda_{\mathrm{rms}}N},
 \label{main:splitting_rms_asymptotic}\\
 \lambda_{\mathrm{rms}}
 &=\frac12\log\frac{\sigma_{t-\Delta}}{E_g}
 +\frac12\log r_+\!\left(\frac{\sigma_\mu^2}{E_g\sigma_{t-\Delta}}\right).
 \nonumber
\end{align}
Note that $\lambda_{\mathrm{rms}}<0$ for the case of weak noise we consider,
i.e., the RMS splitting decreases with chain length.

The typical splitting instead obeys for large $N$ the asymptotic form
\begin{align}
 |\delta\epsilon_N|_{\mathrm{typ}}
 &\propto e^{\lambda_{\mathrm{typ}}N},
 \label{main:splitting_typical}\\
 \lambda_{\mathrm{typ}}
 &=\frac12\log\frac{\sigma_{t-\Delta}}{E_g}
 +L\!\left(\frac{\sigma_\mu^2}{E_g\sigma_{t-\Delta}}\right).
 \nonumber
\end{align}
Within our assumption of weak noise, $\lambda_{\mathrm{typ}}=-1/\xi_{\mathrm M}$,
where $\xi_{\mathrm M}$ is the typical localization length of the Majorana
wavefunction~\cite{hegde_majorana_2016}.
The function $L(u)$ is the Lyapunov exponent of the dimensionless
recurrence. It depends only on the noise ratio $u$ and is computed from
the stationary distribution of the associated Markov chain
(App.~\ref{app:splitting_markov}).
We evaluate $L(u)$ numerically and show its dependence on $u$ in
Fig.~\ref{fig:splitting_coefficients}(a) in the appendix.
Eq.~\eqref{main:splitting_typical} also has a length-independent prefactor, which we
give in App.~\ref{app:splitting_offset}. We show the asymptotic typical splitting in Fig.~\ref{fig:main_splitting}(a)
as a dashed curve, again agreeing well with the numerical results even for short
chain lengths.

We define the difference in the decay rates of the RMS and typical splittings as
\begin{equation} \label{eq:delta_lambda}
 \Delta\lambda=\lambda_{\mathrm{rms}}-\lambda_{\mathrm{typ}}
\end{equation}
This expression sets the exponential separation of the RMS and
typical splitting, $\delta\epsilon_{N,\mathrm{rms}}/|\delta\epsilon_N|_{\mathrm{typ}}
\propto e^{N\Delta\lambda}$. We observed this behavior in Fig.~\ref{fig:main_splitting}(a)
for a specific noise ratio.
In the inset, we show the dependence of $\Delta\lambda$ on the noise ratio $u$.
We find that the difference is positive for all noise ratios $u>0$, so that the typical splitting
always decays faster than the RMS splitting.

Our Markov chain calculation also gives information about the full
distribution of the energy splitting.
In particular, it determines how the distribution broadens around
its typical scale through the central limit theorem. We find
(App.~\ref{app:splitting_log_variance}):
\begin{equation}
 \log\frac{|\delta\epsilon_N|}{|\delta\epsilon_N|_{\mathrm{typ}}}
 \simeq\sqrt{N\mathcal D(u)}\,G,
 \quad G\sim\mathcal N(0,1).
 \label{main:splitting_central_distribution}
\end{equation}
Thus, the logarithmic width grows as $\sqrt N$.
The coefficient $\mathcal D(u)$ determines the rate of this growth and
is again only a function of the noise ratio $u$
and is plotted in Fig.~\ref{fig:splitting_coefficients}(c) in the appendix.
In addition, we find that the skewness of the log-splitting distribution
decreases as $N^{-1/2}$ (App.~\ref{app:splitting_skewness}). Hence, the central part of the splitting-magnitude
distribution indeed approaches a log-normal form.

In the following sections
we use these splitting statistics to compute
Rabi and Ramsey coherence.

\section{Rabi coherence and chain-length protection}
\label{sec:rabi}
\label{sec:scaling}

We now investigate how coherent Rabi oscillations can probe the topological
protection of Kitaev chains. In particular, we analyze the conditions for
experiments to reveal the exponential suppression of the splitting
with chain length, characteristic for topological protection.

As we have seen in Sec.~\ref{sec:minimal}, increasing the middle-dot
detuning $\mu_D$ can lengthen the Rabi dephasing time. However, it also
reduces the Rabi frequency and can reduce the visibility. The number
of coherent oscillations $N_\varphi$ instead has an optimum, providing the best operating point for single-axis control of the
Majorana parity qubit given a specific noise environment. We therefore first consider
this optimum and its scaling with chain length, before discussing
control through a tunnel barrier.

\subsection{Optimal number of coherent oscillations through middle-dot detuning}
\label{sec:rabi_optimization}

\subsubsection{Access to strong coupling}
\label{sec:rabi_strong_coupling}

High-visibility Rabi oscillations require a well defined rotation axis translating into the operating window
\begin{equation}
 |\delta\varepsilon_N|\ll\Gamma_0\ll E_g.
 \label{main:rabi_coupling_window}
\end{equation}
Increasing $N$ suppresses $\delta\varepsilon_N$ and therefore permits high 
visibility in a larger window of $\Gamma_0$ values. 
Additionally, the onset of the weak-coupling regime moves to larger $\mu_D$.
This allows us to apply our arguments for the strong-coupling regime from Sec.~\ref{subsec:strong} in a larger range of $\Gamma_0$. 
In the following, we will therefore focus on the strong-coupling regime for general $N$.

For $|\delta\Gamma|,|\delta\varepsilon_N|\ll\Gamma_0$, we apply the expansion of Eq.~\eqref{strong:omega}.
To probe the suppression of splitting noise through the dephasing time,
we must be able to access a regime where the splitting noise is dominant.
This requires
\begin{equation}
 \sigma_\Gamma\ll\frac{(\delta\varepsilon_N)^2}{2\Gamma_0}.
 \label{main:rabi_splitting_dominance}
\end{equation}
Noise on the middle-dot allows to write
\begin{equation}
 \sigma_\Gamma=\frac{\Gamma_0\sigma_{\mu_D}}{\mu_D}
 =\frac{2\sigma_{\mu_D}\Gamma_0^2}{t_{LD}t_{RD}},
 \label{main:rabi_dot_noise_width}
\end{equation}
through Eq.~\eqref{model:gamma}.
Together, $|\delta\varepsilon_N|\ll\Gamma_0$ and 
$\sigma_\Gamma\ll\frac{(\delta\varepsilon_N)^2}{2\Gamma_0}$ give the coupling window
\begin{equation}
 |\delta\varepsilon_N|\ll\Gamma_0\ll
 \left[\frac{t_{LD}t_{RD}(\delta\varepsilon_N)^2}
 {4\sigma_{\mu_D}}\right]^{1/3}.
 \label{main:rabi_joint_window}
\end{equation}
Its lower boundary scales $\sim(\delta\varepsilon_N)$,
while the upper boundary $\sim(\delta\varepsilon_N)^{2/3}$.
Suppressing $\delta\varepsilon_N$ through larger $N$ increases their
relative separation. While accessing this regime was hard
for $N=2$, the exponential suppression of the splitting with chain
length makes it increasingly possible to satisfy.

\subsubsection{Ideal scaling with chain length}
\label{sec:rabi_ideal_scaling}

We start by considering the behavior of the system when only
splitting noise and coupling noise from middle-dot fluctuations are present.
Other noise channels will be discussed in Sec.~\ref{sec:rabi_optimization_limits}.

We first present a scaling argument that captures the leading growth of the
optimal cycle count $N_{\varphi,\max}$ with chain length. Changing
the middle-dot detuning $\mu_D$ changes the ratio between the two
dephasing mechanisms: for small $\mu_D$, the middle-dot noise dominates,
while for large $\mu_D$, the chain splitting dominates.  The competition
is in fact the same as for $N=2$, Eqs.~\eqref{minimal:strong_linear_dephasing}
and \eqref{minimal:strong_quadratic_dephasing}. To estimate its leading
exponential dependence on chain length, we use the typical single-chain
splitting $|\delta\epsilon_N|_{\mathrm{typ}}$. The two noise channels,
considered separately, give
\begin{align}
 T_{\varphi,D}&\simeq
 \frac{\hbar\mu_D}{\sqrt2\Gamma_0\sigma_{\mu_D}}
 \propto\mu_D^2, \nonumber\\
 T_{\varphi,\mathrm{split}}&\sim
 \frac{\hbar\Gamma_0}{|\delta\epsilon_N|_{\mathrm{typ}}^2}
 \propto\frac{1}{\mu_D|\delta\epsilon_N|_{\mathrm{typ}}^2}.
 \label{main:rabi_time_balance}
\end{align}
$N_\varphi$ takes on its maximum when both are similar. 
Multiplying by Eq.~\eqref{quasi:frequency} gives $N_{\varphi,D}\propto \mu_D$ and $N_{\varphi,\mathrm{split}}\propto \mu_D^{-2}|\delta\epsilon_N|_{\mathrm{typ}}^{-2}$, respectively.
Balancing these estimates yields
$N_{\varphi,\max}\propto|\delta\epsilon_N|_{\mathrm{typ}}^{-2/3}$
at leading exponential order. Since
$\log(|\delta\epsilon_N|_{\mathrm{typ}}/E_g)
\simeq\lambda_{\mathrm{typ}}N$, we obtain
\begin{equation}
 \log N_{\varphi,\max}
 \simeq-\frac23\lambda_{\mathrm{typ}}N.
 \label{main:rabi_qualitative_scaling}
\end{equation}
Hence, $N_{\varphi,\max}$ grows exponentially with chain length.
Importantly, its rate is directly related to the typical
splitting exponent. It is thus a promising tool to probe the topological
protection of the Majorana states.

We now evaluate the strong-coupling phase average, Eq.~\eqref{rabi:phase_average}, using the full
splitting distribution of Sec.~\ref{sec:splitting_statistics}, and from
there find the maximum of $N_\varphi$ with respect to $\mu_D$. The details
of this calculation are given in Apps.~\ref{app:rabi_envelope}--\ref{app:rabi_asymptotic_optimization}.
We find the asymptotic length dependence
\begin{equation}
 N_{\varphi,\max}
 \propto\frac{1}{N^{1/6}}
 \exp\!\left[-\frac23\lambda_{\mathrm{typ}}N
 -\frac{2a_*}{3}\sqrt{N\mathcal D(u)}\right],
 \label{main:rabi_optimal_scaling}
\end{equation}
with $a_*\simeq0.270288$. The prefactor of Eq.~\eqref{main:rabi_optimal_scaling},
is explicitly given in App.~\ref{app:rabi_asymptotic_optimization}.

Notably, the log-width broadening of the splitting distribution gives
a direct contribution $\sqrt{N}$ to the exponent, even though it does not
show in the typical splitting itself. This is a consequence of the fact
that the splitting energy enters non-linearly in the Rabi envelope.
In addition to the exponential behavior, we also find an
algebraic factor $N^{-1/6}$, stemming from the optimization of the cycle
count.

\subsubsection{Limits from detuning and local coupling noise}
\label{sec:rabi_optimization_limits}

The value of $\mu_D$ at which $N_{\varphi}$ becomes maximal, Eq.~\eqref{rabi:optimal_detuning_asymptotic}, moves to \emph{exponentially} larger values as the chain length increases. 
Experiments are however typically restricted to $\mu_D\leq\mu_{D,\max}$ due to e.g. charging energies or other orbitals.
This can prevent this optimum from being reached. 
Even if splitting noise becomes negligible, linear middle-dot noise alone then limits the cycle count to $\mu_D/(\sqrt2\pi\sigma_{\mu_D})$. 
The maximally available detuning therefore sets a finite cycle-count limit.

A potentially even lower limit can arise from additional noise channels such as
fluctuations of Majorana edge states on the inner dots, $\eta_\alpha$, or the tunneling
$t_{\alpha,D}$ to the dot. Suppressing middle-dot noise through larger $\mu_D$
eventually makes these other channels dominant due to their generally different dependence on $\mu_D$.

The coupling in Eq.~\eqref{model:gamma} contains the product of the
two inner Majorana amplitudes. For weak onsite noise, their leading
local correction is $\eta_\alpha\simeq1-(\delta\mu_{\alpha,\mathrm{in}})^2/(2E_g^2)$.
For two independent inner dots, this gives
\begin{equation}
 \delta\Gamma_{\mathrm{in}}
 \simeq-\frac{\Gamma_0}{2E_g^2}
 \left[(\delta\mu_{L,\mathrm{in}})^2+
       (\delta\mu_{R,\mathrm{in}})^2\right],
 \label{rabi:local_interface_correction}
\end{equation}
with noise width $\sigma_{\Gamma,\mathrm{in}}$. 
Notably, this correction is not suppressed through $N$, despite decreasing when increasing $\mu_D$.
Because Rabi oscillations rely on precise control of the Majorana coupling to acquire a specific phase, this is an intrinsically unprotected operation.

Each independent Gaussian fluctuation gives a quadratic-dephasing
factor with exponent $-1/4$~\cite{makhlin_dephasing_2004}.
For the two inner dots, their product yields
\begin{equation}
 \mathcal{C}_{\mathrm{in}}(t)
 =\left[1+\left(\frac{2\Gamma_0\sigma_\mu^2t}
 {E_g^2\hbar}\right)^2\right]^{-1/2}.
 \label{main:rabi_interface_envelope}
\end{equation}
Its $1/e$ crossing yields the cycle count
\begin{equation}
 N_{\varphi,\mathrm{inner}}
 =\frac{E_g^2\sqrt{e^2-1}}{2\pi\sigma_\mu^2}.
 \label{main:rabi_interface_limit}
\end{equation}
In particular, this limit is independent of the chain length and the middle-dot
detuning.

Another source of coupling noise is fluctuations of the tunnel barriers
connecting the chains to the middle dot. For independent
Gaussian noise for this coupling, the leading noise variance is
\begin{equation}
 \frac{\sigma_{\Gamma,\mathrm{link}}^2}{\Gamma_0^2}
 =\frac{\sigma_{t_{LD}}^2}{t_{LD}^2}
 +\frac{\sigma_{t_{RD}}^2}{t_{RD}^2}.
 \label{main:rabi_link_noise}
\end{equation}
Only considering this noise channel, the Gaussian average gives
\begin{equation}
    N_{\varphi,\mathrm{tunnel}}
=\Gamma_0/(\sqrt2\pi\sigma_{\Gamma,\mathrm{link}}).
\end{equation}
Again, this cycle count is independent of chain length and middle-dot detuning.

All of these three effects set a limit on the cycle count that 
cut off the exponential improvement from Eq.~\eqref{main:rabi_optimal_scaling}.

\subsubsection{Crossover from ideal scaling to local noise limits}
\label{sec:rabi_optimization_crossover}

We now illustrate the overall behavior of $N_\varphi$ as a function
of $N$. Guided by Ref.~\cite{zatelli_majorana_2026}, we take
$t=\Delta=20$, $E_g=40$, $t_{LD}=t_{RD}=7.5$, $\sigma_{\mu_D}=4.5$, 
$\sigma_\mu=2.5$, and  $\sigma_{t-\Delta}=0.1$ in units of 
$\mu\mathrm{eV}$.\footnote{These parameters are slightly modified
from~\cite{zatelli_majorana_2026}: we increased the chain noise 
$\sigma_\mu$ to $2.5\,\mu\mathrm{eV}$ to be closer to experimental
$N_{\varphi,max}$. We also set the noise on the tunnel couplings to zero
to simplify the model. The $N=2$ data in~\cite{zatelli_majorana_2026} is not
sufficient to determine all noise parameters, and specifically the 
parameters there were not a fit.} These parameters give a noise ratio
$u=1.5625$, and will be used in the remainder of this manuscript.
Additionally, we assume that the maximum middle-dot detuning is 
$\mu_{D,\max}=10\,\mathrm{meV}$. We compare our analytical predictions with numerical
simulations of a full many-body model (see App.~\ref{app:numerics} for details).

\begin{figure}[!t]
 \centering
 \includegraphics[width=\columnwidth]{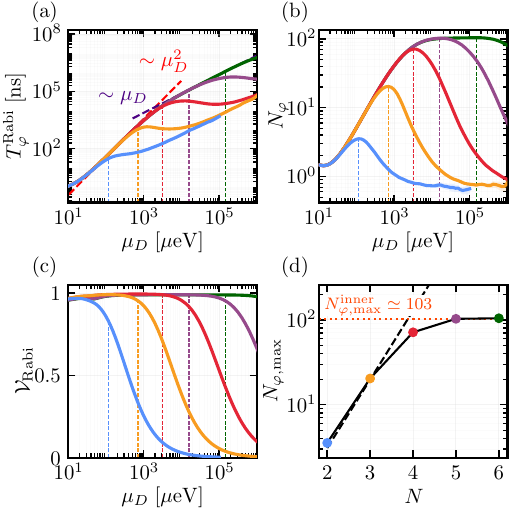}
 \caption{Behavior of the ensemble characteristics of Sec.~\ref{sec:noise} for Rabi oscillations in the set-up of Fig.~\ref{fig:figure_1} (a).
 Parameters are given in Sec.~\ref{sec:rabi_optimization_crossover};
 colors denote $N=2,\ldots,6$.
 (a) Coherence time, (b) number of return cycles, and (c) visibility from ensemble averaged numerical time-evolutions.
 (d) Extracted $N_{\varphi,\mathrm{max}}$ for increasing $N$. 
 Dashed lines in (a) show the strong-coupling limits for middle-dot noise
 alone, $T_\varphi\propto\mu_D^2$, and inner-dot noise alone,
 $T_\varphi\propto\mu_D$. Colored vertical dashed lines in (a)--(c)
 mark the detunings that maximize the cycle count. 
 The transition between the two yields a saturation of the number of return cycles visible in (b) and (d). 
 In (d), the dashed curve is the ideal asymptote of
 Eq.~\eqref{rabi:physical_cycle_count_asymptotic}, and the horizontal
 dotted line is the inner-dot noise limit. Shading gives twice the
 standard error of the numerical estimates.
 }
 \label{fig:rabi_optimization}
\end{figure}

Figure~\ref{fig:rabi_optimization}(a) and (b) show $T_\varphi$ and
$N_\varphi$ as a function of the middle-dot detuning $\mu_D$ for different
chain lengths. For the chosen parameters, the dephasing time for
$N=2$ does not show a clear maximum, but the curves for $N\geq3$ do.
This is due to the fact that the strong coupling regime is more easily
reached in longer chains, as discussed in Sec.~\ref{sec:rabi_strong_coupling}.
As expected, the cycle count $N_\varphi$ shows a clear maximum for all
chain lengths, reflecting the competition between middle-dot noise and splitting noise.

At sufficiently large middle-dot detuning, the longer chains also 
enter the weak-coupling regime discussed in Sec.~\ref{subsec:weak_coupling}.
The oscillation weight then selects nearly resonant realizations, so only 
the splitting density near zero contributes. This density determines 
the visibility (Fig.~\ref{fig:rabi_optimization} (c)), but cancels from the normalized envelope. Consequently, 
the limiting decay has the same form for every chain length, with 
$T_\varphi\propto\mu_D$, only shifted to larger detuning.

In both panels, we also show the asymptotes from middle-dot noise
and inner-dot noise, which agree well with the numerics. 
Already from $N=5$, the maximum cycle count
is limited by the inner-dot noise. We also see that the value of
$\mu_D$ at which the maximum occurs increases exponentially with $N$,
as expected from our theory. At $\mu_{D,\max}=10\,\mathrm{meV}$,
middle-dot noise alone would allow approximately $500$ coherent
oscillations. For these parameters, inner-dot noise limits the cycle
count first.

Figure~\ref{fig:rabi_optimization}(d) compares the numerical maxima
with the ideal scaling asymptote and the inner-dot noise limit.
We use the full expression for the optimal cycle count from
Eq.~\eqref{rabi:physical_cycle_count_asymptotic} in
App.~\ref{app:rabi_asymptotic_optimization}, which shows good
agreement with the numerical maxima for small $N$, without any fitting parameters.
The numerical results initially closely follow the predicted growth, with
about $3.5$ cycles at $N=2$ and $20$ at $N=3$. The value for $N=4$, 
still shows a rapid increase, but only reaches
about $71$ cycles instead of the ideal prediction of $133$. By $N=5$,
the cycle count approaches a plateau near $100$, consistent with
$N_{\varphi,\mathrm{inner}}\simeq103$ from
Eq.~\eqref{main:rabi_interface_limit}.

Thus, we do observe an exponential increase of $N_\varphi$, directly linked to the 
exponential protection of the Majorana splitting, for experimentally relevant parameters. 
However, this exponential increase is visible only over a very limited range of chain lengths
of 2-3 sites before being limited by inner-dot noise. Still, the increase can be
significant, up to a factor of $30$. Improving inner-dot noise may raise the
ceiling, but the exponential increase will always be restricted to a limited
range  of chain lengths.

Here, we explicitly considered the case of inner-dot noise for which
the noise strength comes from onsite potential fluctuations. However, we
note that existing experimental data~\cite{zatelli_majorana_2026}
do not yet determine the dominant limiting noise channel: the minimal
parity qubit has a maximum number of coherent oscillations of $N_\varphi\simeq3.5$ at $N=2$,
and is not sensitive to noise channels with a much higher limit.
We believe that future experiments with longer chains could help distinguish
inner-dot fluctuations from other sources of coupling noise,
including the tunnel-coupling fluctuations neglected here.

\subsection{Control through a tunnel barrier}
\label{sec:rabi_barrier}

So far, we considered coupling the two Kitaev chains through a middle
dot as shown in Fig.~\ref{fig:figure_1}(a).
Alternatively, one can couple the chains also through a direct tunnel barrier as shown in Fig.~\ref{fig:figure_1}(c) \cite{pan_rabi_2025}.
Here, we briefly discuss how our previous arguments change for this device setup.

We describe the tunnel barrier by a direct coupling $\tau_0$, and hence
$\Gamma_0=\tau_0/2$. We assume that the tunnel amplitude depends exponentially
on the barrier voltage $V_B$ as $\tau_0(V_B)\propto\exp(-\alpha V_B)$
\cite{simmons_charge_2009}. For weak Gaussian quasistatic voltage noise,
we linearize this dependence and obtain $\tau\simeq\tau_0(1+\xi_B)$, with
$\xi_B\sim\mathcal N(0,s_B^2)$ and
$s_B=|\alpha|\sigma_{V_B}\ll1$ \cite{miles_braiding_2026}.
Here $s_B$ is a noise scale that generally depends on device details.
We assume there is an interval of barrier voltages over which it is
approximately constant. 

The barrier noise has a width $\sigma_{\Gamma,B}\simeq s_B\Gamma_0$. 
In the strong-coupling regime, this channel alone has a Gaussian
envelope with
\begin{equation}
 T_{\varphi,B}\simeq\frac{\hbar}{\sqrt2\,s_B\Gamma_0},
 \qquad
 N_{\varphi,B}\simeq\frac{1}{\sqrt2\pi s_B}.
 \label{main:rabi_barrier_limit}
\end{equation}
Thus reducing the tunnel coupling increases the barrier-limited coherence time
but leaves its cycle count fixed. This differs significantly from
middle-dot tuning, for which the coupling noise is suppressed faster than
the coupling, and the cycle count shows a maximum.

\begin{figure}[!t]
 \centering
 \includegraphics[width=\columnwidth]{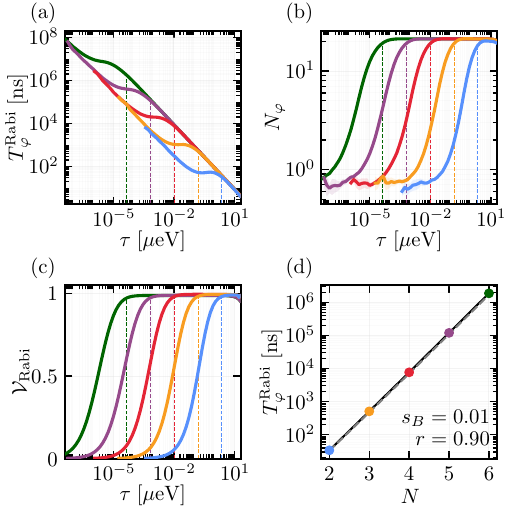}
 \caption{Ensemble characteristics of Sec.~\ref{sec:noise} for Rabi oscillations in the set-up of Fig.~\ref{fig:figure_1} (c) at transverse noise scale $s_B=0.01$ (Sec.~\ref{sec:rabi_barrier}).
  (a) Coherence time, (b) number of return cycles, and (c) visibility depending on the averaged tunnel coupling $\tau=\langle \tau_0(1+\xi_B)\rangle$.
  The number of return cycles is limited by noise on $\tau$ itself and therefore not suited to demonstrate the effect of length protection.
  Instead, reducing $\tau$, the coherence time increases exponentially with $N$.
  (d) Defining a threshold $r=0.9$, we probe the smallest $\tau$ for which
  $N_\varphi(\tau)=rN_{\varphi,\mathrm{max}}$ holds. Colored vertical dashed
  lines in (a)--(c) mark these couplings. Repeating for different $N$
  yields the shown data points. The grey dashed curve is the full asymptote
  of Eq.~\eqref{rabi:barrier_threshold_time}, without a fitted prefactor.
  Shading and error bars give twice the standard error.
}
 \label{fig:rabi_barrier}
\end{figure}

Figure~\ref{fig:rabi_barrier} shows numerics for $s_B=0.01$.
In particular, Fig.~\ref{fig:rabi_barrier}(b) shows a plateau close to the
prediction $N_{\varphi,B}\simeq22.5$, instead of a unique maximum in $N_\varphi$ as for
the model with the central dot. In contrast, the coherence time $T_\varphi$
continues to increase (Fig.~\ref{fig:rabi_barrier}(a)). This increase should not
be confused with a better operating point of the qubit, as the 
cycle count remains fixed. Instead, it is rather a consequence of the
reduction in frequency due to the smaller coupling.

As the coupling decreases, the splitting-induced frequency fluctuations
become comparable to the coupling noise, and $N_\varphi$ departs from its
plateau. At still weaker coupling, the visibility $\mathcal V$ also decreases
as the system approaches the weak-coupling regime (Fig.~\ref{fig:rabi_barrier}).

Increasing the chain length moves the departure from the cycle-count
plateau to weaker coupling due to the exponential suppression of the
splitting. This shift can thus be used to probe the exponential protection
of the Majorana splitting. Qualitatively, the onset of the drop
balances the relative contributions of splitting noise,
$|\delta\epsilon_N|_{\mathrm{typ}}^2/\Gamma_0$, and barrier noise,
$s_B\Gamma_0$. The corresponding coupling therefore
follows the typical splitting at leading exponential order,
$\Gamma_0\propto|\delta\epsilon_N|_{\mathrm{typ}}$.
Since the coherence time increases as
$T_\varphi\propto1/\Gamma_0$ while $N_\varphi$ remains fixed, we find
at the onset of the drop in cycle count
$T_\varphi\propto\exp[-\lambda_{\mathrm{typ}}N]$. Thus, it directly
connects to the exponential protection of the Majorana splitting, similar to
the optimal cycle count in Sec.~\ref{sec:rabi_ideal_scaling}.

To be more quantitative, we need to mathematically define a threshold for departure from
the $N_\varphi$-plateau. We define the threshold coupling $\Gamma_r$ as the value
where $N_\varphi(\Gamma_r)=rN_{\varphi,\mathrm{plateau}}$ ($0<r<1$), i.e., where the
cycle count has dropped to a fraction $r$ of its plateau value.

Assuming that barrier noise dominates the coupling fluctuations, we find
\begin{equation}
 T_\varphi(\Gamma_r)
 \propto\exp\!\left[-\lambda_{\mathrm{typ}}N
                         -a_r\sqrt{N\mathcal D(u)}\right].
 \label{main:rabi_barrier_threshold_time}
\end{equation}
The coefficient $a_r$ depends only on the threshold fraction $r$ and is
given in Eq.~\eqref{rabi:barrier_threshold_loss}.
The derivation and the complete expression, including the
prefactor, are given in App.~\ref{app:rabi_barrier_threshold}.
The log-width broadening thus again contributes a $\sqrt N$ correction to the
leading exponential growth.

Figure~\ref{fig:rabi_barrier}(d) shows numerical results for
$T_\varphi(\Gamma_r)$ for $r=0.9$, together with our asymptotic
expression in Eq.~\eqref{rabi:barrier_threshold_time}.
We find very good agreement with our analytical prediction.

Although barrier tuning does not increase the plateau cycle count, the
growth of the coherence time at its departure provides a probe of length
protection. Along this sequence of decreasing couplings, inner-dot
fluctuations do not impose a length-independent ceiling on $T_\varphi$:
their dephasing time, like that from barrier noise, increases as
$1/\Gamma_0$. Within this model, the accessible range of chain lengths is
therefore limited by how far the tunnel coupling can be reduced while the
fractional-noise description remains valid. Its exponential dependence on
barrier voltage may allow a wider coupling range than middle-dot detuning,
which is constrained by the dot's finite level spacing.

Both the middle-dot and direct-barrier setups allow one to probe 
the exponential protection of the Majorana splitting through changing the
coupling with $N$. For comparison, App.~\ref{app:rabi_fixed_operating_conditions} 
discusses protection at fixed operating conditions, where no clear exponential scaling
is observed.

\section{Ramsey experiments} \label{sec:ramsey}

We now use the Ramsey protocol introduced in
Sec.~\ref{subsec:ramsey_return_probability} to study length protection
and its limitations due to residual interchain coupling. We assume
ideal pulses, as independent pulse-angle fluctuations only reduce the
contrast~\cite{braun_cooperative_2014} without changing the normalized
envelope (App.~\ref{app:ramsey_pulse_errors}).

\subsection{Length protection with decoupled chains}
\label{sec:ramsey_decoupled}

For decoupled chains, the accumulated phase in
Eq.~\eqref{quasi:ramsey_envelope} is linear in $\delta\varepsilon_N$.
An order-one phase therefore accumulates on a
time scale $\hbar/|\delta\epsilon_N|_{\mathrm{typ}}$, giving, at leading
exponential order,
\begin{equation}
 T_\varphi^{\mathrm{Ramsey}}
 \propto \exp[-\lambda_{\mathrm{typ}}N].
 \label{main:ramsey_leading_scaling}
\end{equation}
The Ramsey time in this ideal limit thus directly probes the exponential
suppression of the splitting.

In order to go beyond this estimate, we average over the full 
splitting distribution of
Sec.~\ref{sec:splitting_statistics} using the Ramsey envelope defined
in Eq.~\eqref{quasi:ramsey_envelope}. This gives the asymptotic
length dependence derived in App.~\ref{app:ramsey_length_dependence},
\begin{equation}
 T_\varphi^{\mathrm{Ramsey}}
 \propto\exp\!\left[-\lambda_{\mathrm{typ}}N
                         -a_*\sqrt{N\mathcal D(u)}\right],
 \label{main:ramsey_time_asymptotic}
\end{equation}
with the same coefficient $a_*$ as in the optimized Rabi cycle count,
Eq.~\eqref{main:rabi_optimal_scaling}.
The broadening of the log-splitting distribution again contributes a
$\sqrt N$ correction to the leading exponential scaling.
The length-independent prefactor and finite-length variance correction
are included in Eq.~\eqref{ramsey:time_asymptotic} in the appendix.

Figure~\ref{fig:ramsey_protection}(a) compares the analytic Ramsey dephasing time, Eq.~\eqref{ramsey:time_asymptotic},
with the numerical results for the Ramsey dephasing time, with the
same parameters as in Sec.~\ref{sec:rabi_optimization_crossover} and assuming
zero coupling between the two chains, finding
good agreement. The blue curves in panels (b) and (c) show the return probability
for $N=3$ and $N=4$ in this decoupled case. As expected, the curves show a
nonoscillating approach to $1/2$.

\begin{figure}[t]
 \centering
 \includegraphics[width=\columnwidth]{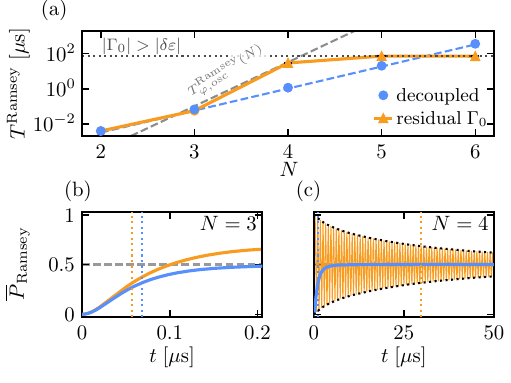}
 \caption{Ramsey coherence and return probability with and without residual
 coupling, for the parameters of Sec.~\ref{sec:rabi_optimization_crossover}
 and ideal preparation and readout pulses.
 (a) Blue symbols show decoupled chains; orange symbols correspond to
 $\mu_D=10\,\mathrm{meV}$. Circles denote the direct return-probability
 threshold, while triangles denote the oscillation-envelope lifetime.
 The blue and grey dashed curves give the decoupled and splitting-only
 carrier asymptotes, Eqs.~\eqref{ramsey:time_asymptotic}
 and~\eqref{ramsey:carrier_time_asymptotic}, respectively.
 The horizontal dotted line is the inner-dot-noise limit; error bars
 give twice the standard error.
 (b), (c) Averaged returns for $N=3,4$, with the same color convention.
 Decoupled chains show nonoscillating decay, while residual coupling
 produces oscillations. Vertical dotted lines mark the corresponding
 dephasing times. Black dotted curves in (c) show the oscillation envelope.
 In the splitting-noise-dominated regime, quadratic sensitivity to the
 splitting slows its decay, Eq.~\eqref{main:ramsey_carrier_time}.
 }
 \label{fig:ramsey_protection}
\end{figure}

\subsection{Ramsey sequence with residual coupling}
\label{sec:ramsey_residual_coupling}

With a residual interchain coupling $\Gamma_0^{\mathrm{min}}$,
the decoupled Ramsey decay can still be observed as long as
\begin{equation}
 \frac{|\Gamma_0^{\mathrm{min}}|T_\varphi^{\mathrm{Ramsey}}}{\hbar}\ll1.
 \label{main:ramsey_isolation}
\end{equation}
At leading exponential order, this requires
$|\Gamma_0^{\mathrm{min}}|\ll|\delta\epsilon_N|_{\mathrm{typ}}$.
Middle-dot detuning $\mu_D$ alone may thus not suffice, since it suppresses the
coupling only algebraically. However, the tunnel couplings $t_{LD}$ and $t_{RD}$
depend exponentially on barrier-gate voltage. We therefore expect
experiments to be able to suppress the residual coupling exponentially.

Even when experiments can achieve the decoupled limit, 
the case of finite residual coupling is still physically
interesting: it changes how the qubit responds to
splitting noise, leading to a distinct regime of Ramsey dynamics.
With residual coupling, the waiting dynamics is precession about an
axis in the $xz$ plane. The initial Bloch vector along $-y$ is
perpendicular to this axis for every realization, giving
\begin{equation}
 P_{\mathrm{Ramsey}}(t)
 =\sin^2\!\left(\frac{\Omega t}{\hbar}\right),
 \label{main:ramsey_residual_return}
\end{equation}
with $\Omega$ from Eq.~\eqref{quasi:omega}. This is essentially
the Rabi return probability, Eq.~\eqref{quasi:abstract_return}, 
however without weight $W$:
each disorder realization has unit visibility.

At fixed residual coupling, increasing $N$ eventually gives
$|\delta\epsilon_N|_{\mathrm{typ}}\ll\Gamma_0^{\mathrm{min}}$.
The return probability changes character from a monotonic rise to $1/2$ to oscillations
with a carrier frequency set by $\Gamma_0^{\mathrm{min}}$.
In this strong-coupling regime, Rabi and Ramsey experiments exhibit the same
envelope of Eq.~\eqref{rabi:phase_average} and its $1/e$ decay time.
We denote the first $1/e$ crossing of
$\mathcal{C}_{\mathrm{Ramsey,osc}}(t)=|\langle e^{-2i\Omega t/\hbar}\rangle|$
by $T_{\varphi,\mathrm{osc}}^{\mathrm{Ramsey}}$.
When splitting noise dominates,
$(\delta\varepsilon_N)^2/(2\Gamma_0^{\mathrm{min}})\gg\sigma_\Gamma$,
the long-chain result is
\begin{equation}
 T_{\varphi,\mathrm{osc}}^{\mathrm{Ramsey}}
 \propto\frac{\Gamma_0^{\mathrm{min}}}{\hbar}
 \bigl(T_\varphi^{\mathrm{Ramsey}}\bigr)^2,
 \label{main:ramsey_carrier_time}
\end{equation}
with the length-independent factor given in
Eq.~\eqref{ramsey:carrier_time_asymptotic} in
App.~\ref{app:ramsey_residual_coupling}.
The quadratic sensitivity to the splitting~\cite{makhlin_dephasing_2004}
doubles both the leading exponent and the $\sqrt N$ correction of
Eq.~\eqref{main:ramsey_time_asymptotic}.
The splitting-dominated window typically spans changes in $N$ of
order one [Eq.~\eqref{rabi:fixed_length_window} in
App.~\ref{app:rabi_fixed_operating_conditions}], leaving little room
to resolve this scaling before coupling noise causes saturation.

Figure~\ref{fig:ramsey_protection}(a) shows this crossover 
for a fixed residual coupling ($\mu_D=10\,\mathrm{meV}$ 
with $t_{LD}$ and $t_{RD}$ unchanged).
For $N=2,3$, we use the return-probability threshold of
Sec.~\ref{sec:ramsey_decoupled}. For $N\geq4$, we use the first
$1/e$ crossing of the oscillation envelope. Different symbols
distinguish these definitions.
The $N=2,3$ results largely follow the decoupled-chain scaling,
although oscillations already appear at $N=3$ as
Eq.~\eqref{main:ramsey_isolation} is not clearly satisfied
[Fig.~\ref{fig:ramsey_protection}(b)].
At $N=4$, strong coupling is reached while splitting noise still
dominates. The lifetime jumps close to the asymptote of
Eq.~\eqref{ramsey:carrier_time_asymptotic} in
App.~\ref{app:ramsey_residual_coupling}, signaling the change from
linear to quadratic sensitivity to the splitting.
Already at $N=5$, coupling noise dominates and the lifetime saturates.

\begin{table}[t]
 \caption{Leading length dependence of observables that probe chain
 protection, before coupling noise or tuning constraints limit the scaling.
 Here $\lambda_{\mathrm{typ}}<0$; subleading factors are omitted.}
 \label{tab:protocol_scaling}
 \centering
 \small
 \setlength{\tabcolsep}{4pt}
 \renewcommand{\arraystretch}{1.15}
 \begin{ruledtabular}
 \begin{tabular}{lll}
 Protocol & Observable & Scaling \\[3pt]
 \hline
 \begin{tabular}[c]{@{}l@{}}Rabi, central dot\\maximize $N_\varphi$\end{tabular}
 & \begin{tabular}[c]{@{}l@{}}$N_{\varphi,\max}$\\[3pt]$T_{\varphi,\mathrm{opt}}^{\mathrm{Rabi}}$\end{tabular}
 & \begin{tabular}[c]{@{}l@{}}$e^{-2\lambda_{\mathrm{typ}}N/3}$\\[3pt]$e^{-4\lambda_{\mathrm{typ}}N/3}$\end{tabular} \\[6pt]
 \begin{tabular}[c]{@{}l@{}}Rabi, tunnel barrier\\fixed plateau fraction\end{tabular}
 & $T_\varphi(\Gamma_r)$ & $e^{-\lambda_{\mathrm{typ}}N}$ \\[6pt]
 \begin{tabular}[c]{@{}l@{}}Ramsey\\decoupled chains\end{tabular}
 & $T_\varphi^{\mathrm{Ramsey}}$ & $e^{-\lambda_{\mathrm{typ}}N}$ \\
 \end{tabular}
 \end{ruledtabular}
\end{table}

\section{Discussion} \label{sec:discussion}

Our analysis reveals that the exponential protection of the
energy splitting \emph{can} be observed using unprotected operations. 
The choice of the operation is important:
for Rabi measurements, following an operating point where splitting and
coupling noise compete provides a systematic way to reveal this protection.
In Ramsey experiments, the exponential protection is revealed by
decoupling the two chains as much as possible.

The exponential protection reveals itself through different observables,
depending on the protocol and device. All of these, however, are directly
linked to the Majorana protection represented by the Lyapunov exponent
$\lambda_{\rm typ}$. Table~\ref{tab:protocol_scaling} summarizes how this
exponent enters each measurement.

Our theory allows for a quantitative prediction of the range over which
an exponential increase can be reliably observed in experiments.
Generally, this depends intimately on the noise environment and on/off 
ratio of the coupling between the chains.
For a Rabi experiment as in Ref.~\cite{zatelli_majorana_2026} but with longer
chains, the cycle count initially follows the predicted exponential scaling
from $N=2$ to $N=3$ and still increases substantially at $N=4$ for experimentally relevant
parameters. At this length, coupling noise that is not suppressed by adding
sites, such as inner-dot noise, begins to limit the growth.
Here, revealing chain protection also increases the number of coherent
rotations available for qubit operation.

In contrast, a Rabi experiment with a direct tunnel barrier can probe
exponential protection through a longer dephasing time without increasing
the number of coherent rotations. Within our noise model, coupling noise
does not impose a length-independent ceiling on this lifetime. The accessible
scaling range instead depends on how far the tunnel coupling can be reduced
while the fractional-noise model remains valid. Finally, Ramsey experiments
directly probe exponential protection through pure dephasing, as long as
residual coupling is sufficiently suppressed. Direct-barrier Rabi
measurements and sufficiently isolated Ramsey storage may therefore probe
protection over a wider range of chain lengths than the optimized middle-dot
cycle count.

Our analysis provides an analytical description of quasistatic dephasing
in the Majorana parity qubit across the competing noise regimes and the associated changes in dephasing and visibility. 
The qualitative agreement of our $N=2$ theory with the key features observed in the recent
experiment~\cite{zatelli_majorana_2026}, in particular the behavior of the
number of coherent oscillations and the dephasing time with detuning,
gives us confidence that our model captures the essential physics of the
experimentally realized device.

Predicting these signatures quantitatively requires accounting for
the non-Gaussian splitting distribution. Approximating this distribution by a Gaussian
with the same variance would predict a leading exponential scaling
governed by the RMS splitting exponent. Accounting for its non-Gaussian
shape instead gives scaling governed by the typical splitting exponent,
changing the leading exponential dependence on chain length.

A preprint by Pan and Das Sarma~\cite{pan_scaling_2026}, posted while we
were finalizing this manuscript, shows that extending
the chain length from $N=2$ to $N=3$  does not automatically improve
Rabi or Ramsey coherence. Their analysis emphasizes that protection of the
chains need not translate into better qubit operation when control introduces
additional noise. Our results agree with this finding. However, we identify
how qubit dynamics can nevertheless reveal the protection of the constituent
chains. For Rabi experiments, we determine
how to adapt the operating point with chain length and which observable
reveals the suppression of splitting noise, even when the number of coherent
oscillations does not improve. The Ramsey protocols also probe different
aspects of the qubit: Ref.~\cite{pan_scaling_2026} studies rotations generated
by deliberately detuning the chains, whereas we consider storage at the sweet
spot with the interchain coupling minimized. Hence, the two works 
address complementary questions.

Finally, we note some limitations of our work. While we have argued that
experimentally motivated parameters allow the onset of exponential
protection to be observed, some uncertainties remain.
In particular, the precise noise ceiling in experiments remains unknown.
The current $N=2$ experiment is not sensitive to a ceiling much
above the observed coherence time or cycle count, e.g. due to interface
weight fluctuations. This limits the accuracy with which noise parameters
can be inferred. Measurements across several chain lengths could help
distinguish the limiting noise channels. Another limitation is our assumption of
quasistatic noise. While this is a reasonable assumption for the current
experiment, it is not clear whether this will remain the case for longer chains
where coherence times increase exponentially. Extending our analysis to
include the dynamics of the noise, for example for a $1/f$ spectrum, is an
interesting direction for future work.

\section{Conclusions} \label{sec:summary}

Our results show that unprotected qubit operations \emph{can} reveal
protection of the underlying Kitaev chains, provided the observable and
operating point are chosen appropriately. We have developed an analytical
theory of dephasing for the Majorana parity qubit with chains of arbitrary
length under quasistatic noise. We identify the corresponding length
scaling and the relevant constraints on coupling noise and tuning that
limit its observation. Nevertheless, current experimental parameters seem
already sufficient to see an onset of exponential protection in Rabi and
Ramsey experiments.

\paragraph{Author contributions}
S.M. and M.W. performed the analytical calculations, J.D.T.L. and S.M. performed the numerical simulations.
S.M. and M.W. wrote the manuscript with input from all authors.
M.W. supervised the project.
All authors contributed to the understanding and interpretation of the results through regular discussions. 

\paragraph{Acknowledgements}
We acknowledge Pablo Cova Fari\~na for helpful discussions. 
This work was supported by the Dutch Organization for Scientific Research (NWO) through No. OCENW.GROOT.2019.004, the Horizon Europe Framework Program of the European Commission through the European Innovation Council Pathfinder Grant No. 101115315 (QuKiT), and Microsoft Station Q.
We acknowledge OpenAI for providing access to the ChatGPT for Academic Researchers program.

\paragraph{Data availability}
All data and analysis notebooks associated with this work are available at Ref.~\cite{zenodo}.

\paragraph{AI disclosure}
GPT-5.5, GPT-5.6 and GPT-6 were used for mathematical derivations closely steered by the authors,
code generation, and assistance with writing this manuscript. 
The authors have reviewed, revised, and edited the content and take full responsibility for their
respective contributions, as specified in the author contribution statement.

\appendix

\section{Coulomb repulsion in the minimal Kitaev chain parity qubit} \label{app:coulomb}

We first consider two $N=2$ chains coupled through a central dot shown in Fig.~\ref{fig:figure_1}(a),
each tuned to the sweet spot,
$\mu_{\alpha i}=0$ and $t_\alpha=\Delta_\alpha=E_g/2$. 
We add Coulomb repulsion between electrons on the central and inner dots as
\begin{equation}
 H_U^{(D)}=U_{LD}n_{L2}n_D+U_{RD}n_{R1}n_D,
 \label{app:coulomb_dot_interaction}
\end{equation}
where $n_i=c_i^\dagger c_i$. 
We construct the many-body Hamiltonian~\cite{second_quantization} perform perturbation theory to obtain Eq.~\eqref{model:h_pq_general}~\cite{araya_day_pymablock_2025}.

The leading correction from Coulomb repulsion requires two tunnel events and first appears at total order $t_{\alpha,i}t_{\beta,i}U$.
With $\Gamma_0=t_{LD}t_{RD}/(2\mu_D)$, we find
\begin{equation}
 \frac{\delta\Gamma}{\Gamma_0}
 =-\frac{(U_{LD}+U_{RD})E_g}
          {2\mu_D(\mu_D+E_g)}.
 \label{app:coulomb_dot_correction}
\end{equation}
Thus repulsion reduces the interchain coupling, with a relative correction proportional to $\mu_D^{-2}$ at large detuning.
At the sweet spot the longitudinal correction vanishes through fourth total order. 
The same-side longitudinal contribution that decays as $\mu_D^{-4}$ at large detuning requires departure from that point.

For the parameters from Sec.~\ref{sec:rabi_optimization_crossover} and $\mu_D=130\,\mu\mathrm{eV}$ we obtain $\Gamma_0=0.2163\,\mu\mathrm{eV}$.
The correction from Eq.~\eqref{app:coulomb_dot_correction} is $\delta\Gamma/\Gamma_0=-(U_{LD}+U_{RD})/(1.105\,m\mathrm{eV})$.
For $U_{LD}=U_{RD}=40\,\mu\mathrm{eV}$, this predicts a $7.24\%$ reduction. 
Comparing to exact diagonalization we find a change of $5.31\%$. 
The distinction reflects the finite-interaction corrections
omitted by the linear-in-$U$ expression: for
$U_{LD}=U_{RD}=4\,\mu\mathrm{eV}$, the perturbative and exact reductions
are $0.724\%$ and $0.688\%$, respectively. At the detuning limit
$\mu_{D,\max}=10\,\mathrm{meV}$ used in the main text,
Eq.~\eqref{app:coulomb_dot_correction} gives only $0.00159\%$
for the $40\,\mu\mathrm{eV}$ interaction benchmark.

We now replace the central dot by the direct tunnel barrier shown in Fig.~\ref{fig:figure_1}(c).
In contrast to Eq.~\eqref{app:coulomb_dot_interaction}, the interaction couples the two inner dots directly
\begin{equation}
 H_U^{(B)}=U n_{L2}n_{R1}.
 \label{app:coulomb_barrier_interaction}
\end{equation}
Up to omitted global shifts, the leading correction is proportional to $\tau_0U$:
\begin{equation}
 \frac{\delta\Gamma}{\Gamma_0}=\frac{U}{4E_g}.
 \label{app:coulomb_barrier_correction}
\end{equation}
Repulsion therefore \emph{enhances} the coupling in both parity sectors. 
Its relative effect is controlled by $U/E_g$, without the suppression through central-dot detuning.

At the full sweet spot, the outer Majoranas commute with the coupling. 
Small detunings of the outer dots, $\mu_{L1}$ and $\mu_{R2}$, give the contribution
\begin{equation}
 \varepsilon_U^{(p)}
 =-\frac{U}{4E_g}(\mu_{L1}+p\mu_{R2}),
 \qquad p=\pm1,
 \label{app:coulomb_splitting_detuning}
\end{equation}
where $p$ is the total parity. 
We recognize that Coulomb repulsion have both quantum dots of each chain contribute to the coupling noise.

For the parameters from Sec.~\ref{sec:rabi_optimization_crossover} and $U=4\,\mu\mathrm{eV}$,
Eq.~\eqref{app:coulomb_barrier_correction} gives a $2.5\%$ enhancement ($2.26\%$ from exact diagonalization). 
The contribution of the outer dots to Eq.~\eqref{app:coulomb_splitting_detuning} also give an additional longitudinal contribution to the root-mean-square width.
For the parameters considered here we find this contribution to be $\sigma_{\varepsilon,U}=U\sigma_\mu/(2\sqrt{2}E_g)=0.0884\,\mu\mathrm{eV}$.

\section{Quasistatic dephasing of the minimal parity qubit}
\label{app:n2}
\label{app:probablity}

We derive the splitting distribution and Rabi response used in
Sec.~\ref{sec:minimal}, using the noise model defined in
Sec.~\ref{sec:noise}.

\subsection{Characteristic function of the splitting}
\label{app:n2_characteristic}

Combining the two chain splittings gives
\begin{align}
 \delta\varepsilon
 &=\frac{\delta(t_L-\Delta_L)+p\,\delta(t_R-\Delta_R)}{2}
 \nonumber\\
 &\quad-\frac{\delta\mu_{L,1}\delta\mu_{L,2}
              +p\,\delta\mu_{R,1}\delta\mu_{R,2}}{2E_g},
 \qquad p=\pm1.
 \label{probability:delta_vareps}
\end{align}
The first term is Gaussian with variance $\sigma_{t-\Delta}^2/2$.
For each onsite contribution, the characteristic function of a product
of two independent centered Gaussian variables gives~\cite{gaunt_basic_2022}
\begin{equation}
 \left\langle e^{-ik\delta\mu_{\alpha,1}\delta\mu_{\alpha,2}/(2E_g)}
 \right\rangle
 =\frac{1}{\sqrt{1+b_2^2k^2}},
 \label{n2:product_characteristic}
\end{equation}
where $b_2=\sigma_\mu^2/(2E_g)=u\sigma_{t-\Delta}/2$.
Since characteristic functions multiply when independent random variables
are added, the two onsite contributions have the combined characteristic
function $(1+b_2^2k^2)^{-1}$, corresponding to a Laplace distribution
of scale $b_2$. Their symmetry makes the result independent of $p$.
Multiplying by the characteristic function of the independent Gaussian
bond contribution gives the characteristic function of the qubit splitting:
\begin{equation}
 \left\langle e^{ik\delta\varepsilon}\right\rangle
 =\frac{e^{-k^2\sigma_{t-\Delta}^2/4}}{1+b_2^2k^2}.
 \label{probability:n2_kernel}
\end{equation}

Its mean vanishes and its variance is
$\sigma_\varepsilon^2=\sigma_{t-\Delta}^2/2+2b_2^2$, consistent with
Eq.~\eqref{minimal:sigma_epsilon}.

\subsection{Strong-coupling envelope from the characteristic function}
\label{app:n2_strong}

For $\sigma_\varepsilon,\sigma_\Gamma\ll\Gamma_0$, expanding the energy
and return weight gives
\begin{align}
 \Omega&=\Gamma_0+\delta\Gamma
 +\frac{\delta\varepsilon^2}{2\Gamma_0}
 -\frac{\delta\Gamma\,\delta\varepsilon^2}{2\Gamma_0^2}
 -\frac{\delta\varepsilon^4}{8\Gamma_0^3}+\cdots,
 \nonumber\\
 W&=1-\frac{\delta\varepsilon^2}{\Gamma_0^2}+\cdots.
 \label{n2:strong_expansion}
\end{align}
We retain the linear coupling shift and the quadratic splitting shift
in $\Omega$, and set $W\simeq1$. The full retained splitting from
Eq.~\eqref{probability:delta_vareps} enters this quadratic phase;
we do not expand it further in the microscopic noise variables.
Independence of the two noise channels then gives
\begin{align}
 \left\langle We^{-2i\Omega t/\hbar}\right\rangle
 &\simeq e^{-2i\Gamma_0t/\hbar}
 e^{-2\sigma_\Gamma^2t^2/\hbar^2}\psi_2(s;u),
 \nonumber\\
 s&=\frac{\sigma_\varepsilon^2t}{\Gamma_0\hbar}.
 \label{n2:strong_signal}
\end{align}

To evaluate the splitting factor directly from
Eq.~\eqref{probability:n2_kernel}, we use the Fourier transform of an
oscillatory Gaussian. Taking the limit of the convergent Gaussian
integral~\cite[Eq.~(7.7.3)]{nist_dlmf} gives, for $s>0$,
\begin{equation}
 \begin{aligned}
 e^{-is\varepsilon^2/\sigma_\varepsilon^2}
 &=\frac{\sigma_\varepsilon}{\sqrt{4\pi is}}
 \lim_{\eta\to0^+}\\
 &\quad\times\int_{-\infty}^{\infty}
 e^{-\eta k^2+ik^2\sigma_\varepsilon^2/(4s)-ik\varepsilon}\,dk.
 \end{aligned}
 \label{n2:fresnel_identity}
\end{equation}
Averaging over $\delta\varepsilon$ before taking the limit replaces
$e^{-ik\varepsilon}$ by the characteristic function. 
For $\sigma_{t-\Delta}>0$, the resulting Gaussian
factor makes the integral absolutely convergent, allowing us to take the limit $\eta\to0^+$ 
inside the integral. We then obtain:
\begin{equation}
 \psi_2(s;u)=\frac{\sigma_\varepsilon}{\sqrt{4\pi is}}
 \int_{-\infty}^{\infty}
 \frac{e^{ik^2\sigma_\varepsilon^2/(4s)-k^2\sigma_{t-\Delta}^2/4}}
 {1+b_2^2k^2}\,dk.
 \label{n2:characteristic_phase_average}
\end{equation}
With $x=b_2k$, we use~\cite[Eq.~(7.7.1)]{nist_dlmf}
\begin{equation}
 \int_{-\infty}^{\infty}\frac{e^{-ax^2}}{1+x^2}\,dx
 =\pi\operatorname{erfcx}(\sqrt a),
 \qquad \operatorname{Re}a>0,
 \label{n2:erfcx_identity}
\end{equation}
Here $\operatorname{erfcx}(z)=e^{z^2}\operatorname{erfc}(z)$, and we set
$a=\sigma_{t-\Delta}^2/(4b_2^2)+\sigma_\varepsilon^2/(4is b_2^2)$.
Expressing $a$ in terms of $u$ using Eq.~\eqref{minimal:sigma_epsilon} yields
\begin{equation}
 \psi_2(s;u)
 =\sqrt{\frac{\pi(1+u^2)}{2is\,u^2}}\,
 \operatorname{erfcx}\!\left[
 \sqrt{\frac{1+u^2+2is}{2is\,u^2}}\right].
 \label{prob:psi}
\end{equation}
At $s=0$, $\psi_2(0;u)=1$ by continuity. With only bond noise ($u=0$),
$\psi_2(s;0)=(1+2is)^{-1/2}$; continuation to zero bond noise gives
$\psi_2(s;\infty)=\sqrt{\pi/(2is)}\operatorname{erfcx}(1/\sqrt{2is})$.

Taking the modulus of Eq.~\eqref{n2:strong_signal} gives
Eq.~\eqref{minimal:strong_envelope}, from which the limiting dephasing
times in Eqs.~\eqref{minimal:strong_linear_dephasing} and
\eqref{minimal:strong_quadratic_dephasing} follow.
For $u=0$, i.e. only bond noise, $s_{1/e}(0)=\sqrt{e^4-1}/2$.
For general $u$, the crossing is obtained numerically.

\subsection{Splitting density}
\label{app:n2_distribution}

For the crossover calculation, we need the full splitting density.
The inverse Fourier transform of the characteristic function in
Eq.~\eqref{probability:n2_kernel} gives
\begin{equation*}
 p_\varepsilon(\varepsilon)
 =\frac{1}{2\pi}\int_{-\infty}^{\infty}
 \frac{e^{-ik\varepsilon-k^2\sigma_{t-\Delta}^2/4}}
 {1+b_2^2k^2}\,dk.
\end{equation*}
For $\sigma_{t-\Delta},b_2>0$, we evaluate this integral by writing
\begin{equation*}
 \frac{1}{1+b_2^2k^2}
 =\int_0^\infty e^{-v(1+b_2^2k^2)}\,dv.
\end{equation*}
The $k$-integral is then Gaussian and is performed first.
Substituting $y=\sqrt{\sigma_{t-\Delta}^2/4+b_2^2v}$ in the remaining
integral gives
\begin{align*}
 p_\varepsilon(\varepsilon)
 &=\frac{e^{\sigma_{t-\Delta}^2/(4b_2^2)}}{\sqrt\pi\,b_2^2}
 \int_{\sigma_{t-\Delta}/2}^{\infty}\\
 &\quad\times\exp\!\left[-\frac{y^2}{b_2^2}
 -\frac{\varepsilon^2}{4y^2}\right]dy.
\end{align*}
Using the standard integral~\cite[Eq.~(7.7.7)]{nist_dlmf}, we obtain
\begin{align}
 p_\varepsilon(\varepsilon)
 &=\frac{e^{\sigma_{t-\Delta}^2/(4b_2^2)}}{4b_2}
 \sum_{\eta=\pm1}e^{-\eta\varepsilon/b_2}
 \nonumber\\
 &\qquad\times\operatorname{erfc}\!\left(
 \frac{\sigma_{t-\Delta}^2/(2b_2)-\eta\varepsilon}
 {\sigma_{t-\Delta}}\right).
 \label{probability:p_varepsilon}
\end{align}
Evaluating this density at $\varepsilon=0$, using
$u=2b_2/\sigma_{t-\Delta}$, gives
\begin{equation}
 p_\varepsilon(0)
 =\frac{\operatorname{erfcx}(1/u)}{2b_2}
 =\frac{E_g}{\sigma_\mu^2}\operatorname{erfcx}(1/u).
 \label{n2:density_zero}
\end{equation}
With only onsite noise, $p_\varepsilon=e^{-|\varepsilon|/b_2}/(2b_2)$;
with only bond noise it is Gaussian with variance
$\sigma_{t-\Delta}^2/2$.

\subsection{Crossover integral}
\label{app:n2_crossover}

We now assume weak coupling noise, $\sigma_\Gamma\ll\Gamma_0$, while
allowing arbitrary $\delta\varepsilon/\Gamma_0$. We first average over
the coupling noise at fixed splitting:
\begin{equation}
 \left\langle We^{-2i\Omega t/\hbar}\right\rangle=\int d\varepsilon\,p_\varepsilon(\varepsilon)
 \left\langle W e^{-2i\Omega t/\hbar}\right\rangle_{\delta\Gamma}.
 \label{n2:oscillatory_average}
\end{equation}
At fixed $\varepsilon$, put
$\Omega_0=\sqrt{\Gamma_0^2+\varepsilon^2}$ and
$W_0=\Gamma_0^2/\Omega_0^2$. Expanding only in coupling noise gives
\begin{align}
 \Omega&=\Omega_0+\frac{\Gamma_0}{\Omega_0}\delta\Gamma
 +\frac{\varepsilon^2}{2\Omega_0^3}(\delta\Gamma)^2+\cdots,
 \label{n2:energy_expansion}\\
 W&=W_0+\frac{2\Gamma_0\varepsilon^2}{\Omega_0^4}\delta\Gamma
 +O\!\left(\frac{(\delta\Gamma)^2}{\Gamma_0^2}\right).
 \label{n2:weight_expansion}
\end{align}
We linearize $\Omega$ and $W$ in $\delta\Gamma$ and keep the full
exponential $e^{-2it\Gamma_0\delta\Gamma/(\hbar\Omega_0)}$ when averaging
over the coupling noise. Neglecting the quadratic term in $\Omega$
requires its contribution to the accumulated phase to remain small,
\begin{equation}
 \frac{t\sigma_\Gamma^2\varepsilon^2}{\hbar\Omega_0^3}\ll1
 \label{n2:crossover_validity}
\end{equation}
for the realizations contributing to the signal.
The required Gaussian averages are
\begin{align}
 \left\langle e^{-ik\delta\Gamma}\right\rangle
 &=e^{-k^2\sigma_\Gamma^2/2},\nonumber\\
 \left\langle\delta\Gamma e^{-ik\delta\Gamma}\right\rangle
 &=-ik\sigma_\Gamma^2e^{-k^2\sigma_\Gamma^2/2}.
 \label{n2:gaussian_phase_averages}
\end{align}
Evaluating these at $k=2t\Gamma_0/(\hbar\Omega_0)$ and substituting
in Eq.~\eqref{n2:oscillatory_average} yields
\begin{align}
 \left\langle We^{-2i\Omega t/\hbar}\right\rangle&\simeq\int_{-\infty}^{\infty}d\varepsilon\,
 p_\varepsilon(\varepsilon)W_0
 \left(1-\frac{4it\sigma_\Gamma^2\varepsilon^2}
 {\hbar\Omega_0^3}\right)\nonumber\\
 &\quad\times\exp\!\left[-\frac{2it\Omega_0}{\hbar}
 -\frac{2t^2\sigma_\Gamma^2\Gamma_0^2}{\hbar^2\Omega_0^2}\right],
 \label{n2:crossover_average}\\
 \mathcal V&\simeq\int_{-\infty}^{\infty}
 p_\varepsilon(\varepsilon)\frac{\Gamma_0^2}{\Gamma_0^2+\varepsilon^2}
 \,d\varepsilon.
 \label{n2:crossover_visibility}
\end{align}
Taking the modulus and normalizing by the visibility gives the envelope
\begin{equation}
 \begin{aligned}
 \mathcal C(t)\simeq\frac{1}{\mathcal V}\Bigg|
 &\int_{-\infty}^{\infty}d\varepsilon\,
 p_\varepsilon(\varepsilon)\frac{\Gamma_0^2}{\Omega_0^2}
 \left(1-\frac{4it\sigma_\Gamma^2\varepsilon^2}{\hbar\Omega_0^3}\right)\\
 &\times\exp\!\left[-\frac{2it\Omega_0}{\hbar}
 -\frac{2t^2\sigma_\Gamma^2\Gamma_0^2}{\hbar^2\Omega_0^2}\right]\Bigg|.
 \end{aligned}
 \label{n2:crossover_envelope}
\end{equation}
$\mathcal C(t)$ and its first $1/e$ crossing are evaluated numerically.

\subsubsection{Long-time oscillations}
\label{app:n2_long_time}

We first neglect coupling noise. Following the
long-time analysis of Koppens et al.~\cite{koppens_universal_2007},
we evaluate the oscillatory average in Eq.~\eqref{n2:crossover_average}
near its stationary point at $\varepsilon=0$,
where $W_0=1$ and
$\Omega_0=\Gamma_0+\varepsilon^2/(2\Gamma_0)+O(\varepsilon^4/\Gamma_0^3)$.
Near $\varepsilon=0$, contributions largely cancel where
the phase $\varepsilon^2t/(\Gamma_0\hbar)$ varies rapidly, so the leading long-time contribution
comes from the region
\begin{equation*}
 |\varepsilon|\lesssim\sqrt{\frac{\Gamma_0\hbar}{t}}.
\end{equation*}
Since $p_\varepsilon$ is finite
and continuous at zero, we also replace $p_\varepsilon(\varepsilon)$
in this window by $p_\varepsilon(0)$, provided the region is narrow
compared with the scale over which the density varies.
In the crossover
$\sigma_\varepsilon\sim\Gamma_0$, requiring this region to be narrow
compared with $\Gamma_0$ gives $t\gg\hbar/\Gamma_0$.

With these approximations, the oscillatory average becomes
\begin{align}
 \left\langle We^{-2i\Omega t/\hbar}\right\rangle
 &\simeq p_\varepsilon(0)e^{-2i\Gamma_0t/\hbar}
 \int_{-\infty}^{\infty}e^{-it\varepsilon^2/(\Gamma_0\hbar)}d\varepsilon
 \nonumber\\
 &=p_\varepsilon(0)\sqrt{\frac{\pi\Gamma_0\hbar}{t}}
 e^{-i(2\Gamma_0t/\hbar+\pi/4)}.
 \label{n2:stationary_phase}
\end{align}
Its real part gives Eq.~\eqref{minimal:crossover_tail}, and its modulus
gives $\mathcal C(t)\simeq[p_\varepsilon(0)/\mathcal V]
\sqrt{\pi\Gamma_0\hbar/t}$. Our non-Gaussian distribution changes
the prefactor but preserves the $t^{-1/2}$ decay and $\pi/4$ phase
shift of Ref.~\cite{koppens_universal_2007}.

Restoring weak coupling noise introduces the factor
$e^{-2\sigma_\Gamma^2t^2/\hbar^2}$ near the stationary point. 
This additional damping is
negligible for $t\ll\hbar/\sigma_\Gamma$. 
Hence, the algebraic tail applies in the
intermediate-time window
$\hbar/\Gamma_0\ll t\ll\hbar/\sigma_\Gamma$.

\subsection{Weak-coupling limit}
\label{app:n2_weak}

We now take the limit of weak coupling, $\Gamma_0\to 0$. In this
limit, coupling noise becomes negligible because
$\sigma_\Gamma/\Gamma_0\to0$ for middle-dot control,
and we consequently set $\sigma_\Gamma=0$.
The return weight
$W_0=\Gamma_0^2/(\Gamma_0^2+\varepsilon^2)$ suppresses realizations
with $|\varepsilon|\gg\Gamma_0$. As $\Gamma_0$ decreases, the
oscillatory signal therefore samples an increasingly narrow region
of the splitting density around zero, as in the weak-drive regime
discussed by Koppens et al.~\cite{koppens_universal_2007}.

To resolve this region, we substitute $\varepsilon=\Gamma_0y$.
The weight becomes $W_0=(1+y^2)^{-1}$, while the phase is
$2\Omega_0t/\hbar=\tau\sqrt{1+y^2}$, where $\tau=2\Gamma_0t/\hbar$.
Equation~\eqref{n2:crossover_average}
then gives
\begin{align*}
 \left\langle We^{-2i\Omega t/\hbar}\right\rangle
 &\simeq\Gamma_0\int_{-\infty}^{\infty}dy\,
 p_\varepsilon(\Gamma_0y)\\
 &\qquad\times\frac{e^{-i\tau\sqrt{1+y^2}}}{1+y^2}.
\end{align*}
In this expression, we can replace
$p_\varepsilon(\Gamma_0y)$ by $p_\varepsilon(0)$ as $\Gamma_0\to0$.
Applying the same substitution to
the visibility in Eq.~\eqref{n2:crossover_visibility} yields
\begin{align}
 \left\langle We^{-2i\Omega t/\hbar}\right\rangle&\simeq\Gamma_0p_\varepsilon(0)
 \int_{-\infty}^{\infty}\frac{e^{-i\tau\sqrt{1+y^2}}}{1+y^2}\,dy,
 \nonumber\\
 \mathcal V&\simeq\pi\Gamma_0p_\varepsilon(0).
 \label{n2:weak_signal}
\end{align}
The common factor $\Gamma_0p_\varepsilon(0)$ cancels from
$\mathcal C(t)$, yielding an envelope
that depends only on the scaled time:
\begin{equation}
 \mathcal C(t)\simeq\left|\frac1\pi\int_{-\infty}^{\infty}
 \frac{e^{-i\tau\sqrt{1+y^2}}}{1+y^2}\,dy\right|.
 \label{minimal:weak_contour}
\end{equation}

Its first $1/e$ crossing is $\tau_{1/e}\simeq4.41632$, giving
\begin{equation}
 T_\varphi\simeq\frac{\hbar\tau_{1/e}}{2\Gamma_0},\qquad
 N_\varphi\simeq\frac{\tau_{1/e}}{2\pi}\simeq0.70288.
 \label{n2:weak_time_cycles}
\end{equation}
The long-time envelope follows from Eq.~\eqref{n2:stationary_phase} as
$\mathcal C\simeq\sqrt{2/(\pi\tau)}$.

\section{Energy splitting and its distribution in a disordered Kitaev chain}
\label{app:splitting}
\label{app:toeplitz}

We derive the distribution of the parity splitting of an isolated
chain subject to quasistatic onsite and bond noise. Near the sweet
spot, a determinant recurrence relates the splitting to the
microscopic disorder. This recurrence provides the starting point
for the finite-length moments and the logarithmic statistics
that determine the typical splitting.

\subsection{Signed parity splitting and the continuant representation}
\label{app:splitting_model}
\label{app:splitting_recurrence}

For one chain, we suppress the chain label and define the signed
parity splitting by
\begin{equation}
 \delta\epsilon_N=E_{e,N}-E_{o,N},
 \label{splitting:parity_convention}
\end{equation}
where $E_{e,N}$ and $E_{o,N}$ are the lowest energies in the even
and odd fermion-parity sectors.

\subsubsection{Majorana representation and the signed gap}

At a fixed disorder realization, the spinless chain Hamiltonian is
\begin{align}
 H&=\sum_{j=1}^{N}\mu_jc_j^\dagger c_j\nonumber\\
 &\quad+\sum_{j=1}^{N-1}\left(
 t_jc_j^\dagger c_{j+1}
 +\Delta_jc_j^\dagger c_{j+1}^\dagger+\mathrm{h.c.}\right),
 \label{splitting:microscopic_hamiltonian}
\end{align}
with real hopping and pairing amplitudes. To compute the spectrum
and its parity, we use the site Majoranas
\begin{equation}
 a_j=c_j+c_j^\dagger,\qquad b_j=i(c_j-c_j^\dagger).
 \label{splitting:site_majoranas}
\end{equation}
The Hamiltonian then takes the form
\begin{align}
 H&=\frac12\sum_j\mu_j-\frac{i}{2}\bm a^T B_N\bm b,
 \label{splitting:majorana_hamiltonian}\\
 B_N&=\begin{pmatrix}
 \mu_1&t_1-\Delta_1&&\\
 t_1+\Delta_1&\mu_2&t_2-\Delta_2&\\
 &\ddots&\ddots&\ddots\\
 &&t_{N-1}+\Delta_{N-1}&\mu_N
 \end{pmatrix}.
 \label{splitting:majorana_block}
\end{align}

Here $\bm a=(a_1,\ldots,a_N)^T$ and $\bm b=(b_1,\ldots,b_N)^T$.
The singular values $0\leq s_1\leq\cdots\leq s_N$ of $B_N$ are the single-quasiparticle
excitation energies. With Kitaev's ground-state
parity expression $P_0=\operatorname{sgn}\det B_N$ \cite{kitaev_unpaired_2001},
we thus find
\begin{equation}
 \delta\epsilon_N
 =-\operatorname{sgn}(\det B_N)s_1
 =-\frac{\det B_N}{\prod_{k=2}^N s_k}.
 \label{splitting:exact_gap}
\end{equation}

At the uniform sweet spot, $\mu_j=0$ and
$t_j=\Delta_j=E_g/2$ with $E_g>0$, all $N-1$ bulk singular values
equal $E_g$. For disorder much weaker than the clean gap,
we can approximate the bulk singular-value product in
Eq.~\eqref{splitting:exact_gap} by its clean value and obtain
\begin{equation}
 \delta\epsilon_N\simeq-\frac{\det B_N}{E_g^{N-1}}.
 \label{toeplitz:delta_determinant}
\end{equation}

\subsubsection{Determinant recurrence and microscopic noise}

Let $K_n=\det B_n$, where $B_n$ is the leading $n\times n$
block of $B_N$. Expanding the determinant along its last row gives
\begin{equation}
 K_n=\mu_nK_{n-1}
 -(t_{n-1}+\Delta_{n-1})(t_{n-1}-\Delta_{n-1})K_{n-2},
 \label{splitting:general_continuant}
\end{equation}
with $K_0=1$ and $K_1=\mu_1$. The tridiagonal determinant $K_n$
is called a continuant.

For weak disorder about the uniform sweet spot, we replace the bond
sum $t_j+\Delta_j$ by its clean value $E_g$, neglecting products of
fluctuations of the bond sum and bond mismatch in
Eq.~\eqref{splitting:general_continuant}. We therefore use
\begin{align}
 \mu_j&=\delta\mu_j=\sigma_\mu G_j,\nonumber\\
 t_j+\Delta_j&\simeq E_g,\qquad
 t_j-\Delta_j=\delta(t_j-\Delta_j)=\sigma_{t-\Delta}H_j.
 \label{splitting:noise_ensemble}
\end{align}
Here $G_j$ and $H_j$ are mutually independent standard normal
variables, and the disorder widths are uniform along the chain.
Within this approximation, Eq.~\eqref{splitting:general_continuant}
reduces to
\begin{align}
 K_n&=\delta\mu_nK_{n-1}
       -E_g\delta(t_{n-1}-\Delta_{n-1})K_{n-2},
 \label{toeplitz:noise_param_continuant}\\
 K_0&=1,\qquad K_1=\delta\mu_1.\nonumber
\end{align}
For $N=2,3$, combining this recurrence with
Eq.~\eqref{toeplitz:delta_determinant} gives
\begin{align}
 \delta\epsilon_2&\simeq\delta(t_1-\Delta_1)
              -\frac{\delta\mu_1\delta\mu_2}{E_g},\nonumber\\
 \delta\epsilon_3&\simeq
 \frac{\delta\mu_3\delta(t_1-\Delta_1)
       +\delta\mu_1\delta(t_2-\Delta_2)}{E_g}\nonumber\\
 &\quad-\frac{\delta\mu_1\delta\mu_2\delta\mu_3}{E_g^2}.
 \label{splitting:short_gaps}
\end{align}
Bond noise thus contributes directly for $N=2$, while for $N=3$
it enters through products with onsite noise. For $N\geq3$, the
polynomial contributions generally share disorder variables and
are not mutually independent.

\subsubsection{Dimensionless recurrence}

For $\sigma_{t-\Delta}>0$, we separate the overall disorder scale
from the relative strength of onsite and bond noise by defining
\begin{equation}
 d=\frac{\sigma_{t-\Delta}}{E_g},\qquad
 u=\frac{\sigma_\mu^2}{E_g\sigma_{t-\Delta}},\qquad
 \delta Y_n=-\frac{K_n}{E_g^n d^{n/2}}.
 \label{splitting:scales}
\end{equation}
Substitution into Eq.~\eqref{toeplitz:noise_param_continuant} yields
\begin{align}
 \delta Y_n&=\sqrt u\,G_n\delta Y_{n-1}-H_{n-1}\delta Y_{n-2},
 \label{splitting:scaled_recurrence}\\
 \delta Y_0&=-1,\qquad \delta Y_1=-\sqrt u\,G_1.\nonumber
\end{align}
The splitting is therefore
\begin{equation}
 \delta\epsilon_N\simeq E_gd^{N/2}\delta Y_N.
 \label{splitting:physical_scale}
\end{equation}
The distribution of $\delta Y_N$ depends only on $N$ and $u$,
while $E_gd^{N/2}$ sets the energy scale. Pure onsite noise,
for which $\sigma_{t-\Delta}=0$, is treated directly with the
unscaled recurrence.

\subsection{Second moment of the energy splitting}
\label{app:splitting_moments}

We compute the second moment
\begin{equation}
 m_N(u)=\bigl\langle(\delta Y_N)^2\bigr\rangle,
 \label{splitting:second_moment_definition}
\end{equation}
where the average is over the disorder ensemble. For $N\geq2$, the new
variables $G_N$ and $H_{N-1}$ are independent of each other and of
$\delta Y_{N-1}$ and $\delta Y_{N-2}$. Their zero means, together with
$\delta Y_1=-\sqrt u\,G_1$, give $\langle\delta Y_N\rangle=0$ for
$N\geq1$, so $m_N$ is also the variance at these lengths.

Squaring Eq.~\eqref{splitting:scaled_recurrence} and averaging gives
\begin{align}
 m_N={}&u\bigl\langle G_N^2(\delta Y_{N-1})^2\bigr\rangle
       +\bigl\langle H_{N-1}^2(\delta Y_{N-2})^2\bigr\rangle\nonumber\\
 &-2\sqrt u\,\bigl\langle
 G_NH_{N-1}\delta Y_{N-1}\delta Y_{N-2}\bigr\rangle.
 \label{splitting:second_moment_expansion}
\end{align}
The cross term vanishes because
\begin{align}
 &\bigl\langle G_NH_{N-1}\delta Y_{N-1}\delta Y_{N-2}\bigr\rangle\nonumber\\
 &\qquad=\langle G_N\rangle\langle H_{N-1}\rangle
          \langle\delta Y_{N-1}\delta Y_{N-2}\rangle=0.
 \label{splitting:moment_cross_term}
\end{align}
Using $\langle G_N^2\rangle=\langle H_{N-1}^2\rangle=1$
in the remaining terms yields
\begin{equation}
 m_N=u\,m_{N-1}+m_{N-2},\qquad m_0=1,\quad m_1=u.
 \label{splitting:moment_recurrence}
\end{equation}

To solve this recurrence, the ansatz $m_N\propto r^N$ gives
\begin{equation}
 r^2-ur-1=0,\qquad r_\pm=\frac{u\pm\sqrt{u^2+4}}{2}.
 \label{splitting:moment_roots}
\end{equation}
The general solution is a linear combination of $r_+^N$ and
$r_-^N$. Imposing the two initial conditions gives
\begin{equation}
 m_N=\frac{r_+^{N+1}-r_-^{N+1}}{r_+-r_-}.
 \label{splitting:finite_second_moment}
\end{equation}
Restoring the energy scale with Eq.~\eqref{splitting:physical_scale},
we obtain the finite-length splitting variance
\begin{equation}
 \operatorname{Var}(\delta\epsilon_N)
 \simeq E_g^2d^N\frac{r_+^{N+1}-r_-^{N+1}}{r_+-r_-}.
 \label{splitting:physical_variance}
\end{equation}

For fixed $u>0$, $|r_-|<r_+$, so the positive root determines the
large-$N$ behavior. We separate this exponential dependence by writing
\begin{align}
 M_N(u)&=\frac{m_N}{r_+^N}
 =\frac{r_+}{r_+-r_-}
 \left[1-\left(\frac{r_-}{r_+}\right)^{N+1}\right],
 \label{splitting:finite_moment_normalization}\\
 M_\infty(u)&=\lim_{N\to\infty}M_N(u)
 =\frac{r_+}{\sqrt{u^2+4}}.
 \label{splitting:moment_normalization_limit}
\end{align}
The root-mean-square splitting then scales as
\begin{align}
 \sqrt{\operatorname{Var}(\delta\epsilon_N)}
 &\simeq E_g\sqrt{M_\infty(u)}\,e^{N\lambda_{\mathrm{rms}}},
 \label{splitting:rms_asymptotic}\\
 \lambda_{\mathrm{rms}}(u)
 &=\frac12\log d+\frac12\log r_+(u).
 \label{splitting:rms_exponent}
\end{align}
Note that the negative root produces a finite-length even--odd correction.
In the extreme limit of only bond disorder, $u=0$, the roots are $r_\pm=\pm1$, so
Eq.~\eqref{splitting:finite_second_moment} gives $m_{2j}=1$ and $m_{2j+1}=0$.

\subsection{Typical energy splitting}
\label{app:splitting_typical}

Already for $N=2$, onsite disorder produces a non-Gaussian splitting
distribution with tails that decay more slowly than Gaussian tails.
The second moment weights large splittings strongly and can be dominated
by rare disorder realizations. To characterize the typical magnitude,
we thus use the geometric mean,
\begin{equation}
 |\delta\epsilon_N|_{\mathrm{typ}}
 =E_g\exp\left\langle\log\frac{|\delta\epsilon_N|}{E_g}\right\rangle.
 \label{splitting:typical_definition}
\end{equation}
Equation~\eqref{splitting:physical_scale}
gives
\begin{equation}
 \left\langle\log\frac{|\delta\epsilon_N|}{E_g}\right\rangle
 \simeq\frac N2\log d+\langle\log|\delta Y_N|\rangle.
 \label{splitting:mean_log_scale}
\end{equation}
For fixed $u>0$, we define the logarithmic growth rate
\begin{equation}
 L(u)=\lim_{N\to\infty}\frac{\langle\log|\delta Y_N|\rangle}{N}.
 \label{splitting:lyapunov_definition}
\end{equation}
We define the physical exponent $\lambda_{\mathrm{typ}}$ through the
large-$N$ behavior of the geometric mean,
\begin{equation}
 |\delta\epsilon_N|_{\mathrm{typ}}
 =E_g\exp\bigl[N\lambda_{\mathrm{typ}}+o(N)\bigr].
 \label{splitting:typical_exponent_definition}
\end{equation}
A negative $\lambda_{\mathrm{typ}}$ corresponds to exponential suppression
with chain length; the subleading term allows for a prefactor.
Within the splitting approximation,
Eqs.~\eqref{splitting:mean_log_scale} and
\eqref{splitting:lyapunov_definition} give
\begin{equation}
 \lambda_{\mathrm{typ}}=\frac12\log d+L(u).
 \label{splitting:typical_exponent}
\end{equation}
Thus $L$ depends only on $u$, while the physical exponent also depends
on the overall noise scale $d$. We first evaluate the two pure-disorder limits, then compute $L(u)$
for mixed disorder.

\subsubsection{Pure onsite and pure bond disorder}

Both limits reduce the splitting to a product of independent Gaussian
variables. For $G\sim\mathcal N(0,1)$, we have
$\log|G|=\tfrac12\log(G^2)$ with $G^2\sim\chi_1^2$.
The mean logarithm of a chi-square variable~\cite{pav_moments_2015}
therefore gives
\begin{equation}
 \langle\log|G|\rangle=-c_G,\qquad
 c_G=\frac{\gamma_{\mathrm E}+\log2}{2}\simeq0.63518,
 \label{splitting:log_gaussian_mean}
\end{equation}
where $\gamma_{\mathrm E}$ is Euler's constant.

For pure onsite disorder, $\sigma_{t-\Delta}=0$, the unscaled recurrence
in Eq.~\eqref{toeplitz:noise_param_continuant} gives
$K_N=\prod_{j=1}^N\delta\mu_j$. Thus
\begin{align}
 \delta\epsilon_N&\simeq
 -E_g\left(\frac{\sigma_\mu}{E_g}\right)^N\prod_{j=1}^NG_j,
 \label{splitting:onsite_product}\\
 \left\langle\log\frac{|\delta\epsilon_N|}{E_g}\right\rangle
 &\simeq N\log\frac{\sigma_\mu}{E_g}
       +\sum_{j=1}^N\langle\log|G_j|\rangle\nonumber\\
 &=N\left(\log\frac{\sigma_\mu}{E_g}-c_G\right).
 \label{splitting:onsite_mean_log}
\end{align}
Since $\langle G_j^2\rangle=1$, the RMS splitting is
$E_g(\sigma_\mu/E_g)^N$ within the same approximation. The typical
and RMS exponents are therefore
\begin{equation}
 \lambda_{\mathrm{typ}}^{\mathrm{onsite}}
 =\log\frac{\sigma_\mu}{E_g}-c_G,\qquad
 \lambda_{\mathrm{rms}}^{\mathrm{onsite}}
 =\log\frac{\sigma_\mu}{E_g}.
 \label{splitting:onsite_exponents}
\end{equation}
The statistics of the corresponding finite-length Lyapunov exponent
for onsite disorder have recently been studied using large-deviation
theory by Fortin \textit{et al.}~\cite{fortin_finite_2026}.

For pure bond disorder, $u=0$, Eq.~\eqref{splitting:scaled_recurrence}
reduces to $\delta Y_n=-H_{n-1}\delta Y_{n-2}$. The two initial values
give
\begin{equation}
 \delta Y_{2j}=(-1)^{j+1}\prod_{k=1}^{j}H_{2k-1},\qquad
 \delta Y_{2j+1}=0.
 \label{splitting:bond_product}
\end{equation}
There is one Gaussian factor per two sites, so
$\langle\log|\delta Y_{2j}|\rangle=-jc_G$. Using the second moment
$m_{2j}=1$ derived above, the exponents along the even subsequence are
\begin{equation}
 \lambda_{\mathrm{typ}}^{\mathrm{bond,even}}
 =\frac12\log d-\frac{c_G}{2},\qquad
 \lambda_{\mathrm{rms}}^{\mathrm{bond,even}}=\frac12\log d.
 \label{splitting:bond_exponents}
\end{equation}
At $u=0$, the logarithm is $-\infty$ for odd lengths;
the value $-c_G/2$ is the dimensionless rate along even lengths,
not a finite limit over all $N$.

Importantly, in both limits the typical magnitude is
\emph{exponentially smaller} than the RMS.

\subsubsection{Typical splitting for mixed disorder}
\label{app:splitting_markov}

Wright and Trefethen~\cite{wright_computing_2001} computed Lyapunov constants
for Gaussian random recurrences from discretized invariant measures.
Here we formulate the corresponding stationary problem for the
splitting recurrence at arbitrary $u>0$.

For mixed disorder, $u>0$, we express the logarithm as a sum of
successive increments. Define
\begin{equation}
 r_n=\frac{\delta Y_n}{\delta Y_{n-1}},\qquad q_n=\log|r_n|.
 \label{splitting:log_ratio}
\end{equation}
The ratios are defined almost surely. Since $|\delta Y_0|=1$,
\begin{equation}
 \sum_{n=1}^Nq_n
 =\sum_{n=1}^N\bigl(\log|\delta Y_n|-\log|\delta Y_{n-1}|\bigr)
 =\log|\delta Y_N|.
 \label{splitting:log_ratio_sum}
\end{equation}
Dividing Eq.~\eqref{splitting:scaled_recurrence} by $\delta Y_{n-1}$
gives
\begin{equation}
 r_n=\sqrt u\,G_n-\frac{H_{n-1}}{r_{n-1}}.
 \label{splitting:ratio_recurrence}
\end{equation}
Conditioned on $r_{n-1}$, the two new noise variables $G_n$ and
$H_{n-1}$ remain independent standard normals. Their linear combination
is therefore centered Gaussian, with
\begin{equation}
 \operatorname{Var}(r_n\mid r_{n-1})=u+\frac{1}{r_{n-1}^2}.
 \label{splitting:conditional_ratio_variance}
\end{equation}
We can represent this conditional distribution as
\begin{equation}
 r_n\mid r_{n-1}\ \overset{d}{=}
 \sqrt{u+r_{n-1}^{-2}}\,G'_n,
 \label{splitting:conditional_ratio_gaussian}
\end{equation}
where $G'_n$ is a standard normal variable independent of the preceding
chain, and $\overset{d}{=}$ denotes equality in distribution.
Since $r_{n-1}^{-2}=e^{-2q_{n-1}}$, taking the logarithm of the absolute
value gives the equivalent Markov recursion
\begin{align}
 q_n&=\frac12\log\bigl(u+e^{-2q_{n-1}}\bigr)+\xi_n\nonumber\\
 &=h_u(q_{n-1})+\xi_n,
 \label{splitting:log_ratio_recurrence}
\end{align}
where $h_u(q)=\tfrac12\log(u+e^{-2q})$ and
$\xi_n=\log|G'_n|$. Thus the current log-ratio sets the scale of the
next ratio, while the independent $\xi_n$ supplies its random increment.

Since $(G'_n)^2\sim\chi_1^2$ and
$\xi_n=\tfrac12\log[(G'_n)^2]$, rescaling the log-chi-square
density~\cite{pav_moments_2015} gives
\begin{equation}
 f_\xi(\xi)=\sqrt{\frac2\pi}
 \exp\left(\xi-\frac{e^{2\xi}}2\right),\qquad -\infty<\xi<\infty.
 \label{splitting:log_gaussian_density}
\end{equation}
At fixed $q_n$, the next log-ratio is obtained by adding $h_u(q_n)$
to this random increment. Its conditional density is therefore
\begin{equation}
 p(q_{n+1}=q'\mid q_n)=f_\xi\bigl(q'-h_u(q_n)\bigr).
 \label{splitting:conditional_log_ratio_density}
\end{equation}
If $p_n(q)$ is the density of $q_n$, averaging this conditional density
over the preceding value gives the transfer equation
\begin{equation}
 p_{n+1}(q')=(T_up_n)(q')
 =\int_{-\infty}^{\infty}f_\xi\bigl(q'-h_u(q)\bigr)p_n(q)\,dq.
 \label{splitting:density_operator}
\end{equation}
For fixed $u>0$, the process approaches a stationary density $p_u$
satisfying
\begin{equation}
 T_up_u=p_u,\qquad \int_{-\infty}^{\infty}p_u(q)\,dq=1.
 \label{splitting:stationary_density}
\end{equation}
Taking expectations in Eq.~\eqref{splitting:log_ratio_sum} and dividing
by $N$ then gives
\begin{equation}
 L(u)=\lim_{N\to\infty}\frac1N\sum_{n=1}^N\langle q_n\rangle
 =\int_{-\infty}^{\infty}q\,p_u(q)\,dq.
 \label{splitting:stationary_mean}
\end{equation}

Numerically, we represent the density on a uniform grid over $[-Q,Q]$
and iterate $T_u$ from a normalized initial density. Each step moves
the probability mass under $h_u$, interpolates it onto the grid, and
convolves it with $f_\xi$. We renormalize the truncated density and
stop when the change between successive iterates is below the chosen
tolerance. The stationary mean is then evaluated by quadrature.
Accuracy is checked by reducing the grid spacing and increasing $Q$;
convergence of the iteration alone does not control either error.
Repeating this calculation for each $u$ determines the typical exponent
through Eq.~\eqref{splitting:typical_exponent}.

\subsubsection{Finite-length offset of the mean logarithm}
\label{app:splitting_offset}

The physical initial condition $\delta Y_1=-\sqrt u\,G_1$ corresponds to
\begin{equation}
 q_1=\frac12\log u+\log|G_1|,\qquad
 p_1(q)=f_\xi\left(q-\frac12\log u\right).
 \label{splitting:initial_log_ratio_density}
\end{equation}
This density is generally not stationary. Its approach to $p_u$ leaves
a constant offset in the mean logarithm,
\begin{equation}
 b_Y(u)=\lim_{N\to\infty}
 \left[\langle\log|\delta Y_N|\rangle-NL(u)\right].
 \label{splitting:boundary_offset}
\end{equation}
Propagating $p_n=T_u^{n-1}p_1$ and using
Eq.~\eqref{splitting:log_ratio_sum}, we obtain
\begin{align}
 \langle\log|\delta Y_N|\rangle-NL
 &=\sum_{n=1}^N\bigl(\langle q_n\rangle-L\bigr),\nonumber\\
 b_Y&=\sum_{n=1}^{\infty}\int_{-\infty}^{\infty}
       (q-L)p_n(q)\,dq.
 \label{splitting:boundary_offset_sum}
\end{align}
For $u>0$, the transient decays sufficiently rapidly for this sum to
converge. We compute $b_Y$ by propagating the physical density $p_1$
and summing the deviations from the stationary mean, checking convergence
with the number of steps as well as the grid spacing and interval.

Combining this offset with Eq.~\eqref{splitting:mean_log_scale} gives
\begin{align}
 \langle\log|\delta Y_N|\rangle&=NL(u)+b_Y(u)+o(1),
 \label{splitting:mean_log_asymptotic}\\
 |\delta\epsilon_N|_{\mathrm{typ}}
 &\simeq E_g e^{b_Y(u)}e^{N\lambda_{\mathrm{typ}}}[1+o(1)].
 \label{splitting:typical_prefactor}
\end{align}

\subsection{Fluctuations around the typical splitting}
\label{app:splitting_fluctuations}

The coefficients $L(u)$ and $b_Y(u)$ determine the mean logarithm and
hence the typical splitting. The variance and higher cumulants do not
enter this mean. However,  disorder-averaged qubit coherence depends on
nonlinear functions of the splitting and therefore on its \emph{full distribution}.
To prepare for this later calculation, we calculate the variance
and the third cumulant, which describes its leading asymmetry.

\subsubsection{Variance of the log-splitting}
\label{app:splitting_log_variance}

The log-splitting is a sum of correlated increments,
Eq.~\eqref{splitting:log_ratio_sum}. For sufficiently rapid mixing,
its central fluctuations obey a central limit theorem, giving an
approximately log-normal distribution of the splitting magnitude.
Such log-normal statistics of Majorana end-state energies, including
the square-root growth of logarithmic fluctuations with wire length,
were obtained for a related continuum model of disordered p-wave
wires by Brouwer \textit{et al.}~\cite{brouwer_probability_2011}. Here
we perform the explicit calculation for the discrete Kitaev chain with
mixed disorder around the sweet spot.

The central-limit approximation gives
\begin{equation}
 \log\frac{|\delta\epsilon_N|}{E_g}
 \simeq N\lambda_{\mathrm{typ}}+b_Y+\sqrt{N\mathcal D}\,G,
 \qquad G\sim\mathcal N(0,1),
 \label{splitting:central_log_distribution}
\end{equation}
where $\mathcal D(u)$ is the variance growth rate per site,
\begin{equation}
 \mathcal D(u)=\lim_{N\to\infty}
 \frac{\operatorname{Var}(\log|\delta Y_N|)}{N}.
 \label{splitting:log_variance_rate}
\end{equation}
The standard deviation of the logarithm therefore grows as $\sqrt N$.
This approximation describes central fluctuations; it does not
determine the second moment calculated in
Sec.~\ref{app:splitting_moments}, which is sensitive to rare splittings.

The sum in Eq.~\eqref{splitting:log_ratio_sum} contains correlated
increments, so its variance is not the sum of the individual variances.
At fixed $u>0$, the initial transient does not affect the long-chain
variance growth rate, provided its correlations with subsequent
increments are summable. We therefore calculate $\mathcal D$ using
the stationary process with density $p_u$. Its covariances depend
only on the separation $k$, giving
\begin{equation}
 c_k=\operatorname{Cov}(q_0,q_k)
 =\bigl\langle(q_0-L)(q_k-L)\bigr\rangle.
 \label{splitting:stationary_covariance}
\end{equation}
For this stationary process, expanding the variance gives
\begin{align}
 \operatorname{Var}\!\left(\sum_{j=1}^{N}q_j\right)
 &=\sum_{i,j=1}^{N}\operatorname{Cov}(q_i,q_j)\nonumber\\
 &=Nc_0+2\sum_{k=1}^{N-1}(N-k)c_k.
 \label{splitting:finite_log_variance}
\end{align}
There are $N-k$ pairs at separation $k$; exchanging the two sites
accounts for the factor of two. Dividing by $N$ and taking the
long-chain limit yields
\begin{equation}
 \mathcal D(u)=c_0+2\sum_{k=1}^{\infty}c_k.
 \label{splitting:covariance_sum}
\end{equation}
For chains much longer than the correlation length, $c_k$ is
appreciable only for $k\ll N$. We can therefore replace $1-k/N$
by one in the sum.

\subsubsection{Computing the correlations}

The transfer operator in Eq.~\eqref{splitting:density_operator} also
propagates the information needed for the covariances. Write
$f(q)=q-L$. Using the conditional density from
Eq.~\eqref{splitting:conditional_log_ratio_density}, the first
covariance is
\begin{equation}
 c_1=\int dq'\,f(q')\int dq\,
 f_\xi\bigl(q'-h_u(q)\bigr)f(q)p_u(q).
 \label{splitting:one_step_covariance}
\end{equation}
The inner integral applies $T_u$ to $f p_u$. We therefore store the
initial centered value as a weight in
\begin{equation}
 a_0(q)=f(q)p_u(q),\qquad a_{k+1}=T_u a_k.
 \label{splitting:covariance_propagation}
\end{equation}
Each propagation integrates over one intermediate log-ratio while
preserving the weight attached to $q_0$. Multiplying by the final
centered value gives
\begin{equation}
 c_k=\int f(q)a_k(q)\,dq,
 \qquad c_0=\int f(q)^2p_u(q)\,dq.
 \label{splitting:covariance_quadrature}
\end{equation}
Thus covariances at arbitrary separations require successive
one-dimensional integral operations.

\subsubsection{Third cumulant and skewness}
\label{app:splitting_skewness}

The variance describes the width of the log-splitting distribution.
Its leading asymmetry is measured by the third cumulant,
\begin{equation}
 \operatorname{cum}_3(A)=\left\langle
 (A-\langle A\rangle)^3\right\rangle,
 \label{splitting:third_cumulant_definition}
\end{equation}
and we define its rate per site by
\begin{equation}
 \kappa_3(u)=\lim_{N\to\infty}
 \frac{\operatorname{cum}_3(\log|\delta Y_N|)}{N}.
 \label{splitting:third_cumulant_rate}
\end{equation}
As for the variance, we use the stationary process and assume that
correlations decay over a range much shorter than the chain. Writing
$f_j=f(q_j)=q_j-L$, we have $\langle f_j\rangle=0$, so the third
cumulant is the third moment of the centered sum. Expanding it gives
\begin{align}
 \left\langle\left(\sum_{j=1}^N f_j\right)^3\right\rangle
 &=\sum_{i=1}^N\langle f_i^3\rangle\nonumber\\
 &\quad+3\sum_{i<j}\bigl[\langle f_i^2f_j\rangle
                           +\langle f_if_j^2\rangle\bigr]\nonumber\\
 &\quad+6\sum_{i<j<k}\langle f_if_jf_k\rangle.
 \label{splitting:third_moment_expansion}
\end{align}
The three terms distinguish contributions from one, two, and three
distinct sites; the factors three and six count the permutations of
the indices. Stationarity makes each expectation depend only on the
separations. There are $N-k$ placements of a pair separated by $k$
and $N-k-\ell$ placements of a triple with successive separations
$k$ and $\ell$. For the separations that contribute appreciably,
we can replace both counts by $N$. Dividing by $N$ gives
\begin{align}
 \kappa_3
 &=\langle f_0^3\rangle
   +3\sum_{k=1}^{\infty}
     \bigl[\langle f_0^2f_k\rangle+\langle f_0f_k^2\rangle\bigr]
     \nonumber\\
 &\quad+6\sum_{k,\ell=1}^{\infty}
      \langle f_0f_kf_{k+\ell}\rangle.
 \label{splitting:third_cumulant_correlation_sum}
\end{align}
Thus the third cumulant grows linearly with $N$, just as the variance
does. For independent increments, the terms involving distinct sites
vanish, leaving only the single-increment third cumulant.

The transfer calculation extends directly from two to three factors:
\begin{equation}
 \langle f_0f_kf_{k+\ell}\rangle
 =\int f(q)\,T_u^\ell\!\left[f\,T_u^k(fp_u)\right](q)\,dq.
 \label{splitting:three_point_transfer}
\end{equation}
Starting from $fp_u$, we propagate through $k$ sites, multiply by
$f$ at the intermediate site, and propagate through another $\ell$
sites before integrating against the final $f$. The two-site terms
in Eq.~\eqref{splitting:third_cumulant_correlation_sum} use the same
procedure as Eq.~\eqref{splitting:covariance_quadrature}, with one of
the two factors $f$ replaced by $f^2$.

The standardized skewness measures asymmetry in units of the
distribution's own width:
\begin{align}
 \frac{\operatorname{cum}_3(\log|\delta Y_N|)}
 {\operatorname{Var}(\log|\delta Y_N|)^{3/2}}
 &=\frac{\kappa_3(u)}{\mathcal D(u)^{3/2}\sqrt N}
   +o(N^{-1/2}).
 \label{splitting:standardized_skewness}
\end{align}
The denominator is the cube of the standard deviation, so this ratio
is unchanged by a positive rescaling of the centered variable. It vanishes as
$N^{-1/2}$ even when $\kappa_3$ is nonzero, consistently with the
Gaussian central limit. The offset $b_Y$ shifts the mean logarithm;
$\kappa_3$ describes asymmetry about that mean.

\subsubsection{Limiting cases and numerical evaluation}

The product formulas in Eqs.~\eqref{splitting:onsite_product} and
\eqref{splitting:bond_product} provide independent checks. For each
Gaussian factor, $\xi=\tfrac12\log(G^2)$ with $G^2\sim\chi_1^2$.
Using the log-chi-square cumulants~\cite{pav_moments_2015} gives
\begin{equation}
 \operatorname{Var}(\xi)=\frac{\pi^2}{8},\qquad
 \operatorname{cum}_3(\xi)=-\frac74\zeta(3),
 \label{splitting:log_gaussian_higher_cumulants}
\end{equation}
where $\zeta$ is the Riemann zeta function. The factors $1/4$ and
$1/8$ relative to the cumulants of $\log(G^2)$ follow from the
factor $1/2$ in $\xi$.

For pure onsite disorder, the logarithm contains $N$ independent
copies of $\xi$. Their cumulants add, giving the rates per site
\begin{equation}
 \mathcal D_{\mathrm{onsite}}=\frac{\pi^2}{8},\qquad
 \kappa_{3,\mathrm{onsite}}=-\frac74\zeta(3).
 \label{splitting:onsite_cumulant_rates}
\end{equation}
For pure bond disorder and $N=2j$, there are only $j$ independent
factors, hence
\begin{equation}
 \mathcal D_{\mathrm{bond,even}}=\frac{\pi^2}{16},\qquad
 \kappa_{3,\mathrm{bond,even}}=-\frac78\zeta(3).
 \label{splitting:bond_cumulant_rates}
\end{equation}
These bond-disorder rates apply only to the even subsequence. At
$u=0$ the odd splittings vanish and the adjacent log-ratio process
is not defined. The endpoint calculation therefore does not establish
the $u\downarrow0$ limits of the mixed-disorder rates, which take
$N\to\infty$ first at fixed positive $u$.

Numerically, we reuse the stationary density, grid, and implementation
of $T_u$ from Sec.~\ref{app:splitting_typical}. We evaluate the
correlation sums by accumulating propagated densities. The covariance
calculation already supplies
\begin{equation}
 h_1=\sum_{k=0}^{\infty}T_u^k(fp_u)
     =\sum_{k=0}^{\infty}a_k,
 \label{splitting:first_moment_density_limit}
\end{equation}
so Eq.~\eqref{splitting:covariance_sum} becomes
\begin{equation}
 \mathcal D=\int\bigl[f^2p_u+2fT_uh_1\bigr]dq.
 \label{splitting:variance_density_formula}
\end{equation}
For the third cumulant, we group the first two factors in each
correlation before summing over the distance to the final site:
\begin{equation}
 h_2=\sum_{\ell=0}^{\infty}T_u^\ell
       \bigl[f^2p_u+2fT_uh_1-\mathcal Dp_u\bigr].
 \label{splitting:second_moment_density_limit}
\end{equation}
The term $f^2p_u$ accounts for two factors at the same site, while
$2fT_uh_1$ sums over their positive separation. Their combined
integral is $\mathcal D$ by
Eq.~\eqref{splitting:variance_density_formula}. Subtracting
$\mathcal Dp_u$ therefore removes the stationary component and lets
the propagated source decay. It does not change its contribution to
the third cumulant, since the final integration against $f$ gives
$\int fp_u\,dq=0$. Regrouping
Eq.~\eqref{splitting:third_cumulant_correlation_sum} then gives
\begin{equation}
 \kappa_3=\int\bigl[f^3p_u+3f^2T_uh_1+3fT_uh_2\bigr]dq.
 \label{splitting:third_cumulant_density_formula}
\end{equation}
Expanding $h_1$ and $h_2$ reproduces the two single sums and the
double sum, including their factors three and six. We compute $h_1$
by accumulating the iterates $a_k$, evaluate $\mathcal D$, and then
propagate the source in Eq.~\eqref{splitting:second_moment_density_limit}
and sum its iterates to obtain $h_2$.

Interpolation and truncation can introduce a small nonzero integral
into a propagated weighted density. After each step we remove this
error by replacing a propagated function $v$ with
$v-p_u\int v(q)\,dq$. This prevents an artificial stationary component
from accumulating in the sums. We retain both signs of the weights
and stop the propagation when the integrated absolute value of the
next summand is below the chosen tolerance. Convergence is checked
by tightening this tolerance, refining the grid, and enlarging its
interval.

\begin{figure}[t]
 \centering
 \includegraphics[width=\columnwidth]{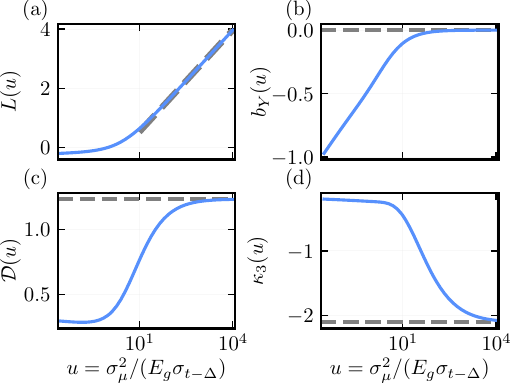}
 \caption{Single-chain log-splitting coefficients as functions of the
 disorder ratio $u$: (a) growth rate $L(u)$, (b) boundary correction
 $b_Y(u)$, (c) variance rate $\mathcal D(u)$, and (d) third-cumulant
 rate $\kappa_3(u)$. Solid curves follow from the transfer operator
 and its transient and weighted-density iterations. Dashed lines show the
 onsite asymptotes $\tfrac12\log u-c_G$, $0$, $\pi^2/8$, and
 $-7\zeta(3)/4$, respectively. The curves take $N\to\infty$ at fixed
 $u>0$; the pure-bond even-subsequence rates are separate endpoint
 results.}
 \label{fig:splitting_coefficients}
\end{figure}

Figure~\ref{fig:splitting_coefficients} shows the four coefficients
as functions of $u$. The large-$u$ behavior approaches the independent
onsite product: $L(u)\simeq\tfrac12\log u-c_G$, $b_Y(u)\to0$, and
the cumulant rates approach Eq.~\eqref{splitting:onsite_cumulant_rates}.

\subsection{Distribution of the single-chain splitting}
\label{app:splitting_distribution}

The preceding sections characterize the \emph{magnitude of the splitting}
through its second moment and the cumulants of its logarithm. For the
subsequent qubit calculations, we however need the \emph{signed splitting}.
Below, we express the distribution of the signed splitting as a mixture of
Gaussians.

\subsubsection{Conditional Gaussian representation}

For $u>0$ and $N\geq2$, fix the preceding amplitudes
$\delta Y_{N-1}$ and $\delta Y_{N-2}$. Then
Eq.~\eqref{splitting:scaled_recurrence} is a linear combination of the
two fresh independent Gaussian variables $G_N$ and $H_{N-1}$.
Its conditional mean is zero and its conditional variance is
\begin{align}
 Q_N&=\operatorname{Var}(\delta Y_N\mid
           \delta Y_{N-1},\delta Y_{N-2})\nonumber\\
 &=u(\delta Y_{N-1})^2+(\delta Y_{N-2})^2.
 \label{splitting:conditional_variance}
\end{align}
This variance is itself random because it depends on the preceding
chain. The conditional Gaussian distribution depends on that chain
only through $Q_N$, so
\begin{equation}
 \delta Y_N\mid Q_N\sim\mathcal N(0,Q_N),\qquad
 \delta Y_N\overset{d}{=}\sqrt{Q_N}\,G_N',
 \label{splitting:conditional_splitting_gaussian}
\end{equation}
where $G_N'$ is a standard normal independent of $Q_N$. We write
$\langle\cdot\rangle_{Q_N}$ for the average over the distribution of
the conditional variance, and $\langle\cdot\rangle_{G_N'}$ for the
Gaussian average at fixed $Q_N$. At fixed $Q_N$, the Gaussian
second moment is $Q_N$, since $\langle(G_N')^2\rangle_{G_N'}=1$.
Averaging over the conditional variance therefore gives
\begin{equation}
 m_N=\langle(\delta Y_N)^2\rangle=\langle Q_N\rangle_{Q_N}.
 \label{splitting:mean_conditional_variance}
\end{equation}
Thus $Q_N$ is the variance for one conditional Gaussian, whereas $m_N$
is the variance of the full ensemble.

Averaging the conditional Gaussian density gives the probability
density of $\delta Y_N$,
\begin{equation}
 P_N(y)=\left\langle
 \frac{1}{\sqrt{2\pi Q_N}}
 \exp\left(-\frac{y^2}{2Q_N}\right)
 \right\rangle_{Q_N}.
 \label{splitting:gaussian_mixture_density}
\end{equation}
Each preceding chain contributes a Gaussian of a different width.
The mixture is generally non-Gaussian.

This representation allows us to perform the Gaussian
average at fixed conditional variance before averaging over the
distribution of the variance. In the qubit coherence calculations,
the remaining average is conveniently expressed in terms of the
log-width $W_N$ introduced below.

\subsubsection{Statistics of the conditional width}
\label{app:conditional_log_width}

The logarithm of the random Gaussian width is
\begin{equation}
 W_N=\log\sqrt{Q_N}=\frac12\log Q_N.
 \label{splitting:conditional_log_width}
\end{equation}
Taking the absolute value and logarithm in
Eq.~\eqref{splitting:conditional_splitting_gaussian} gives the
independent decomposition
\begin{equation}
 \log|\delta Y_N|=W_N+\xi_N,\qquad
 \xi_N=\log|G_N'|.
 \label{splitting:log_width_decomposition}
\end{equation}
This is the same Gaussian increment used in the log-ratio process:
factoring $(\delta Y_{N-1})^2$ out of $Q_N$ gives
$W_N=\log|\delta Y_{N-1}|+h_u(q_{N-1})$, so subtracting
$\log|\delta Y_{N-1}|$ reproduces
Eq.~\eqref{splitting:log_ratio_recurrence}.

Since $W_N$ and $\xi_N$ are independent, their cumulants add.
Using Eqs.~\eqref{splitting:log_gaussian_mean} and
\eqref{splitting:log_gaussian_higher_cumulants} together with the
long-chain results above gives
\begin{align}
 \langle W_N\rangle
 &=\langle\log|\delta Y_N|\rangle+c_G\nonumber\\
 &=NL(u)+b(u)+o(1),
 \label{splitting:width_mean_asymptotic}\\[3pt]
 \operatorname{Var}(W_N)
 &=\operatorname{Var}(\log|\delta Y_N|)-\frac{\pi^2}{8}\nonumber\\
 &=N\mathcal D(u)+o(N),
 \label{splitting:width_variance_asymptotic}\\[3pt]
 \operatorname{cum}_3(W_N)
 &=\operatorname{cum}_3(\log|\delta Y_N|)+\frac74\zeta(3)\nonumber\\
 &=N\kappa_3(u)+o(N),
 \label{splitting:width_third_cumulant_asymptotic}
\end{align}
where $b(u)=b_Y(u)+c_G$. Thus $W_N$ has the same drift, variance
rate, and third-cumulant rate as the log-splitting, but a different
mean offset $b$. Its standardized
skewness has the same leading coefficient as
Eq.~\eqref{splitting:standardized_skewness}.

The Gaussian central limit for the log-splitting also determines the
central distribution of $W_N$. Using $b=b_Y+c_G$, the independent
decomposition in Eq.~\eqref{splitting:log_width_decomposition} gives
\begin{equation}
 \begin{aligned}
 \frac{W_N-NL-b}{\sqrt{N\mathcal D}}
 &=\frac{\log|\delta Y_N|-NL-b_Y}{\sqrt{N\mathcal D}}\\
 &\quad-\frac{\xi_N+c_G}{\sqrt{N\mathcal D}}.
 \end{aligned}
 \label{splitting:width_central_limit}
\end{equation}
The first term on the right approaches a standard Gaussian by
Eq.~\eqref{splitting:central_log_distribution}. The second term has
zero mean and variance $\pi^2/(8N\mathcal D)$, which tends to zero
at fixed $u>0$. Its fluctuations therefore become negligible on the
$\sqrt{N\mathcal D}$ scale, so the normalized log-width on the left
has the same Gaussian limit.

\section{Rabi coherence and optimal operating point}
\label{app:rabi}

\subsection{Rabi envelope of two coupled chains}
\label{app:rabi_envelope}

We consider two independent chains of equal length $N\geq2$ with the
same disorder parameters $d$ and $u>0$, coupled through the central
dot. In the strong-coupling regime discussed in the main
text, the oscillations have approximately unit visibility and the
qubit energy is
\begin{equation}
 \Omega\simeq\Gamma_0+\delta\Gamma
       +\frac{(\delta\varepsilon_N)^2}{2\Gamma_0},
 \label{rabi:strong_coupling_energy}
\end{equation}
where $\Gamma_0>0$ and the expansion assumes
$|\delta\Gamma|,|\delta\varepsilon_N|\ll\Gamma_0$.
Removing the deterministic phase $2\Gamma_0t/\hbar$ leaves the
normalized envelope
\begin{equation}
 \mathcal C_{\mathrm{Rabi}}(t)\simeq
 \left|\left\langle
 e^{-2it\delta\Gamma/\hbar}
 e^{-it(\delta\varepsilon_N)^2/(\Gamma_0\hbar)}
 \right\rangle\right|.
 \label{rabi:phase_average}
\end{equation}
We evaluate the splitting contribution using the conditional
Gaussian representation of App.~\ref{app:splitting_distribution}.

The $\sigma_z$-term combines the signed splittings of the two
chains,
\begin{equation}
 \delta\varepsilon_N
 =\frac{\delta\epsilon_{L,N}+p\,\delta\epsilon_{R,N}}{2},
 \label{rabi:qubit_detuning}
\end{equation}
where $p=+1$ and $p=-1$ correspond to even and odd total parity,
respectively. Using the expressions for the single-chain splittings,
Eqs.~\eqref{splitting:physical_scale} and
\eqref{splitting:conditional_splitting_gaussian} give
$\delta\epsilon_{\alpha,N}\mid Q_{\alpha,N}
\sim\mathcal N(0,E_g^2d^NQ_{\alpha,N})$ for $\alpha=L,R$.
At fixed $Q_{L,N}$ and $Q_{R,N}$, the two splittings are independent
Gaussians, so their variances add:
\begin{equation}
 \delta\varepsilon_N\mid V_N\sim\mathcal N(0,V_N),\qquad
 V_N=\frac{E_g^2d^N}{4}(Q_{L,N}+Q_{R,N}).
 \label{rabi:conditional_variance}
\end{equation}
The sign $p$ drops out of this variance, giving the same distribution
in both parity sectors. Averaging over the conditional variances and
using Eq.~\eqref{splitting:mean_conditional_variance}, we obtain
\begin{equation}
 \sigma_{\varepsilon,N}^2
 =\langle(\delta\varepsilon_N)^2\rangle
 =\langle V_N\rangle_{V_N}
 =\frac{E_g^2d^N}{2}m_N,
 \label{rabi:ensemble_variance}
\end{equation}
where $m_N$ is the finite-length second moment of the dimensionless
single-chain splitting, Eq.~\eqref{splitting:finite_second_moment}.

At fixed $V_N$, the splitting contribution to the phase average is
an elementary Gaussian integral:
\begin{align}
 &\left\langle e^{-it(\delta\varepsilon_N)^2/(\Gamma_0\hbar)}
       \mid V_N\right\rangle\nonumber\\
 &\quad=\frac{1}{\sqrt{2\pi V_N}}
 \int_{-\infty}^{\infty}
 \exp\left[-\left(\frac{1}{2V_N}
             +\frac{it}{\Gamma_0\hbar}\right)z^2\right]dz
 \nonumber\\
 &\quad=\left(1+\frac{2itV_N}{\Gamma_0\hbar}\right)^{-1/2}.
 \label{rabi:conditional_phase_integral}
\end{align}
We choose the square root so that the conditional phase average equals
unity at $t=0$.

To separate the overall splitting scale from its distribution, we
introduce the normalized conditional variance and dimensionless time,
\begin{equation}
 v_N=\frac{V_N}{\sigma_{\varepsilon,N}^2}
     =\frac{Q_{L,N}+Q_{R,N}}{2m_N},\qquad
 s=\frac{\sigma_{\varepsilon,N}^2t}{\Gamma_0\hbar}.
 \label{rabi:dimensionless_variables}
\end{equation}
The remaining average then takes the form
\begin{equation}
 g_N(s;u)=\left\langle(1+2isv_N)^{-1/2}\right\rangle_{v_N}.
 \label{rabi:splitting_kernel}
\end{equation}
Note that although $\langle v_N\rangle_{v_N}=1$, the average depends on the
full distribution of $v_N$.

For fixed contact amplitudes and positive central-dot detuning,
leading-order dot elimination gives
$\Gamma_0=t_{LD}t_{RD}/(2\mu_D)$. Linearizing in the Gaussian
dot-potential noise gives $\delta\Gamma=-\Gamma_0\delta\mu_D/\mu_D$,
with
\begin{equation}
 \sigma_\Gamma
 =\frac{\Gamma_0\sigma_{\mu_D}}{\mu_D}
 =\frac{2\sigma_{\mu_D}\Gamma_0^2}{t_{LD}t_{RD}}.
 \label{rabi:bridge_tuning}
\end{equation}
Its phase average is
$\langle e^{-2it\delta\Gamma/\hbar}\rangle
=e^{-2\sigma_\Gamma^2t^2/\hbar^2}$.
Assuming the coupling noise to be independent of both chains,
Eq.~\eqref{rabi:phase_average} factorizes and yields
\begin{equation}
 \mathcal C_{\mathrm{Rabi}}(t)\simeq
 e^{-2\sigma_\Gamma^2t^2/\hbar^2}
 \left|g_N\!\left(
 \frac{\sigma_{\varepsilon,N}^2t}{\Gamma_0\hbar};u
 \right)\right|.
 \label{rabi:envelope}
\end{equation}

\subsection{Optimizing the number of coherent oscillations}
\label{app:rabi_optimization}

For each coupling $\Gamma_0$, we define the coherence time
$T_\varphi^{\mathrm{Rabi}}$ by $\mathcal C_{\mathrm{Rabi}}(T_\varphi^{\mathrm{Rabi}})=e^{-1}$ and count the
oscillations at the bare frequency $\Gamma_0/(\pi\hbar)$,
\begin{equation}
 N_\varphi=\frac{\Gamma_0T_\varphi^{\mathrm{Rabi}}}{\pi\hbar}.
 \label{rabi:cycle_count}
\end{equation}
We want to find the maximum number of coherent oscillations accessible
by varying the positive central-dot detuning
$\mu_D$, while the other parameters of the parity qubit are kept fixed.
Increasing $\Gamma_0$ reduces the quadratic
response to splitting noise, but increases the coupling noise as
$\Gamma_0^2$ according to Eq.~\eqref{rabi:bridge_tuning}.

We define the contribution of the energy splitting to the logarithmic decay
of the Rabi envelope as
\begin{equation}
 \ell_N(s)=-\log|g_N(s;u)|,
 \label{rabi:splitting_loss}
\end{equation}
and denote the dimensionless time at the $1/e$ crossing by
$s_\varphi=\sigma_{\varepsilon,N}^2T_\varphi^{\mathrm{Rabi}}/(\Gamma_0\hbar)$.
Taking minus the logarithm of Eq.~\eqref{rabi:envelope} at the $1/e$ crossing
gives
\begin{equation}
 \frac{2\sigma_\Gamma^2\bigl(T_\varphi^{\mathrm{Rabi}}\bigr)^2}{\hbar^2}
 +\ell_N(s_\varphi)=1.
 \label{rabi:crossing_condition}
\end{equation}
Thus, noise in the energy splitting contributes $\ell_N(s_\varphi)$ to the
logarithmic decay of the Rabi envelope, while the remaining amount
$1-\ell_N(s_\varphi)$
comes from coupling noise.

We now eliminate $T_\varphi^{\mathrm{Rabi}}$ and $\Gamma_0$ in favor of the cycle
count and the crossing value $s_\varphi$. From their definitions, we have
\begin{equation*}
 T_\varphi^{\mathrm{Rabi}}=\frac{\hbar\Gamma_0s_\varphi}{\sigma_{\varepsilon,N}^2},
 \qquad
 N_\varphi=\frac{\Gamma_0^2s_\varphi}{\pi\sigma_{\varepsilon,N}^2}.
\end{equation*}
Substituting these relations and Eq.~\eqref{rabi:bridge_tuning} into
Eq.~\eqref{rabi:crossing_condition} yields
\begin{equation}
 N_\varphi^3
 =\frac{t_{LD}^2t_{RD}^2}
 {8\pi^3\sigma_{\mu_D}^2\sigma_{\varepsilon,N}^2}
 s_\varphi\bigl[1-\ell_N(s_\varphi)\bigr].
 \label{rabi:cycle_count_at_crossing}
\end{equation}
In particular, the prefactor of the right-hand side is constant with
respect to changing the coupling strength. Instead, the coupling strength
enters through $s_\varphi$.

Let $s_{1/e,N}$ denote the $1/e$ crossing of the splitting-only
envelope $|g_N(s;u)|$ in Eq.~\eqref{rabi:splitting_kernel}, so
$\ell_N(s_{1/e,N})=1$.
With middle-dot noise, the full envelope reaches $e^{-1}$ earlier, so
$0<s_\varphi<s_{1/e,N}$. Each value in this interval corresponds
to a different coupling $\Gamma_0$ and corresponding number of coherent
oscillations given by Eq.~\eqref{rabi:cycle_count_at_crossing}.
Maximizing over $s_\varphi$ therefore corresponds to finding the
optimal coupling for maximizing the number of coherent oscillations.

We can thus express the maximum number of coherent oscillations
as a fixed prefactor times a shape factor $B_N(u)$ that depends only
on the energy splitting distribution,
\begin{align}
 N_\varphi^{\max}
 &=\left(\frac{t_{LD}t_{RD}}{2\sigma_{\mu_D}\sigma_{\varepsilon,N}}\right)^{2/3}
 B_N(u),
 \label{rabi:optimized_cycle_count}\\
 B_N(u)
 &=\left[\frac{1}{2\pi^3}
 \max_{0<s_\varphi<s_{1/e,N}}s_\varphi\bigl(1-\ell_N(s_\varphi)\bigr)\right]^{1/3}.
 \label{rabi:shape_factor}
\end{align}
Since $v_N$ is normalized by the splitting variance, $B_N(u)$
depends only on the shape of the normalized splitting distribution.
We will retain the exact finite-length variance in the prefactor of
Eq.~\eqref{rabi:optimized_cycle_count} and approximate only the effect
of the non-Gaussian shape by its long-chain expression.
To calculate this shape factor, we will use the single-chain log-width
statistics of $W_N$ derived in App.~\ref{app:splitting_distribution}.

The objective to be maximized vanishes at both ends of the interval,
so its maximum lies in the interior. Differentiating with respect to $s_\varphi$
gives the stationarity condition
\begin{equation}
 1-\ell_N(s_{\varphi,\mathrm{opt}})
 =s_{\varphi,\mathrm{opt}}\ell_N'(s_{\varphi,\mathrm{opt}}).
 \label{rabi:stationary_optimum}
\end{equation}
In the following, we will use this condition together with the asymptotic
long-chain approximation.

\subsection{Long-chain approximation to the splitting envelope}
\label{app:rabi_long_chain_envelope}

To evaluate $B_N(u)$ analytically, we approximate the splitting
average $g_N(s;u)$ for long chains at fixed $u>0$. The single-chain
log-width $W_N$ has a spread $w_N=\sqrt{N\mathcal D(u)}$ according to
Eq.~\eqref{splitting:width_variance_asymptotic}. In contrast, the
conditional phase average changes from near unity to a small value
over an order-one interval of $\log(sv_N)$. For $w_N\gg1$, this
separation of scales allows us to expand the integral corresponding to
the average over $v_N$. Here, we perform this expansion for general $s$,
and will return to the optimization over $s_\varphi$ in the next section.

\subsubsection{Expansion of the smooth kernel average}

We define logarithmic time $x=\log s$ and logarithmic
variance $z=\log v_N$. Writing $p_{z,N}$ and $F_{z,N}$ for the density
and cumulative distribution of $\log v_N$, respectively,
Eq.~\eqref{rabi:splitting_kernel} becomes
\begin{equation}
 \begin{aligned}
 g_N(e^x;u)&=\int_{-\infty}^{\infty}k(x+z)p_{z,N}(z)\,dz,\\
 k(w)&=(1+2ie^w)^{-1/2}.
 \end{aligned}
 \label{rabi:log_kernel_average}
\end{equation}
For $w\ll0$, $k(w)\simeq1$, whereas for $w\gg0$,
$k(w)\simeq e^{-w/2-i\pi/4}/\sqrt2$. Thus conditional Gaussians
with $sv_N\ll1$ contribute strongly to the integral, while those with
$sv_N\gg1$ contribute little.

To isolate this transition, we add and subtract the step function
$\Theta(-w)$, where $\Theta(-w)=1$ for $w<0$ and zero for $w>0$.
Changing variables to $w=x+z$ gives the exact identity
\begin{align}
 g_N(e^x;u)&=F_{z,N}(-x)\nonumber\\
 &\quad+\int_{-\infty}^{\infty}
 [k(w)-\Theta(-w)]p_{z,N}(w-x)\,dw.
 \label{rabi:step_decomposition}
\end{align}
The first term, $F_{z,N}(-x)$, is the probability $\Pr(v_N<1/s)$.
The second term accounts for the fact that the kernel $k(w)$ is smooth
rather than a step function.

The difference $k(w)-\Theta(-w)$ decays exponentially at both ends.
Across its order-one transition width, the broad central density is
approximately constant, $p_{z,N}(w-x)\simeq p_{z,N}(-x)$.
Taking it outside the integral gives
\begin{equation}
 g_N(e^x;u)\simeq F_{z,N}(-x)+Jp_{z,N}(-x),
 \label{rabi:kernel_smoothing_expansion}
\end{equation}
where
\begin{equation}
 J=\int_{-\infty}^{\infty}[k(w)-\Theta(-w)]\,dw
 =\log2-\frac{i\pi}{2}.
 \label{rabi:kernel_smoothing_constant}
\end{equation}
Since the normalized density is spread over an interval of order $w_N$,
its central value is of order $w_N^{-1}$. The smoothing term
$Jp_{z,N}(-x)$ is therefore of this order.
In the following, we use the single-chain log-width distribution to
approximate $F_{z,N}(-x)$ and $p_{z,N}(-x)$ for long chains.

\subsubsection{From single-chain widths to the splitting envelope}

We now express the cumulative probability $F_{z,N}(-x)$ through
the single-chain statistics of App.~\ref{app:splitting_distribution}. Since
$Q_{\alpha,N}=e^{2W_{\alpha,N}}$, Eq.~\eqref{rabi:dimensionless_variables}
gives
\begin{equation}
 v_N=\frac{e^{2W_{L,N}}+e^{2W_{R,N}}}{2m_N}.
 \label{rabi:variance_from_log_widths}
\end{equation}
For brevity, we write $W_L$ and $W_R$ for the two independent log-widths.
Then the cumulative probability is defined by
\begin{equation}
 F_{z,N}(-x)
 =\Pr\!\left(e^{2W_L}+e^{2W_R}<2m_Ne^{-x}\right).
 \label{rabi:two_chain_threshold_event}
\end{equation}
When the log-widths are broad, the larger of the two exponential
terms usually dominates their sum. Replacing the sum by this term
requires both single-chain widths to lie below the same threshold,
\begin{align}
 &F_{z,N}(-x)\nonumber\\
 &\quad\simeq\Pr\!\left(\max(W_L,W_R)<\frac{\log(2m_N)-x}{2}\right)
 \nonumber\\
 &\quad=\left[\Pr\!\left(W_N<\frac{\log(2m_N)-x}{2}\right)\right]^2.
 \label{rabi:independent_thresholds}
\end{align}
The square follows from independence and the fact that we consider
identical statistics for the two chains.

To estimate the error from replacing the sum by its larger term, write
\begin{align}
 \log(e^{2W_L}+e^{2W_R})&=2\max(W_L,W_R)\nonumber\\
 &\quad+\log\!\left(1+e^{-2|W_L-W_R|}\right).
 \label{rabi:sum_max_correction}
\end{align}
The last term is at most $\log2$ and decreases exponentially with
the separation of the widths. To change the threshold event appreciably, both
widths must be close to each other and to the threshold. Their
independence makes the probability of this order-one region of
order $w_N^{-2}$. Our expansion in
Eq.~\eqref{rabi:kernel_smoothing_expansion} retains corrections through
order $w_N^{-1}$, so this order-$w_N^{-2}$ error can be neglected.

We measure the single-chain threshold relative to its mean and
standard deviation by defining
\begin{equation}
 \zeta_N=\frac{W_N-NL-b}{w_N},\qquad w_N=\sqrt{N\mathcal D},
 \label{rabi:standardized_log_width}
\end{equation}
using Eqs.~\eqref{splitting:width_mean_asymptotic} and
\eqref{splitting:width_variance_asymptotic}. The condition
$W_N<\frac{\log(2m_N)-x}{2}$
in Eq.~\eqref{rabi:independent_thresholds} is equivalently
$\zeta_N<a_N(x)$, where
\begin{equation}
 a_N(x)=\frac{\log(2m_N)-x-2NL-2b}{2w_N}.
 \label{rabi:threshold_coordinate}
\end{equation}
Thus, if $F_{\zeta,N}$ denotes the cumulative distribution of
$\zeta_N$,
\begin{equation}
 F_{z,N}(-x)\simeq F_{\zeta,N}\!\left(a_N(x)\right)^2.
 \label{rabi:standardized_threshold_probability}
\end{equation}

At leading order, $F_{\zeta,N}$ is the standard Gaussian cumulative
distribution by Eq.~\eqref{splitting:width_central_limit}. To be
consistent with the expansion in Eq.~\eqref{rabi:kernel_smoothing_expansion},
we need to include the first asymmetry correction, which is also
of order $w_N^{-1}$. This follows from the third-cumulant rate in
Eq.~\eqref{splitting:width_third_cumulant_asymptotic}. The leading
standardized third cumulant is
\begin{equation}
 \gamma_N=\frac{\kappa_3}{\mathcal D^{3/2}\sqrt N}.
 \label{rabi:log_width_skewness}
\end{equation}
Using the first Edgeworth correction to the Gaussian cumulative
distribution~\cite[Sec.~2.2, Eqs.~(2.15) and (2.17)]{hall_edgeworth_1992},
we approximate
\begin{align}
 F_{\zeta,N}\!\left(a_N(x)\right)
 &\simeq\Phi\!\left(a_N(x)\right)\nonumber\\
 &\quad+\frac{\gamma_N}{6}[1-a_N(x)^2]
 \varphi_{\mathrm G}\!\left(a_N(x)\right),
 \label{rabi:central_cumulative_expansion}
\end{align}
where $\varphi_{\mathrm G}(y)=e^{-y^2/2}/\sqrt{2\pi}$ and
$\Phi(y)=\int_{-\infty}^y\varphi_{\mathrm G}(y')\,dy'$ are the
standard Gaussian density and cumulative distribution.
This is an expansion for the central distribution.

Squaring Eq.~\eqref{rabi:central_cumulative_expansion} and retaining
the term linear in $\gamma_N\sim1/w_N$ gives
\begin{align}
 F_{z,N}(-x)&\simeq\Phi\!\left(a_N(x)\right)^2\nonumber\\
 &\quad+\frac{\gamma_N}{3}[1-a_N(x)^2]
 \Phi\!\left(a_N(x)\right)\varphi_{\mathrm G}\!\left(a_N(x)\right).
 \label{rabi:two_chain_cumulative_expansion}
\end{align}
From this we can also obtain the density $p_{z,N}$ as
\begin{equation}
 p_{z,N}(-x)=-\frac{d}{dx}F_{z,N}(-x)
 \simeq\frac{\Phi\!\left(a_N(x)\right)
 \varphi_{\mathrm G}\!\left(a_N(x)\right)}{w_N}.
 \label{rabi:threshold_density}
\end{equation}
We can neglect the skewness correction to this density as it is of higher order.
Substituting into Eq.~\eqref{rabi:kernel_smoothing_expansion}
therefore yields up to order $w_N^{-1}$,
\begin{align}
 g_N(e^x;u)&\simeq\Phi\!\left(a_N(x)\right)^2\nonumber\\
 &\quad+\Phi\!\left(a_N(x)\right)\varphi_{\mathrm G}\!\left(a_N(x)\right)
 \nonumber\\
 &\qquad\times\left\{\frac{\gamma_N}{3}[1-a_N(x)^2]
                         +\frac{J}{w_N}\right\}.
 \label{rabi:long_chain_kernel}
\end{align}

The leading term is real and positive. The imaginary part of $J$
therefore affects the modulus only at order $w_N^{-2}$. We thus find
\begin{align}
 \ell_N(e^x)&\simeq-2\log\Phi\!\left(a_N(x)\right)\nonumber\\
 &\quad-\frac{\varphi_{\mathrm G}\!\left(a_N(x)\right)}
                  {\Phi\!\left(a_N(x)\right)}
 \left\{\frac{\gamma_N}{3}[1-a_N(x)^2]
                      +\frac{\log2}{w_N}\right\}.
 \label{rabi:long_chain_loss}
\end{align}

\subsection{Crossing and optimal shape factor}
\label{app:rabi_asymptotic_optimization}

\subsubsection{Crossing and optimization}

We now use Eq.~\eqref{rabi:long_chain_loss} to evaluate the maximum
in Eq.~\eqref{rabi:shape_factor}. We first locate $x_{1/e,N}=\log s_{1/e,N}$,
where $\ell_N(e^{x_{1/e,N}})=1$. At this point, the splitting
contribution alone reaches the $1/e$ threshold. With coupling noise,
the full envelope crosses earlier, so the optimum lies at $x<x_{1/e,N}$.

The corrections in Eq.~\eqref{rabi:long_chain_loss} can be absorbed
into a small shift of the Gaussian argument. Taylor expansion
gives
\begin{equation*}
 -2\log\Phi(a+\delta a)
 \simeq-2\log\Phi(a)-2\frac{\varphi_{\mathrm G}(a)}{\Phi(a)}\,\delta a.
\end{equation*}
Matching the correction terms therefore gives
\begin{align}
 \ell_N(e^x)&\simeq-2\log\Phi\!\left(a_N(x)+\delta a_N(x)\right),
 \label{rabi:shifted_gaussian_loss}\\
 \delta a_N(x)&=\frac{\gamma_N}{6}[1-a_N(x)^2]
                    +\frac{\log2}{2w_N}.
 \label{rabi:argument_shift}
\end{align}

For $\ell_N(e^{x_{1/e,N}})=1$, the shifted argument must equal
\begin{equation}
 a_* = \Phi^{-1}(e^{-1/2})\simeq0.270288.
 \label{rabi:crossing_quantile}
\end{equation}
Within the skewness correction, we replace $a_N(x_{1/e,N})$ by its
leading value $a_*$, thus neglecting terms of order $w_N^{-2}$.
The crossing condition becomes
\begin{equation}
 a_N(x_{1/e,N})+\frac{\gamma_N}{6}(1-a_*^2)
             +\frac{\log2}{2w_N}\simeq a_*.
 \label{rabi:shifted_crossing_coordinate}
\end{equation}
Substituting Eq.~\eqref{rabi:threshold_coordinate} and using
$w_N\gamma_N=\kappa_3/\mathcal D$, we obtain directly
\begin{align}
 x_{1/e,N}
 &=\log(4m_N)-2NL-2b-2a_*\sqrt{N\mathcal D}\nonumber\\
 &\quad+\frac{\kappa_3}{3\mathcal D}(1-a_*^2)+o(1).
 \label{rabi:corrected_crossing}
\end{align}

Writing $x=\log s_\varphi=x_{1/e,N}-y$, with $y>0$, we Taylor
expand the logarithmic loss about the crossing, where
$\ell_N(e^{x_{1/e,N}})=1$:
\begin{equation}
 1-\ell_N(e^{x_{1/e,N}-y})
 \simeq y\left.\frac{d\ell_N(e^x)}{dx}\right|_{x=x_{1/e,N}}.
 \label{rabi:loss_near_crossing}
\end{equation}
The derivative of the argument correction is only of order
$w_N^{-2}$, so the leading Gaussian term suffices for this slope.
Differentiating $-2\log\Phi(a_N(x))$ gives
\begin{equation}
 \left.\frac{d\ell_N(e^x)}{dx}\right|_{x=x_{1/e,N}}
 \simeq\frac{h_*}{w_N},\qquad
 h_* = \frac{\varphi_{\mathrm G}(a_*)}{\Phi(a_*)}\simeq0.634152.
 \label{rabi:crossing_slope}
\end{equation}
Substituting into the maximization objective of Eq.~\eqref{rabi:shape_factor},
we obtain
\begin{equation}
 e^x[1-\ell_N(e^x)]
 \simeq e^{x_{1/e,N}}\frac{h_*}{w_N}\,y e^{-y}.
 \label{rabi:local_optimization_objective}
\end{equation}
This expression is maximal at $y=1$. Hence
\begin{align}
 \log s_{\varphi,\mathrm{opt}}&=x_{1/e,N}-1+o(1),
 \label{rabi:optimal_log_time}\\
 B_N(u)&\simeq
 \left[\frac{e^{x_{1/e,N}}h_*}{2\pi^3 e\sqrt{N\mathcal D}}\right]^{1/3}.
 \label{rabi:shape_factor_near_crossing}
\end{align}
The optimal displacement $y=1$ is small compared with the envelope's
variation scale $w_N$ for long chains, consistent with the Taylor
expansion.
The width of the log-distribution thus enters in the exponent, but it
also produces a factor $N^{-1/6}$ in $B_N$.

\subsubsection{Long-chain expression}

To display the length dependence explicitly, we now use
$m_N=M_\infty r_+^N[1+o(1)]$ from
Eqs.~\eqref{splitting:finite_moment_normalization} and
\eqref{splitting:moment_normalization_limit}. Substituting
Eq.~\eqref{rabi:corrected_crossing} into
Eq.~\eqref{rabi:shape_factor_near_crossing} gives
\begin{align}
 \log B_N(u)
 &=\frac{2}{3}(\lambda_{\mathrm{rms}}-\lambda_{\mathrm{typ}})N
 -\frac{2a_*}{3}\sqrt{N\mathcal D}\nonumber\\
 &\quad-\frac16\log N+C_B(u)+o(1),
 \label{rabi:asymptotic_shape_factor}\\
 C_B(u)
 &=\frac{\log(2M_\infty h_*)-1-2b}{3}
 +\frac{\kappa_3(1-a_*^2)}{9\mathcal D}\nonumber\\
 &\quad-\frac16\log\mathcal D-\log\pi.
 \label{rabi:shape_constant}
\end{align}
We used $\lambda_{\mathrm{rms}}-\lambda_{\mathrm{typ}}
=\frac12\log r_+-L$, so this difference depends only on $u$.

To recover the physical cycle count, we substitute
$\sigma_{\varepsilon,N}^2=E_g^2d^Nm_N/2$ from
Eq.~\eqref{rabi:ensemble_variance} into
Eq.~\eqref{rabi:optimized_cycle_count}, keeping the exact finite-$N$
moment $m_N$ in this prefactor. Using the asymptotic shape factor in
Eq.~\eqref{rabi:asymptotic_shape_factor} gives
\begin{align}
 N_\varphi^{\max}
 &\simeq\left(\frac{t_{LD}t_{RD}}{2\sigma_{\mu_D}E_g}\right)^{2/3}
 \left(\frac{2r_+^N}{m_N}\right)^{1/3}e^{C_B(u)}N^{-1/6}
 \nonumber\\
 &\quad\times\exp\!\left[-\frac23\lambda_{\mathrm{typ}}N
                   -\frac{2a_*}{3}\sqrt{N\mathcal D}\right].
 \label{rabi:physical_cycle_count_asymptotic}
\end{align}
The factor $r_+^N/m_N=1/M_N$ retains both characteristic roots
through Eq.~\eqref{splitting:finite_second_moment}.
As $N\to\infty$ at fixed $u>0$, it approaches $1/M_\infty$.
The leading exponential is set by the typical-splitting exponent;
its rms counterpart cancels between the physical scale and $B_N$.
This expression combines the exact moment in the prefactor with
the long-chain approximation to the shape factor.

The same optimum fixes the central-dot detuning and coherence time.
Combining Eqs.~\eqref{rabi:cycle_count},
\eqref{rabi:crossing_condition}, and \eqref{rabi:bridge_tuning} gives
\begin{equation}
 1-\ell_N(s_{\varphi,\mathrm{opt}})
 =\frac{2\pi^2\sigma_{\mu_D}^2(N_\varphi^{\max})^2}
 {\mu_{D,\mathrm{opt}}^2}.
 \label{rabi:optimal_detuning_identity}
\end{equation}
At $y=1$, Eqs.~\eqref{rabi:loss_near_crossing} and
\eqref{rabi:crossing_slope} give
$1-\ell_N(s_{\varphi,\mathrm{opt}})\simeq h_*/w_N$.
Solving for the positive detuning and using
$\Gamma_{0,\mathrm{opt}}=t_{LD}t_{RD}/(2\mu_{D,\mathrm{opt}})$,
we obtain
\begin{align}
 \mu_{D,\mathrm{opt}}
 &\simeq\sqrt2\,\pi\sigma_{\mu_D}
 \sqrt{\frac{w_N}{h_*}}\,N_\varphi^{\max},
 \label{rabi:optimal_detuning_asymptotic}\\
 T_{\varphi,\mathrm{opt}}^{\mathrm{Rabi}}
 &=\frac{2\pi\hbar}{t_{LD}t_{RD}}
 \mu_{D,\mathrm{opt}}N_\varphi^{\max}\nonumber\\
 &\simeq\frac{2\sqrt2\,\pi^2\hbar\sigma_{\mu_D}}{t_{LD}t_{RD}}
 \sqrt{\frac{w_N}{h_*}}\,(N_\varphi^{\max})^2.
 \label{rabi:optimal_coherence_time_asymptotic}
\end{align}
Here $w_N=\sqrt{N\mathcal D}$, and $N_\varphi^{\max}$ is given
by Eq.~\eqref{rabi:physical_cycle_count_asymptotic}.
The time $T_{\varphi,\mathrm{opt}}^{\mathrm{Rabi}}$ is evaluated at maximum cycle
count; it is not obtained by separately maximizing the coherence
time. These long-chain results hold at fixed $u>0$, provided the
optimal operating point remains in the dot-elimination, weak-noise,
and strong-coupling regimes assumed above.

\subsection{Rabi dephasing time at fixed coupling}
\label{app:rabi_fixed_coupling}

We now keep $\Gamma_0$ fixed as the chain length changes, within the
strong-coupling regime. If middle-dot noise is negligible at the
$1/e$ crossing, the coherence time follows directly from the
splitting-only crossing calculated above. Using the definition of
$s$ in Eq.~\eqref{rabi:dimensionless_variables}, we obtain
\begin{equation}
 T_\varphi^{\mathrm{Rabi}}\simeq
 \frac{\hbar\Gamma_0}{\sigma_{\varepsilon,N}^2}e^{x_{1/e,N}}.
 \label{rabi:fixed_coupling_time}
\end{equation}
As for the optimized cycle count, we retain the exact finite-length
variance in the prefactor and approximate the dimensionless crossing
factor by its long-chain expression. We therefore use
$m_N\simeq M_\infty r_+^N$ in Eq.~\eqref{rabi:corrected_crossing},
while keeping $\sigma_{\varepsilon,N}^2=E_g^2d^Nm_N/2$ from
Eq.~\eqref{rabi:ensemble_variance} in Eq.~\eqref{rabi:fixed_coupling_time}.
Writing $m_N=r_+^NM_N$, the factors $r_+^N$ cancel and leave
$M_\infty/M_N$. With $\lambda_{\mathrm{typ}}=\frac12\log d+L$, this gives
\begin{align}
 T_\varphi^{\mathrm{Rabi}}
 &\simeq\frac{8\hbar\Gamma_0}{E_g^2}\frac{M_\infty}{M_N}
 \exp\!\left[-2b+\frac{\kappa_3(1-a_*^2)}{3\mathcal D}\right]
 \nonumber\\
 &\quad\times\exp\!\left[-2\lambda_{\mathrm{typ}}N
                         -2a_*\sqrt{N\mathcal D}\right].
 \label{rabi:fixed_coupling_time_asymptotic}
\end{align}
Note that this formula is derived within the strong coupling regime, which
for a given coupling typically is not valid for all chain lengths $N$. Moreover,
at some point the middle-dot noise will become relevant, and the coherence time
will be limited by the middle-dot noise rather than the splitting noise. Hence,
the leading exponential growth $e^{-2\lambda_{\mathrm{typ}}N}$ applies only
over a limited range of chain lengths.

\subsection{Rabi dephasing time at a threshold coupling}
\label{app:rabi_barrier_threshold}

We now consider a direct tunnel barrier as in
Sec.~\ref{sec:rabi_barrier}. For barrier-dominated coupling noise,
the strong-coupling approximation gives the negative logarithm of the
Gaussian barrier-noise envelope as
\begin{equation}
 \ell_B(q)=2s_B^2q^2,
 \label{rabi:barrier_coupling_loss}
\end{equation}
with $q=\Gamma_0t/\hbar$.
Together with the negative logarithm of the splitting envelope,
$\ell_N(s)=-\ln|g_N(s;u)|$ from Eq.~\eqref{rabi:splitting_loss},
this gives the $1/e$ condition
\begin{equation*}
 \ell_B(q)+\ell_N(s)=1.
\end{equation*}
Without splitting noise, $\ell_B(q_p)=1$ fixes
$q_p=1/(\sqrt2s_B)$, giving the plateau
$N_{\varphi,B}=q_p/\pi$ of Eq.~\eqref{main:rabi_barrier_limit}.
Tuning to a fraction $0<r<1$ of this plateau fixes $q_r=rq_p$ and
hence $\ell_B(q_r)=r^2$. The splitting envelope must therefore satisfy
$\ell_N(s_r)=1-r^2$, so that
\begin{align}
 q_r&=\frac{r}{\sqrt2s_B},\nonumber\\
 a_r&=\Phi^{-1}\!\left(e^{-(1-r^2)/2}\right).
 \label{rabi:barrier_threshold_loss}
\end{align}
Thus $a_r$ depends only on the threshold fraction $r$.

The crossing calculation in
App.~\ref{app:rabi_asymptotic_optimization} now applies with $a_*$
replaced by $a_r$ in Eq.~\eqref{rabi:corrected_crossing}.
As in Apps.~\ref{app:rabi_asymptotic_optimization} and
\ref{app:rabi_fixed_coupling}, we use $m_N\simeq M_\infty r_+^N$
in the dimensionless crossing while retaining the exact variance
$\sigma_{\varepsilon,N}^2=E_g^2d^Nm_N/2$ in the physical prefactor.
Using $s_r=\sigma_{\varepsilon,N}^2q_r/\Gamma_r^2$ and
$T_\varphi(\Gamma_r)=\hbar q_r/\Gamma_r$ gives
$T_\varphi(\Gamma_r)=\hbar\sqrt{q_rs_r}/\sigma_{\varepsilon,N}$.
Writing $m_N=r_+^NM_N$, we obtain
\begin{align}
 T_\varphi(\Gamma_r)
 &\simeq\frac{\hbar\sqrt{8q_r}}{E_g}\sqrt{\frac{M_\infty}{M_N}}
 \nonumber\\
 &\quad\times\exp\!\left[-b+\frac{\kappa_3(1-a_r^2)}{6\mathcal D}\right]
 \nonumber\\
 &\quad\times\exp\!\left[-\lambda_{\mathrm{typ}}N
                         -a_r\sqrt{N\mathcal D}\right].
 \label{rabi:barrier_threshold_time}
\end{align}

\section{Ramsey coherence}
\label{app:ramsey}

\subsection{Ramsey signal and disorder average}
\label{app:ramsey_signal}

We consider the same two independent chains as in
App.~\ref{app:rabi}, with quasistatic disorder, and first set the
coupling to zero during the waiting interval. The waiting
Hamiltonian is then $H_{\mathrm q}=\delta\varepsilon_N\sigma_z$,
with $\delta\varepsilon_N$ defined in Eq.~\eqref{rabi:qubit_detuning}.
Starting from $\lvert0\rangle$, we apply an ideal instantaneous
$R_x(\pi/2)$ pulse, wait for a time $t$, and apply a second pulse
of the same sign before measuring the return to $\lvert0\rangle$.
Here $R_x(\theta)=e^{-i\theta\sigma_x/2}$.
The first pulse rotates the Bloch vector from $+z$ to $-y$;
free evolution rotates it about $z$ by
$2\delta\varepsilon_Nt/\hbar$, and the second pulse maps its
$y$ component onto $z$. Thus, for a fixed realization,
\begin{equation}
 P_{\mathrm{Ramsey}}(t\mid\delta\varepsilon_N)
 =\frac{1-\cos(2\delta\varepsilon_Nt/\hbar)}{2}.
 \label{ramsey:single_return}
\end{equation}
At zero waiting time, the two pulses form a $\pi$ rotation and
the return is zero.

Averaging over the splitting distribution gives
\begin{align}
 \mathcal A_{\mathrm{Ramsey}}(t)
 &=\left\langle e^{-2i\delta\varepsilon_Nt/\hbar}\right\rangle,
 \label{ramsey:coherence}\\
 \overline P_{\mathrm{Ramsey}}(t)
 &=\frac{1-\operatorname{Re}\mathcal A_{\mathrm{Ramsey}}(t)}{2}.
 \label{ramsey:averaged_return}
\end{align}

To evaluate this average, we use the conditional Gaussian law
$\delta\varepsilon_N\mid V_N\sim\mathcal N(0,V_N)$ from
Eq.~\eqref{rabi:conditional_variance}. At fixed $V_N$, the phase
average is
\begin{align}
 &\left\langle e^{-2i\delta\varepsilon_Nt/\hbar}\mid V_N\right\rangle
 \nonumber\\
 &\quad=\frac{1}{\sqrt{2\pi V_N}}
 \int_{-\infty}^{\infty}
 e^{-z^2/(2V_N)-2izt/\hbar}\,dz
 =e^{-2V_Nt^2/\hbar^2}.
 \label{ramsey:conditional_phase_integral}
\end{align}
Averaging over the conditional variance therefore yields
\begin{equation}
 \mathcal A_{\mathrm{Ramsey}}(t)
 =\left\langle e^{-2V_Nt^2/\hbar^2}\right\rangle_{V_N}.
 \label{ramsey:variance_average}
\end{equation}
The full distribution of $V_N$ enters this expression; replacing
$V_N$ by its mean $\sigma_{\varepsilon,N}^2$ would give a single
Gaussian decay and discard the non-Gaussian splitting statistics.

For the disorder model considered here, $V_N>0$ with probability
one, so Eq.~\eqref{ramsey:variance_average} is real, positive,
and decreases from one to zero. The averaged return consequently
rises from zero to $1/2$ without carrier oscillations. We define
$\mathcal C_{\mathrm{Ramsey}}(t)=|\mathcal A_{\mathrm{Ramsey}}(t)|$
and the Ramsey dephasing time by
\begin{equation}
 \mathcal C_{\mathrm{Ramsey}}(T_\varphi^{\mathrm{Ramsey}})=e^{-1}.
 \label{ramsey:crossing_condition}
\end{equation}

\subsection{Chain-length dependence of the Ramsey dephasing time}
\label{app:ramsey_length_dependence}

We now evaluate the crossing in Eq.~\eqref{ramsey:crossing_condition}
for long chains at fixed $u>0$. Using the normalized conditional
variance $v_N=V_N/\sigma_{\varepsilon,N}^2$ from
Eq.~\eqref{rabi:dimensionless_variables}, we introduce the logarithmic
time coordinate
\begin{equation}
 x(t)=\log\frac{2\sigma_{\varepsilon,N}^2t^2}{\hbar^2}.
 \label{ramsey:log_time}
\end{equation}
Writing $x_{1/e,N}^{\mathrm{Ramsey}}=x(T_\varphi^{\mathrm{Ramsey}})$,
we recover the physical time from
\begin{equation}
 T_\varphi^{\mathrm{Ramsey}}
 =\frac{\hbar}{\sqrt2\,\sigma_{\varepsilon,N}}
 e^{x_{1/e,N}^{\mathrm{Ramsey}}/2}.
 \label{ramsey:time_from_crossing}
\end{equation}
We keep the finite-length variance in this prefactor and approximate
the dimensionless crossing factor $e^{x_{1/e,N}^{\mathrm{Ramsey}}/2}$
for long chains. This is the same separation of energy scale and
shape used for the optimized Rabi result in
Eq.~\eqref{rabi:physical_cycle_count_asymptotic}.

With $z=\log v_N$ and its density $p_{z,N}$ defined in
App.~\ref{app:rabi_long_chain_envelope}, the coherence becomes
\begin{equation}
 \mathcal A_{\mathrm{Ramsey}}(t)
 =\int_{-\infty}^{\infty}k_{\mathrm{Ramsey}}(x(t)+z)
 p_{z,N}(z)\,dz,
 \label{ramsey:log_kernel_average}
\end{equation}
where $k_{\mathrm{Ramsey}}(w)=e^{-e^w}$.
This kernel changes over an order-one interval of $w$, while the
log-width distribution broadens as $w_N=\sqrt{N\mathcal D}$.
We can therefore use the expansion of
Eqs.~\eqref{rabi:step_decomposition} and
\eqref{rabi:kernel_smoothing_expansion}, replacing only the kernel:
\begin{equation}
 \mathcal A_{\mathrm{Ramsey}}(t)
 \simeq F_{z,N}(-x(t))+J_{\mathrm{Ramsey}}p_{z,N}(-x(t)).
 \label{ramsey:kernel_smoothing_expansion}
\end{equation}
Substituting $y=e^w$ into the smoothing integral gives
\begin{align}
 J_{\mathrm{Ramsey}}
 &=\int_{-\infty}^{\infty}
 [e^{-e^w}-\Theta(-w)]\,dw\nonumber\\
 &=\int_0^1\frac{e^{-y}-1}{y}\,dy
   +\int_1^\infty\frac{e^{-y}}{y}\,dy
 =-\gamma_E,
 \label{ramsey:kernel_smoothing_constant}
\end{align}
where $\gamma_E\simeq0.577216$ is Euler's constant.

The width distribution is the same as for Rabi. We therefore use
Eqs.~\eqref{rabi:two_chain_cumulative_expansion} and
\eqref{rabi:threshold_density}, including the first skewness
correction, to obtain
\begin{align}
 \mathcal A_{\mathrm{Ramsey}}(t)
 &\simeq\Phi\!\left(a_N(x(t))\right)^2\nonumber\\
 &\quad+\Phi\!\left(a_N(x(t))\right)
          \varphi_{\mathrm G}\!\left(a_N(x(t))\right)\nonumber\\
 &\qquad\times\left\{\frac{\gamma_N}{3}
       [1-a_N(x(t))^2]-\frac{\gamma_E}{w_N}\right\}.
 \label{ramsey:long_chain_coherence}
\end{align}
Here $a_N(x)$ and $\gamma_N$ are given by
Eqs.~\eqref{rabi:threshold_coordinate} and
\eqref{rabi:log_width_skewness}. The expansion retains terms through
order $w_N^{-1}$ in the central distribution, as in the Rabi calculation.

Absorbing the correction into a shift of the Gaussian argument,
as in Eq.~\eqref{rabi:shifted_gaussian_loss}, the $1/e$ condition is
\begin{equation}
 a_N\!\left(x_{1/e,N}^{\mathrm{Ramsey}}\right)
 +\frac{\gamma_N}{6}(1-a_*^2)-\frac{\gamma_E}{2w_N}
 \simeq a_*.
 \label{ramsey:shifted_crossing_coordinate}
\end{equation}
Here $a_*=\Phi^{-1}(e^{-1/2})$ is the same as in
Eq.~\eqref{rabi:crossing_quantile}. As in
Eq.~\eqref{rabi:shifted_crossing_coordinate}, we evaluate the small
skewness correction at $a_*$ to this order.
To express the entire dimensionless crossing factor in its long-chain
form, we use $m_N\simeq M_\infty r_+^N$ from
Eqs.~\eqref{splitting:finite_moment_normalization} and
\eqref{splitting:moment_normalization_limit} in $a_N(x)$.
Substituting into Eq.~\eqref{ramsey:shifted_crossing_coordinate} and
using $w_N\gamma_N=\kappa_3/\mathcal D$ gives
\begin{align}
 x_{1/e,N}^{\mathrm{Ramsey}}
 &=N\log r_+-2NL-2a_*\sqrt{N\mathcal D}\nonumber\\
 &\quad+\log(2M_\infty)-2b-\gamma_E\nonumber\\
 &\quad+\frac{\kappa_3}{3\mathcal D}(1-a_*^2)+o(1).
 \label{ramsey:corrected_crossing}
\end{align}

Finally, we insert this crossing into
Eq.~\eqref{ramsey:time_from_crossing}. Writing
$m_N=r_+^NM_N$, the factors $r_+^{N/2}$ cancel, while the ratio
$\sqrt{M_\infty/M_N}$ remains. With
$\lambda_{\mathrm{typ}}=\tfrac12\log d+L$, we obtain
\begin{align}
 T_\varphi^{\mathrm{Ramsey}}
 &\simeq\frac{\sqrt2\,\hbar}{E_g}\sqrt{\frac{M_\infty}{M_N}}
 \exp\!\left[-b-\frac{\gamma_E}{2}
              +\frac{\kappa_3(1-a_*^2)}{6\mathcal D}\right]
 \nonumber\\
 &\quad\times\exp\!\left[-\lambda_{\mathrm{typ}}N
                         -a_*\sqrt{N\mathcal D}\right].
 \label{ramsey:time_asymptotic}
\end{align}

\subsection{Residual coupling during the Ramsey wait}
\label{app:ramsey_residual_coupling}

We now allow a fixed residual coupling $\Gamma_0^{\mathrm{min}}>0$ during the
waiting interval, keeping the ideal pulses of
App.~\ref{app:ramsey_signal}. The waiting Hamiltonian and qubit
energy are
\begin{equation}
 H_{\mathrm q}=\Gamma_0^{\mathrm{min}}\sigma_x+\delta\varepsilon_N\sigma_z,
 \qquad \Omega=\sqrt{(\Gamma_0^{\mathrm{min}})^2+(\delta\varepsilon_N)^2}.
 \label{ramsey:residual_hamiltonian}
\end{equation}
The first pulse prepares a Bloch vector along $-y$, perpendicular
to the waiting axis in the $xz$ plane. Rotation about that axis
through $2\Omega t/\hbar$ gives $r_y=-\cos(2\Omega t/\hbar)$.
The second pulse maps this component onto $z$, so
\begin{equation}
 P_{\mathrm{Ramsey}}(t\mid\delta\varepsilon_N,\Gamma_0^{\mathrm{min}})
 =\frac{1-\cos(2\Omega t/\hbar)}{2}
 =\sin^2(\Omega t/\hbar).
 \label{ramsey:residual_return}
\end{equation}

The residual coupling is negligible over the ideal Ramsey decay
when
\begin{equation}
 \frac{\Gamma_0^{\mathrm{min}}T_\varphi^{\mathrm{Ramsey}}}{\hbar}\ll1.
 \label{ramsey:negligible_residual_coupling}
\end{equation}
As the chain length increases, splitting noise is exponentially suppressed,
and $\Omega$ is dominated by $\Gamma_0^{\mathrm{min}}$. The return probability then exhibits
a resolved carrier oscillation with angular frequency approximately
$2\Gamma_0^{\mathrm{min}}/\hbar$.

In this case, we describe the decay of the return probability
by the complex average and envelope
\begin{align}
 \mathcal F_{\mathrm{Ramsey}}(t)
 &=\left\langle e^{-2i\Omega t/\hbar}\right\rangle,
 \label{ramsey:carrier_average}\\
 \mathcal C_{\mathrm{Ramsey,osc}}(t)
 &=|\mathcal F_{\mathrm{Ramsey}}(t)|.
 \label{ramsey:carrier_envelope}
\end{align}
The averaged return is
\begin{equation}
 \overline P_{\mathrm{Ramsey}}(t)
 =\frac{1-\mathcal C_{\mathrm{Ramsey,osc}}(t)
 \cos\!\left[\arg\mathcal F_{\mathrm{Ramsey}}(t)\right]}{2}.
 \label{ramsey:return_with_envelope}
\end{equation}
We denote the first $1/e$ crossing of $\mathcal C_{\mathrm{Ramsey,osc}}(t)$ by
$T_{\varphi,\mathrm{osc}}^{\mathrm{Ramsey}}$.
Note that this definition differs from the ideal Ramsey decay:
at $\Gamma_0=0$, $\mathcal F_{\mathrm{Ramsey}}$ averages the phase
of $|\delta\varepsilon_N|$, so its modulus generally differs from
$\mathcal C_{\mathrm{Ramsey}}(t)$.

If the residual coupling dominates the splitting, we expand
\begin{equation}
 \Omega=\Gamma_0^{\mathrm{min}}+\frac{(\delta\varepsilon_N)^2}{2\Gamma_0^{\mathrm{min}}}
 +O\!\left(\frac{(\delta\varepsilon_N)^4}{(\Gamma_0^{\mathrm{min}})^3}\right).
 \label{ramsey:quadratic_energy}
\end{equation}
Besides the deterministic phase due to the residual coupling, the splitting
contributes a quadratic phase. The conditional Gaussian integral is then
precisely Eq.~\eqref{rabi:conditional_phase_integral}, so
\begin{align}
 \mathcal F_{\mathrm{Ramsey}}(t)
 &\simeq e^{-2i\Gamma_0^{\mathrm{min}}t/\hbar}
 g_N\!\left(\frac{\sigma_{\varepsilon,N}^2t}{\Gamma_0^{\mathrm{min}}\hbar};u\right),
 \label{ramsey:quadratic_carrier}\\
 \mathcal C_{\mathrm{Ramsey,osc}}(t)
 &\simeq\left|g_N\!\left(
 \frac{\sigma_{\varepsilon,N}^2t}{\Gamma_0^{\mathrm{min}}\hbar};u\right)\right|.
 \label{ramsey:quadratic_envelope}
\end{align}
The carrier dephasing time is therefore given by the fixed-coupling
Rabi result, Eq.~\eqref{rabi:fixed_coupling_time_asymptotic}.

We can relate this time directly to the ideal Ramsey time.
Within the strong-coupling expansion of
Eq.~\eqref{ramsey:quadratic_energy}, Eqs.~\eqref{rabi:fixed_coupling_time}
and \eqref{ramsey:time_from_crossing} give
\begin{equation}
 \frac{T_{\varphi,\mathrm{osc}}^{\mathrm{Ramsey}}}
 {(T_\varphi^{\mathrm{Ramsey}})^2}
 \simeq\frac{2\Gamma_0^{\mathrm{min}}}{\hbar}
 \exp\!\left[x_{1/e,N}-x_{1/e,N}^{\mathrm{Ramsey}}\right].
 \label{ramsey:carrier_time_ratio}
\end{equation}
The long-chain crossing expressions differ only in their constants, $\operatorname{Re}J=\log2$ and
$J_{\mathrm{Ramsey}}=-\gamma_E$. Thus, we have
$x_{1/e,N}-x_{1/e,N}^{\mathrm{Ramsey}}=\log2+\gamma_E+o(1)$,
and hence
\begin{equation}
 T_{\varphi,\mathrm{osc}}^{\mathrm{Ramsey}}
 \simeq4e^{\gamma_E}\frac{\Gamma_0^{\mathrm{min}}}{\hbar}
 \bigl(T_\varphi^{\mathrm{Ramsey}}\bigr)^2,
 \qquad 4e^{\gamma_E}\simeq7.1243,
 \label{ramsey:carrier_time_asymptotic}
\end{equation}
where $\gamma^E$ is the Euler-Mascheroni constant.

\subsection{Imperfect preparation and readout pulses}
\label{app:ramsey_pulse_errors}

We model pulse imperfections by instantaneous rotations
$R_x(\theta_1)$ and $R_x(\theta_2)$, with
$\theta_j=\pi/2+\delta\theta_j$. We take the angular errors to be
independent centered Gaussians with standard deviations
$\sigma_1,\sigma_2$, independent also of the waiting Hamiltonian.
We further assume that the errors are independent of the waiting time.
We first set the coupling to zero. After preparation the Bloch
vector is $(0,-\sin\theta_1,\cos\theta_1)$, and the wait rotates
its transverse component by $2\delta\varepsilon_Nt/\hbar$.
The second pulse gives
\begin{align}
 P_{\mathrm{Ramsey}}(t)
 =\frac12\Bigl[&1+\sin\delta\theta_1\sin\delta\theta_2\nonumber\\
 &-\cos\delta\theta_1\cos\delta\theta_2
 \cos(2\delta\varepsilon_Nt/\hbar)\Bigr].
 \label{ramsey:pulse_error_return}
\end{align}
Since $\langle\sin\delta\theta_j\rangle=0$ and
$\langle\cos\delta\theta_j\rangle=e^{-\sigma_j^2/2}$,
independence gives
\begin{equation}
 a_\theta=\langle\cos\delta\theta_1\cos\delta\theta_2\rangle
 =e^{-(\sigma_1^2+\sigma_2^2)/2}.
 \label{ramsey:gaussian_pulse_amplitude}
\end{equation}
The pulse and waiting-phase averages therefore factorize:
\begin{equation}
 \overline P_{\mathrm{Ramsey}}(t)
 =\frac{1-a_\theta \mathcal C_{\mathrm{Ramsey}}(t)}{2}.
 \label{ramsey:pulse_averaged_return}
\end{equation}
The baseline remains $1/2$, while the coherence amplitude is
reduced by $a_\theta$. Normalizing by the initial amplitude gives
\begin{equation}
 \frac{1-2\overline P_{\mathrm{Ramsey}}(t)}
 {1-2\overline P_{\mathrm{Ramsey}}(0)}
 =\mathcal C_{\mathrm{Ramsey}}(t),
 \label{ramsey:pulse_normalized_coherence}
\end{equation}
so the dephasing time $T_\varphi^{\mathrm{Ramsey}}$ is unchanged.

With a residual coupling, applying the waiting rotation about
$(\Gamma_0,0,\delta\varepsilon_N)/\Omega$ and averaging the same
independent angular errors gives
\begin{equation}
 \overline P_{\mathrm{Ramsey}}(t)
 =\frac{1-a_\theta\operatorname{Re}\mathcal F_{\mathrm{Ramsey}}(t)}{2},
 \label{ramsey:pulse_residual_return}
\end{equation}
where $\mathcal F_{\mathrm{Ramsey}}$ is the ideal-pulse average in
Eq.~\eqref{ramsey:carrier_average}. The oscillation amplitude is
reduced by the same constant $a_\theta$, so the normalized envelope
$\mathcal C_{\mathrm{Ramsey,osc}}$ and its dephasing time
$T_{\varphi,\mathrm{osc}}^{\mathrm{Ramsey}}$ are unchanged as well.

\section{Probing protection with Rabi oscillations at fixed coupling}
\label{app:rabi_fixed_operating_conditions}

Determining $N_{\varphi,\mathrm{max}}$ requires tuning $\mu_D$, as done recently for $N=2$ experimentally~\cite{zatelli_majorana_2026}.
Alternatively, a Rabi experiment could be performed at fixed coupling strength $\Gamma_0$ with fixed noise width $\sigma_\Gamma$.

In this case, changing the chain length $N$ will change $\delta \varepsilon_N$
exponentially. Thus, the system moves between weak and strong coupling
regimes. When $|\delta \varepsilon_N| \gg \Gamma_0$, the system is in the
weak coupling regime with suppressed visibility. As $N$ increases, the system
enters the strong coupling regime $|\delta \varepsilon_N| \ll \Gamma_0$,
where visibility approaches unity.

Particularly interesting in this regime
is the regime of splitting dominated dephasing when
\begin{equation}
 \sqrt{2\Gamma_0\sigma_\Gamma}
 <|\delta\varepsilon_N|<\Gamma_0.
 \label{rabi:fixed_splitting_window}
\end{equation}
 as this regime naturally probes the
 exponential suppression of the splitting with chain length. In fact, in this
 regime, the coherence time scales as
 \begin{equation}
    T_\varphi\propto \exp(-2\lambda_{\mathrm{typ}}N),
    \label{rabi:fixed_scaling}
\end{equation}
i.e., scaling faster than the maximum cycle count in Eq.~\eqref{main:rabi_optimal_scaling}
with length. The full asymptotic expression is derived in
App.~\ref{app:rabi_fixed_coupling}.
However, this regime is only accessible for very few chain lengths:
From Eq.~\eqref{rabi:fixed_splitting_window} and retaining only
the leading exponential dependence $|\delta\varepsilon_N|_\mathrm{typ}
\propto e^{\lambda_{\mathrm{typ}}N}$, we obtain a window of chain length,
\begin{equation}
 \Delta N\simeq
 \frac{\log[\Gamma_0/(2\sigma_\Gamma)]}
 {2|\lambda_{\mathrm{typ}}|}.
 \label{rabi:fixed_length_window}
\end{equation}
For the parameters in Sec.~\ref{sec:rabi_optimization_crossover} at a detuning
of $\mu_{D,\mathrm{max}}=10\,\mathrm{meV}$, the inner-dot noise dominates the
coupling noise width $\sigma_\Gamma$, giving
$\Delta N\lesssim
 \frac{\log[E_g/(\sqrt2\sigma_\mu)]}
 {|\lambda_{\mathrm{typ}}|}$.
For the chosen parameters, $|\lambda_{\mathrm{typ}}|\simeq2.93$
and $\Delta N\simeq0.83$, leaving room for at most one chain length within
this estimate. The exponential scaling therefore cannot be resolved.

As chain length increases, the system then enters a regime where coupling noise
dominates, $|\delta \varepsilon_N| \ll \sqrt{2\Gamma_0\sigma_\Gamma}$, and coherence
is limited by coupling noise. In this regime, the visibility remains high,
but the coherence time no longer scales with chain length.

\begin{figure}[!t]
 \centering
 \includegraphics[width=\columnwidth]{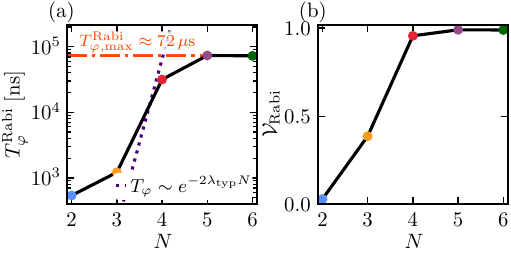}
 \caption{Coherence time and visibility at fixed $\Gamma_0$ for $\mu_D=10\mathrm{meV}$.
  (a) Symbols show the coherence time extracted from the averaged time evolution.
  The dotted line shows the ideal strong-coupling asymptote of
  Eq.~\eqref{rabi:fixed_coupling_time_asymptotic}; the horizontal
  dash-dotted line gives the combined middle-dot and inner-dot noise limit.
  With larger $N$, the system moves from weak to strong coupling.
  This increases the coherence time up to a maximal time depending on the coupling configuration.
  (b) Consequently, the visibility approaches unity.
  Error bars give twice the standard error.
  }
 \label{fig:rabi_fixed_coupling}
\end{figure}

Figure~\ref{fig:rabi_fixed_coupling}(a) and (b) show numerical results
for the dephasing time $T_\varphi$ and visibility as a function of chain length,
for a fixed detuning $\mu_D=\mu_{D,\mathrm{max}}$. As $N$ increases,
the system moves from close to zero visibility at $N=2$ to near unity
at larger $N$, signifying the crossover from weak to strong coupling.
The coherence time also increases, and saturates at a constant value
determined by the coupling noise. We obtain this limit from the $1/e$ crossing
of the product of the Gaussian middle-dot envelope and
$\mathcal{C}_{\mathrm{in}}(t)$ in Eq.~\eqref{main:rabi_interface_envelope}.

We also show the ideal strong-coupling asymptote of
Eq.~\eqref{rabi:fixed_coupling_time_asymptotic} in
Fig.~\ref{fig:rabi_fixed_coupling}(a). Only one chain length lies close to
this asymptote, so the numerical results do not resolve the exponential
scaling. Still, there is a clear improvement in coherence time, although
visibilities close to zero make its experimental extraction difficult.
The experimentally accessible increase is therefore limited to a small
range of chain lengths. This increase reflects both the suppression of the
splitting and the crossover into strong coupling, and therefore does not
directly measure the asymptotic exponential rate. Optimizing $N_\varphi$
(Sec.~\ref{sec:rabi_optimization}) balances coupling and splitting noise,
without requiring splitting noise to dominate. Moreover, the ideal optimum
moves deeper into strong coupling as $N$ increases. Hence, the optimization
of $N_\varphi$ is a more robust way to probe the exponential protection of
the Majorana splitting.

\section{Numerical methods}
\label{app:numerics}

We validate the analytical results by simulating the full Kitaev-chain parity
qubit described by Eq.~\eqref{model:MZM_hamiltonian} for $N=2,\ldots,6$.
We work in the even-parity Hilbert space and define the logical states
\begin{equation}
    |ee\rangle \equiv |e_L 0_D e_R\rangle,
    \qquad
    |oo\rangle \equiv |o_L 0_D o_R\rangle.
\end{equation}
For each disorder realization \(r\), we construct and diagonalize the
time-independent Hamiltonian,
\begin{equation}
    H^{(r)}|E_n^{(r)}\rangle
    =
    E_n^{(r)}|E_n^{(r)}\rangle.
\end{equation}
We use SciPy~\cite{virtanen_scipy_2020} for diagonalization, Sobol sampling,
and root finding.
We use complete dense diagonalization for $N\leq4$; for $N=5,6$, we use
sparse diagonalization and retain the eight lowest eigenstates for
dot-mediated Rabi and the two lowest for direct-tunnel Rabi and Ramsey.
We refine the sparse eigensystems by orthonormalizing the returned
eigenvectors and diagonalizing the Hamiltonian within their span
(Rayleigh--Ritz refinement).
The spectral sums below run over the retained states, without renormalizing
their overlaps with the logical states.

At each operating point, we use $262\,144$ Gaussian disorder realizations
obtained by inverse-CDF transformation of 16 independently scrambled Sobol
sequences~\cite{sobol_distribution_1967,matousek_scrambling_1998}, each
containing $16\,384$ points. We reuse the same realizations
across control settings and keep the disorder fixed during each evolution.

For Rabi oscillations, the logical states are products of the clean
isolated-chain ground states in their respective parity sectors.
For direct tunnelling, we omit the middle dot and couple the inner sites
by normal hopping.
For an initial state \(|\psi_0\rangle=|ee\rangle\), the return and transfer
probabilities are
\begin{align}
    P_{ee}^{(r)}(t)
    &=
    \left|
    \sum_n p_n^{(r)}
    e^{-iE_n^{(r)}t/\hbar}
    \right|^2,
    \\
    P_{oo}^{(r)}(t)
    &=
    \left|
    \sum_n a_n^{(r)}
    e^{-iE_n^{(r)}t/\hbar}
    \right|^2,
\end{align}
where
\begin{equation}
    p_n^{(r)}
    =
    \left|
    \langle E_n^{(r)}|ee\rangle
    \right|^2,
    \qquad
    a_n^{(r)}
    =
    \langle oo|E_n^{(r)}\rangle
    \langle E_n^{(r)}|ee\rangle.
\end{equation}

In the full system, several eigenstates can contribute to the logical
dynamics. Expanding the transfer probability gives
\begin{equation}
    P_{oo}^{(r)}(t)
    =
    \sum_n |a_n^{(r)}|^2
    +
    \operatorname{Re}
    \sum_{n>m}
    c_{mn}^{(r)}
    e^{-i2\pi f_{mn}^{(r)}t},
\end{equation}
with transition frequencies and complex coefficients
\begin{equation}
    f_{mn}^{(r)}
    =
    \frac{E_n^{(r)}-E_m^{(r)}}{2\pi\hbar},
    \qquad
    c_{mn}^{(r)}
    =
    2a_n^{(r)}a_m^{(r)*}.
\end{equation}
We use the transfer probability to compute the qubit properties discussed in the main text.

We quantify visibility by the maximum transfer probability reached within
eight periods of the dominant oscillation, averaged over disorder realizations.
\begin{equation}
    \mathcal{V}
    =
    \left\langle
    \max_{0\leq t\leq 8/f_r}
    P_{oo}^{(r)}(t)
    \right\rangle_r,
\end{equation}
where \(\langle\cdots\rangle_r\) denotes the average over disorder realizations pooled across all scrambles.
We use 256 time steps per period.

Second, we compute the Rabi dephasing time \(T_\varphi^{\mathrm{Rabi}}\) from the oscillating terms in the transfer probability.
For each realization, we calculate the oscillating terms in the transfer probability.
We combine terms with the same frequency by adding their coefficients, then use the frequency $f_r$ and coefficient $c_r$ of the largest resulting term to calculate the dephasing envelope.
We define the normalized empirical envelope
\begin{equation}
    \mathcal C(t)
    =
    \frac{
    \left|
    \left\langle
    c_r e^{-i2\pi f_rt}
    \right\rangle_r
    \right|
    }{
    \left|
    \left\langle c_r\right\rangle_r
    \right|
    }.
\end{equation}
We average the unnormalized complex signals over all realizations before
normalizing the envelope and extracting its dephasing time.
This normalization gives \(\mathcal C(0)=1\). We define the Rabi-envelope decay time as
the first \(1/e\) crossing,
\begin{equation}
    T_\varphi^{\mathrm{Rabi}}
    =
    \inf\left\{
    t>0:
    \mathcal C(t)\leq e^{-1}
    \right\}.
\end{equation}
We find this crossing using an adaptive time search followed by a bracketed
root finder, without fitting a decay function.

Finally, we calculate the nominal cycle count using the frequency
$f_q=\Gamma_0/(\pi\hbar)$,
\begin{equation}
    N_\varphi=f_q T_\varphi^{\mathrm{Rabi}}
    =\frac{\Gamma_0 T_\varphi^{\mathrm{Rabi}}}{\pi\hbar},
\end{equation}
where $\Gamma_0=t_{LD}t_{RD}/(2\mu_D)$ for dot-mediated coupling and
$\Gamma_0=\tau_0/2$ for direct tunnelling.

In the weak-coupling regime, only a small fraction of disorder realizations
have appreciable oscillation amplitudes: most realizations are far from
resonance and contribute little to the averaged signal. A large total
sample count can therefore correspond to few effectively contributing
realizations. We quantify this using
\begin{equation}
    N_{\mathrm{eff}}
    =\frac{\left(\sum_r|c_r|\right)^2}{\sum_r|c_r|^2}.
\end{equation}
For the direct-tunnel Rabi curves, we retain only operating points with
$N_{\mathrm{eff}}\geq1000$ and a resolved dephasing-time crossing.

For Ramsey evolution, we construct $|ee\rangle$ and $|oo\rangle$ for each
disorder realization from the lowest even- and odd-parity states of the
noisy isolated chains, with the middle dot empty.
We initialize in $|ee\rangle$ and apply an ideal $R_x(\pi/2)$ pulse,
preparing $|-y\rangle=(|ee\rangle-i|oo\rangle)/\sqrt2$.
The same disorder realization remains fixed during the waiting interval.
A second ideal $R_x(\pi/2)$ pulse followed by measurement of the
$|ee\rangle$ population corresponds to readout onto
$|+y\rangle=(|ee\rangle+i|oo\rangle)/\sqrt2$ before that pulse.
We compare disconnected chains with a residual dot-mediated coupling at
$\mu_D=10\,\mathrm{meV}$ and $t_{LD}=t_{RD}=7.5\,\mu\mathrm{eV}$.
For disconnected waits and connected waits with $N\leq3$, we extract the
first $1/e$ crossing of the averaged contrast
$1-2\overline P_{\mathrm{Ramsey}}(t)$; for connected waits with $N\geq4$,
we use the normalized envelope of the averaged complex carrier.
These lifetime estimates use the lowest eigenstate pair, while the
$N=3,4$ return-probability traces use the complete spectrum.

\bibliography{bib}

\end{document}